\documentclass[preprint,prd,aps,showpacs,nofootinbib]{revtex4}
\usepackage{inputenc}
\usepackage{epsfig,float}
\usepackage{amsfonts}
\usepackage[hypertex]{hyperref}
\usepackage{amsmath}

\input{epsf}
\newcommand{\im}{{\rm Im}}
\newcommand{\re}{{\rm Re}}
\newcommand{\be}{\begin{eqnarray}}
\newcommand{\ee}{\end{eqnarray}}
\newcommand{\ba}{\begin{array}}
\newcommand{\ea}{\end{array}}

\restylefloat{figure}
\restylefloat{table}

\begin{document}

\title{
A consistent model for $\pi N$ transition distribution
amplitudes and backward pion electroproduction}

\author{J.~P.~Lansberg$^1$, B.~Pire$^2$,  K.~Semenov-Tian-Shansky$^{2,3,4}$, L.~Szymanowski$^{5}$ }
\affiliation{$^1$ IPNO,   Universit\'{e} Paris-Sud, CNRS/IN2P3, 91406 Orsay, France \\
$^2$ CPhT, \'{E}cole Polytechnique, CNRS,  91128 Palaiseau, France  \\
$^3$ LPT,  Universit\'{e} Paris-Sud, CNRS, 91404 Orsay, France \\
$^4$ IFPA, d\'{e}partement AGO,  Universit\'{e} de  Li\`{e}ge, 4000 Li\`{e}ge,  Belgium \\
$^5$ National Center for Nuclear Research (NCBJ), Warsaw, Poland
}

\preprint{CPHT-RR100.1111, LPT-11-114}

\begin{abstract}
The extension of the concept of generalized parton
distributions leads to the  introduction of baryon to meson
transition distribution amplitudes (TDAs), non-diagonal matrix
elements of the nonlocal three quark operator between a nucleon and
a meson state.
We present a general framework for  modelling nucleon to pion ($\pi N$) TDAs.
Our main tool is the spectral representation for
$\pi N$
TDAs in terms of quadruple distributions. We propose a
factorized Ansatz for quadruple distributions with input from the soft-pion theorem for
$\pi N$
TDAs.  The spectral representation is complemented with a
$D$-term like contribution from  the nucleon exchange in the cross channel.
We then study  backward pion electroproduction
in the QCD collinear factorization approach in which the non-perturbative
part of the amplitude involves $\pi N$ TDAs.
Within our two component model for $\pi N$ TDAs we update previous leading-twist estimates of the unpolarized cross
section.
Finally, we compute the transverse target single spin asymmetry as a function of skewness. We find it to be sizable
in the valence region and sensitive to the phenomenological input of our $\pi N$ TDA model.

\end{abstract}

\maketitle
\thispagestyle{empty}
\renewcommand{\thesection}{\arabic{section}}

\renewcommand{\thesubsection}{\arabic{subsection}}

\section{Introduction}
\label{Section_Intro}

\begin{figure}[h]
 \begin{center}
 \epsfig{figure=  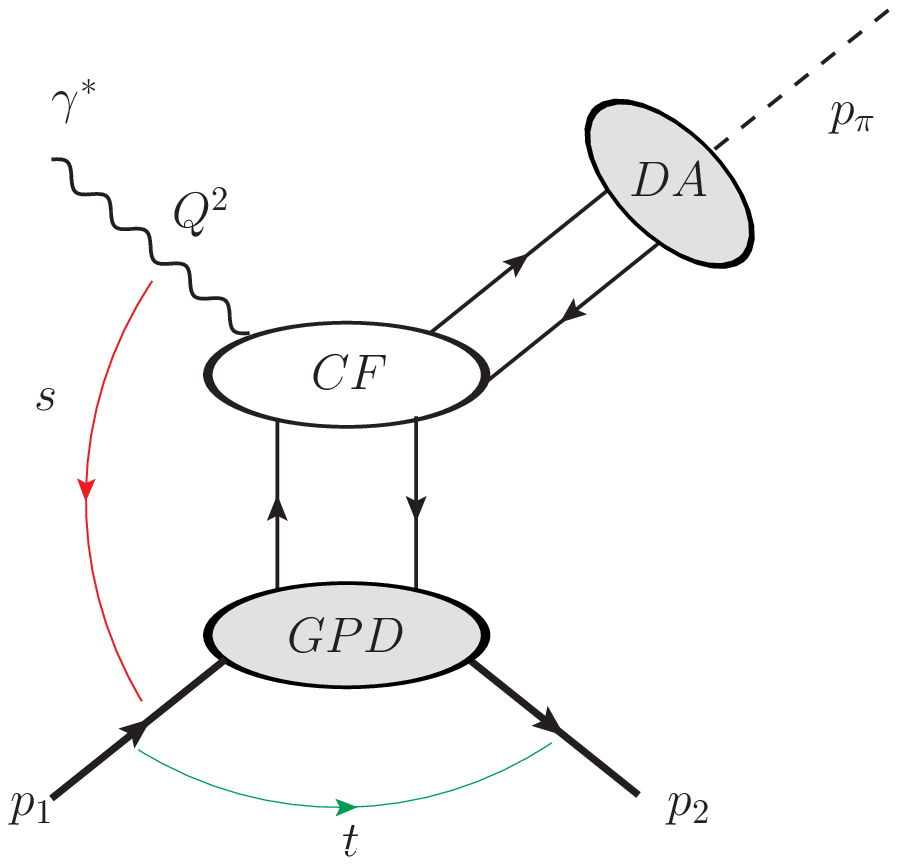 , height=6cm} \ \ \ \ \ \ \ \
  \epsfig{figure= 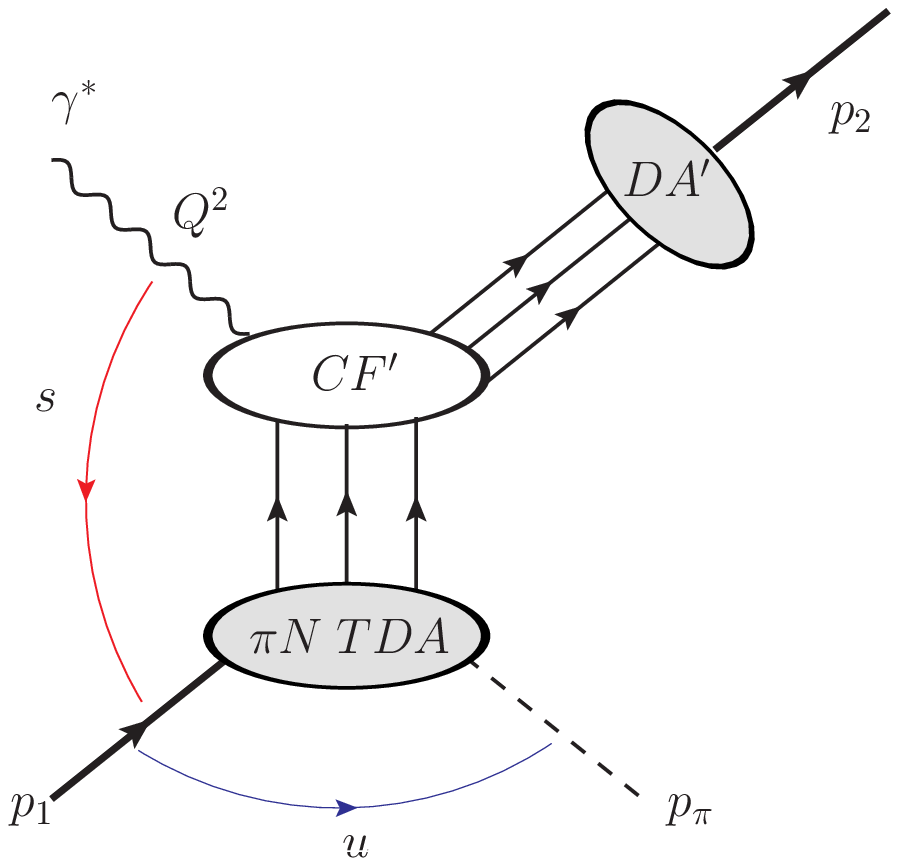 , height=6cm}
     \caption{{\bf Left:} Collinear factorization for hard production of pions in the
     conventional hard meson production kinematics.
    {\bf Right:} Collinear factorization for hard production of pions off nucleons in the
     backward kinematics.
      }
\label{Fig1}
\end{center}
\end{figure}

The familiar collinear factorization theorem
\cite{Collins:1996fb,Radyushkin:1997ki}
for exclusive electroproduction of pions off nucleons
\be
e(k)+ N(p_1) \rightarrow \big( \gamma^*(q) + N(p_1) \big) +e(k')
  \rightarrow  e(k')+ \pi(p_\pi) + N'(p_2),
\label{reaction}
\ee
valid in the generalized Bjorken limit
(large $Q^2=-q^2$ and $s \equiv (p_1+q)^2$;
$x_{B}=\frac{Q^2}{2 p_1 \cdot q}$ and skewness variable $\xi= -\frac{(p_2-p_1) \cdot n}{(p_1+p_2) \cdot n}$  being fixed%
\footnote{
$n$
is the conventional light-cone vector occurring in the Sudakov decomposition of the relevant momenta.};
and small
$-t \equiv (p_2-p_1)^2$)
gives rise to the description of this reaction in terms
of the generalized parton distributions (GPDs) (see left panel of
Fig.~\ref{Fig1}) .

According to a conjecture made in
\cite{Frankfurt:1999fp,Frankfurt:2002kz},
a similar collinear factorization theorem for the reaction
(\ref{reaction})
should be valid in the following complementary kinematical regime:
\begin{itemize}
\item large $Q^2$
and
$s$;

\item fixed $x_{B}$
and   skewness variable
$\xi$,
which is now defined with respect to the
$u$-channel momentum transfer:
\be
\xi= -\frac{(p_\pi-p_1) \cdot n}{ ( p_\pi+p_1) \cdot n};
\label{Def_xi_TDA}
\ee
%
\item the $u$-channel momentum transfer squared
$u \equiv (p_\pi-p_1)^2$
(rather than
$t$)
is small compared to
$Q^2$
and
$s$.
\end{itemize}
Under these assumptions, the amplitude of the reaction
(\ref{reaction})
factorizes as it is shown on the right panel of Fig.\ref{Fig1}.
This requires the introduction of supplementary non-perturbative objects in addition
to GPDs -- nucleon to pion transition distribution amplitudes
($\pi N$ TDAs).
Technically, $\pi N$
TDAs are defined through  the
$\pi N$
matrix element of the tri-local three quark operator on the light cone
  \cite{Radyushkin:1977gp,Efremov:1978rn,Lepage:1980,Chernyak:1983ej,Chernyak_Nucleon_wave}:
\be
&&
\widehat{O}^{\alpha \beta \gamma}_{\rho \tau \chi}( \lambda_1 n,\, \lambda_2 n, \, \lambda_3 n)
\nonumber \\ &&
 =
\varepsilon_{c_{1} c_{2} c_{3}}
\Psi^{c_1 \alpha}_\rho(\lambda_1 n) [\lambda_1 n;\lambda_0 n ] \Psi^{c_2 \beta}_\tau(\lambda_2 n) [\lambda_2 n;\lambda_0 n ] \Psi^{c_3 \gamma}_\chi (\lambda_3 n) [\lambda_3 n;\lambda_0 n ]\,.
\label{oper}
\ee
Here
$\alpha$, $\beta$, $\gamma$
stand for quark flavor indices and
$\rho$, $\tau$, $\chi$
denote the Dirac spinor indices. Antisymmetrization stands over the color group indices $c_{1,2,3}$. Gauge links may be omitted in  the light-like gauge
$A \cdot n=0$.

The detailed account of this approach is presented in
Refs.~\cite{Lansberg:2007ec,Pire:2010if,Pire:2011xv}.
Apart from the description of hard exclusive pion
electroproduction off a nucleon in the backward region,
the same non-perturbative objects appear  in the
collinear factorized description of different
exclusive reactions. Prominent examples are
baryon-antibaryon annihilation into a pion and a lepton pair in the
forward and backward directions
\cite{Pire:2005mt,Pire:2005ax,LPS}.


The physical picture encoded in baryon to meson TDAs is conceptually close to that
contained in baryon GPDs and baryon distribution amplitudes (DAs). Baryon to meson TDAs  are matrix elements of a three
quark operator ({\it i.e.} with baryonic number one)
and characterize partonic correlations inside a baryon. This gives access to the momentum distribution
of the baryonic number inside a nucleon. The same operator also defines the nucleon DA
which can be seen as a limiting case of baryon to meson TDAs with the meson state replaced by the vacuum.
In the language of the Fock state decomposition, baryon to meson TDAs are not restricted to the lowest Fock state
as DAs. They rather probe the non-minimal Fock components with additional
quark-antiquark pair:
\be
&&
| {\rm Nucleon} \rangle= |\Psi \Psi \Psi \rangle+ |\Psi \Psi \Psi; \,  \bar{\Psi} \Psi \rangle+....\;; \nonumber \\ &&
| {\rm Meson} \rangle= |\bar{\Psi}\Psi \rangle+ |\bar{\Psi}\Psi; \, \bar{\Psi} \Psi \rangle+....\;.
\ee

\begin{figure}[h]
 \begin{center}
 \epsfig{figure=  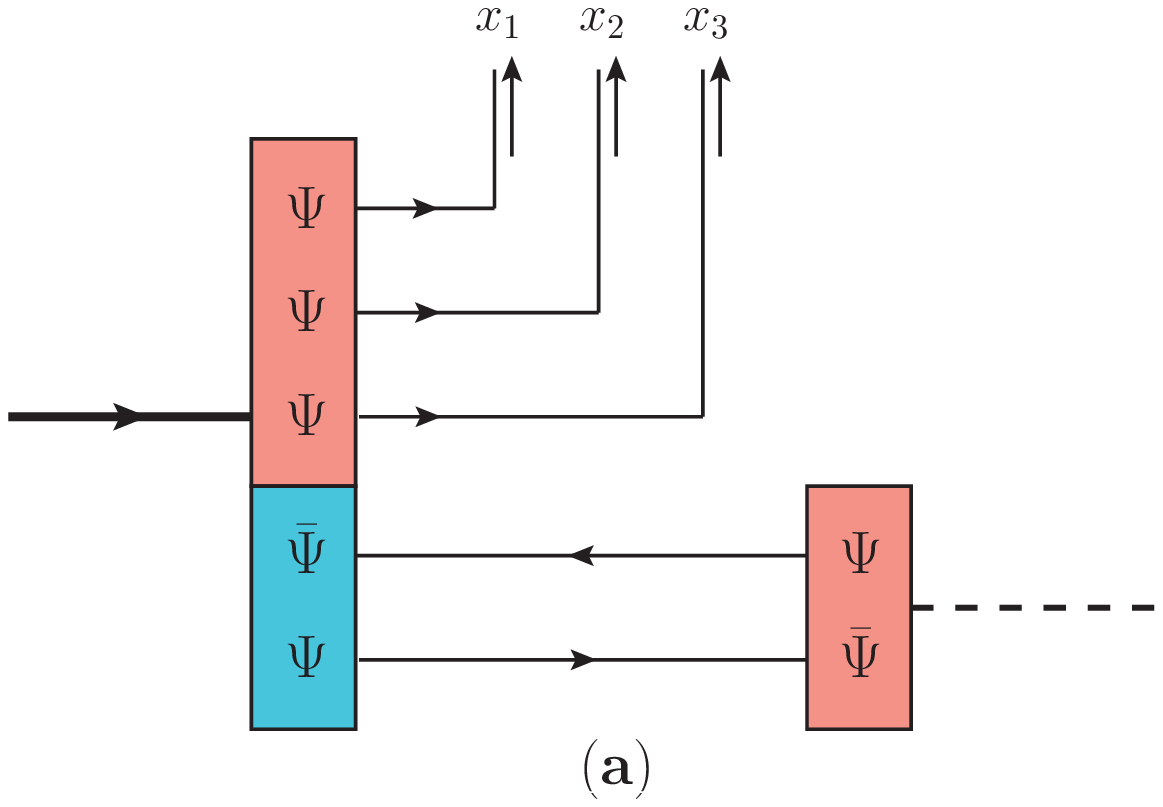 , height=5cm}   \ \ \ \ \ \ \
 \epsfig{figure=  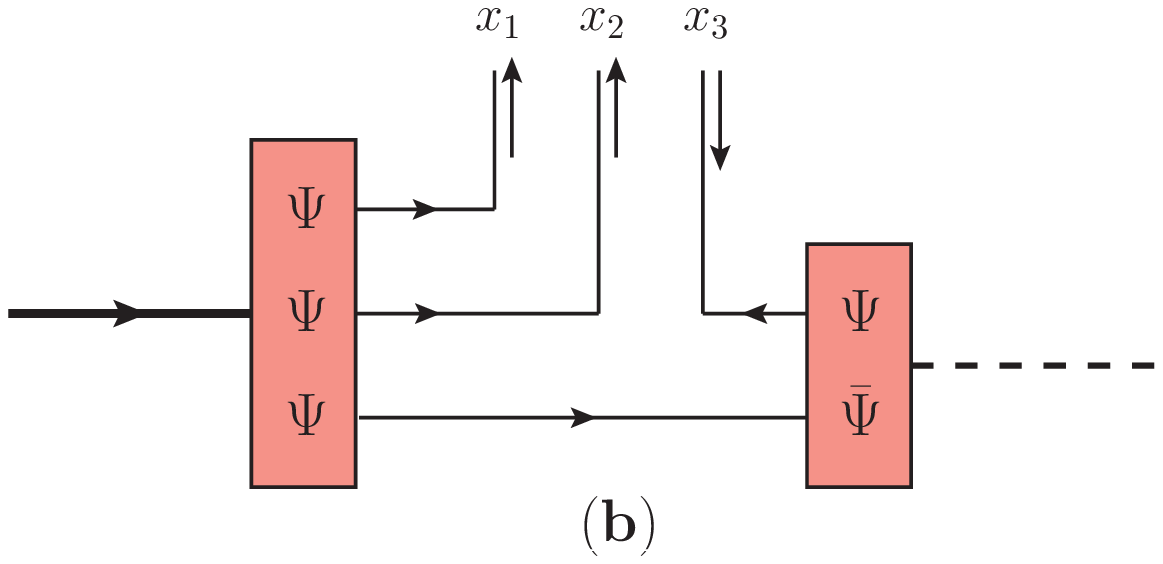 , height=3.5cm}  \\
    \epsfig{figure=  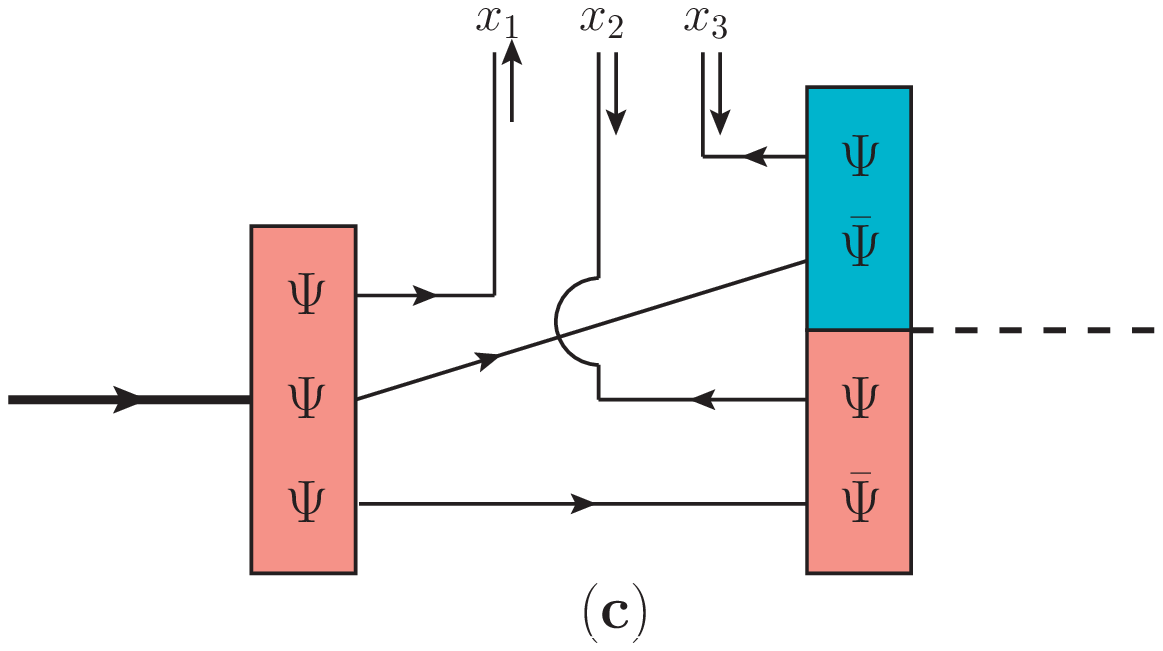 , height=4cm}
\end{center}
    \caption{Interpretation of $\pi N$ TDAs within the light-cone quark model \cite{Pasquini:2009ki}.
    Small vertical arrows show the flow of the momentum.
     {\bf  (a):} Contribution in the ERBL region (all $x_i$ are positive);
    {\bf  (b):} Contribution in the DGLAP I region (one of $x_i$  is negative).
    {\bf  (c):} Contribution in the DGLAP II region (two  $x_i$  are negative).
     }
 \label{Fig_X}
\end{figure}

For baryon to meson TDAs
one may distinguish the
Efremov-Radyushkin-Brodsky-Lepage (ERBL)-like domain in which all three  momentum
fractions of quarks
are positive and two kinds of Dokshitzer-Gribov-Lipatov-Altarelli-Parisi (DGLAP)-like regions
in which either one or two momentum
fractions of quarks are negative.
On Fig.~\ref{Fig_X}, we show the interpretation of $\pi N$ TDAs in the
ERBL-like and in DGLAP-I,II region within the light-cone quark model \cite{Pasquini:2009ki}.
As one can see from Fig.\ref{Fig_X}~(a) the ERBL part is probing the non-minimal Fock components of the
nucleon wave function. In the DGLAP-II Fig.~\ref{Fig_X}~(c) region one rather probes the non-minimal Fock components
of the meson state, while in the DGLAP-I Fig.~\ref{Fig_X}~(b) region there is a non-vanishing contribution of the
minimal Fock states of baryon and meson. This interpretation, obviously, is justified only at a very low normalization
scale. The evolution effects may significantly  change it at higher scales.

Similarly to GPDs
\cite{Burkardt:2000za,Burkardt:2002ks,Diehl:2002he},
by Fourier transforming $\pi N$ TDAs to the
impact parameter space
($\Delta_T \to b_T$),
one obtains additional insight on the nucleon structure
in the transverse plane.
This allows one to perform the femto-photography of hadrons
\cite{Ralston:2001xs}
from a new perspective.
In particular, there are hints
\cite{Strikman:2009bd}
that $\pi N$ TDAs
may be used as a tool to perform spatial imaging of the structure of the meson cloud of the nucleon.
This point, which still awaits a detailed exploration, opens a fascinating window for the investigation of
the various facets of the nucleon interior.

Our paper is organized as follows:
In Sec.~\ref{Sec_general_prop},
we provide a short summary of the basic properties of
$\pi N$
TDAs.
In Sec.~\ref{Section_Spectral_Rep}, we review the spectral representation
for $\pi N$ TDAs and propose a factorized Ansatz
for the quadruple distributions constrained by the soft pion theorem for $\pi N$ TDAs.
We complement the spectral representation with a $D$-term like contribution and build
a two component model for $\pi N$ TDAs.
Sec.~\ref{Section_amplitude} contains the details of calculation of $\gamma^* N \to \pi N$ amplitude
in the backward regime.
In Sec.~\ref{Sec_CS_AS},
we compute the unpolarized cross section and
transverse target single spin asymmetry
of backward $\pi^+$ and $\pi^0$ electroproduction  off protons
using our two component model for $\pi N$ TDAs.
Many technical details  are relegated to Appendices A-D.
Our conclusions are presented in Sec.~\ref{Section_conclusions}.

\section{General properties of $\pi N$ TDAs}
\label{Sec_general_prop}

At the leading twist-$3$, the parametrization of the Fourier transform
of $\pi N$ matrix elements of the three-local light-cone quark operator (\ref{oper})
can be written as
\be
&&
4(P \cdot n)^3 \int   \left[ \prod_{j=1}^3 \frac{d \lambda_j}{2 \pi}   \right]
e^{i \sum_{k=1}^3 x_k \lambda_k (P \cdot n)}
\langle \pi_a(p_\pi)| \widehat{O}_{\rho \, \tau \, \chi}^{\alpha \beta \gamma} (\lambda_1 n, \,\lambda_2 n, \, \lambda_3 n )| N_\iota(p_1)  \rangle
\nonumber \\ &&
=\delta(x_1+x_2+x_3-2 \xi) \; \sum_{s.f.} (f_a)^{\alpha \beta \gamma}_\iota s_{\rho \, \tau, \, \chi} H^{(\pi N)}_{s.f.}(x_1,x_2,x_3,\xi, \Delta^2;\mu^2_F)\,,
\label{Formal_definition_TDA}
\ee
where
$P=\frac{p_1+p_\pi}{2}$ is the average momentum
and
$\Delta=p_\pi-p_1$
is the
$u$-channel momentum transfer.
The spin-flavor ($s.f.$) sum in
(\ref{Formal_definition_TDA})
stands over all independent
flavor structures
$(f_a)^{\alpha \beta \gamma}_\iota$
and the  Dirac structures
$s_{\rho \, \tau, \, \chi}$  relevant at the leading twist;
$\iota$ ($a$) is the nuclon (pion) isotopic index.
The invariant amplitudes
$H^{(\pi N)}_{s.f.}$,
which are often referred to as the leading twist $\pi N$ TDAs,
are functions of  the light-cone momentum fractions
$x_i$ ($i=1,2,3$),
the skewness variable
$\xi$
(\ref{Def_xi_TDA}),
the $u$-channel momentum-transfer squared
$\Delta^2$
and the factorization scale $\mu_F$.

For given isotopic contents (say proton to $\pi^0$ TDA),
the   parametrization (\ref{Formal_definition_TDA}) of twist-$3$
$\pi N$
TDAs involves
$8$
invariant
functions
$V_{1,2}^{(\pi N)}$, $A_{1,2}^{(\pi N)}$, $T_{1,2,3,4}^{(\pi N)}$ (see Eq.~(\ref{Decomposition_piN_TDAs_new})).
However, not all of them are independent. Taking into account  the isotopic and permutation symmetries
(see \cite{Pire:2011xv}),
one may check that in order to describe all isotopic channels of the reaction
(\ref{reaction}),
it suffices to introduce eight independent
$\pi N$ TDAs:
four in the isospin-$\frac{1}{2}$
channel and four
in the isospin-$\frac{3}{2}$ channel.
This result is analogous to the case of leading twist nucleon DAs: initially,
the parametrization
\cite{Chernyak_Nucleon_wave}
involves three proton DAs
$V^p$, $A^p$ and $T^p$.
However, due to the isotopic  and permutation symmetries, these three functions may be expressed through the
unique leading twist nucleon DA
$\phi_N  \equiv V^p-A^p$.
Neutron DAs
are expressed as
$\left\{ V^n,\,A^n,\,T^n \right\}= \left\{ -V^p,\,-A^p,\,-T^p \right\}$.

The support domain of
baryon to meson
TDAs in the light-cone momentum fractions
$x_i$  ($\sum_i x_i=2\xi$)
was established in
\cite{Pire:2010if}.
It is given by the intersection of three stripes
$-1+\xi \le x_i \le 1+\xi$.
Instead of dealing with three dependent light-cone momentum fractions $x_i$,
one can switch to the independent variables.
A convenient choice of independent variables is the use
of the so-called quark-diquark coordinates \cite{Pire:2010if}.
Due to the symmetry of the support of baryon to meson TDAs under rotations by the
$\frac{2 \pi}{3}$ angle,
%
there exist three equivalent choices of
quark-diquark coordinates ($i=\{1,\,2,\,3\}$):
\be
w_i= x_i-\xi; \ \ \ v_i= \frac{1}{2} \sum_{k,l=1}^3 \varepsilon_{ikl} \, x_k,
\label{Quark-diquark_coordinates_compact}
\ee
where
$\varepsilon_{ikl}$ is the totally antisymmetric tensor.
The support domain of baryon to meson TDAs in terms
of quark-diquark coordinates can be parameterized as:
\be
-1 \le w_i \le 1\,; \ \ \ \ -1+|\xi-\xi'_i| \le v_i \le 1- |\xi - \xi'_i|\,,
\label{The_whole_domain}
\ee
where
$\xi'_i \equiv \frac{\xi-w_i}{2}$.

As pointed out in
\cite{Pire:2005ax},
the scale dependence of
$\pi N$
TDAs is described by the appropriate generalization of the ERBL/DGLAP evolution equations.
Splitting functions in this case turn out to be much more complicated, as they include   different pieces
in different (ERBL-like and DGLAP-like) kinematical regions.

Exactly as for the case of the usual parton distributions and GPDs,
evolution of $\pi N$ TDAs can also be  treated in terms of renormalization
of local operators corresponding to the Mellin  moments of
$\pi N$ TDAs in
$x_i$.
Evolution properties of the local operators in question were extensively studied in
connection with  the scale dependence of nucleon DAs (see Refs.
\cite{Braun:1999te,Stefanis:1999wy}).
The conformal partial wave expansion of $\pi N$ TDAs over the conformal basis 
of
the Appel polynomials  or the Jacobi and Gegenbauer polynomials
represents an alternative strategy for the parametrization of $\pi N$ TDAs in the
spirit of the dual representation of GPDs
\cite{DualVSRad,arXiv:1001.2711}
or complex conformal partial wave expansion
\cite{hep-ph/0509204}.

Similarly to the GPD case, the  polynomiality property in $\xi$ of the
Mellin moments of
$\pi N$
TDAs in the light-cone momentum fractions $x_i$
is the direct consequence of the underlying Lorentz symmetry.
For
$n_1+n_2+n_3=N$
the highest power of
$\xi$ occurring in
the $(n_1,n_2,n_3)$-th Mellin moment
of  $V_{1,2}^{\pi N}$,  $A_{1,2}^{\pi N}$, $T_{1,2}^{\pi N}$ is $N+1$, while for
$T_{3,4}^{\pi N}$ it is $N$.
Consequently, TDAs
$V_{1,2}^{\pi N}$,  $A_{1,2}^{\pi N}$, $T_{1,2}^{\pi N}$
include an analogue of the $D$-term contribution
\cite{Polyakov:1999gs},
which generates the highest possible power of
$\xi$.

The most direct way to ensure both the polynomiality and the support properties for
$\pi N$
TDAs is to employ the spectral representation in terms of quadruple distributions,
generalizing the familiar Radyushkin's double distribution representation for GPDs
\cite{RDDA1,RDDA2,RDDA3,RDDA4}.
In phenomenological applications, an inviting strategy, which proved to be successful in the case of GPDs,
is to construct a factorized Ansatz for the corresponding
spectral densities.
However, contrary to the GPD case,
$\pi N$
TDAs lack a comprehensible forward
limit,
$\xi\to0$.
This hampers the construction of the hypothetic factorized Ansatz for
quadruple distributions with input at
$\xi=0$  as suggested in \cite{Pire:2010if}.

In this paper, we  build up a consistent model for $\pi N$ TDAs relying
on their chiral properties. Chiral dynamics constrains $\pi N$ TDAs in the opposite limit, $\xi \to 1$.
Indeed, $\pi N$ TDAs are conceptually much related to pion-nucleon generalized distribution amplitudes
($\pi N$ GDAs),
which are defined through the cross-conjugated matrix element of the same three quark operator
(\ref{oper}):
\be
\langle 0| \widehat{O}_{\rho \, \tau \, \chi}^{\alpha \beta \gamma}
(\lambda_1 n, \,\lambda_2 n, \, \lambda_3 n )| N_\iota(p_1) \pi_a(-p_\pi)  \rangle\,.
\ee
A similar correspondence was previously established
between pion GPDs and
$2 \pi$
GDAs
\cite{Polyakov:1999gs,Polyakov:1998ze}.
For simplicity, let us consider the pion to be massless
($m=0$).
In this case the point
$\xi=1$, $\Delta^2=M^2$,
where
$M$
stands for the nucleon mass,
belongs  simultaneously  to both physical regions: that of
$\pi N$
GDAs and that of  $\pi N$ TDAs (see discussion in \cite{Pire:2011xv}).
Moreover, it is at this very point that the soft-pion theorem
\cite{Pobylitsa:2001cz}
applies for
$\pi N$
GDAs. As argued in
\cite{BLP1,BLP2},
this allows us to express
$\pi N$
GDAs at the threshold in terms of the nucleon DAs
$V^p$, $A^p$ and $T^p$.
In the chiral limit, the  soft-pion theorem for GDAs also constrains
$\pi N$
TDAs similarly to the way
\cite{Kivel:2002ia}
the  soft-pion theorem
\cite{Polyakov:1998ze}
for
$2 \pi$
GDAs in the chiral limit links the isovector pion GPD at
$\xi=1$, $\Delta^2=0$
to the pion DA
$\varphi_\pi$.

In the chiral limit, the soft-pion theorem thus provides  the normalization point for
$\pi N$
TDAs.
The explicit form of the soft-pion theorem for $\pi N$ TDAs was established in
\cite{Pire:2011xv}.
In this paper, we use this information as input for  realistic modelling of
$\pi N$
TDAs based on the spectral representation in terms of quadruple distributions.

\section{Spectral representation for $\pi N$ TDAs, factorized Ansatz for quadruple distributions and $D$-term }
\label{Section_Spectral_Rep}

In this section, we propose a two component model for  $\pi N$ TDAs involving the following contributions:
\begin{enumerate}
\item a spectral representation with input fixed at $\xi=1$ from the soft-pion theorem;
\item a nucleon-pole exchange in the $u$-channel which is a pure $D$-term like contribution
complementary to the spectral representation.
\end{enumerate}
To do so, we formulate the spectral representation constructed in
\cite{Pire:2010if}
in a way suitable  for the implementation of chiral constraints for $\pi N$ TDAs.
This allows us to propose a factorized Ansatz for quadruple distributions with input from the soft-pion theorem.

\subsection{Toy exercise:   GPD case}
\label{Subsection_Toy}

To give an idea of the new type of factorized Ansatz for quadruple distributions,
let us first consider how one can constrain a GPD model from the
$\xi=1$
limit rather than from the forward limit
$\xi=0$.
Let us consider the standard double distribution (DD) representation for GPDs \cite{RDDA1,RDDA2,RDDA3,RDDA4}:
\be
H(x,\,\xi)= \int_{ \Omega} d \beta d \alpha \, \delta(x-\beta-\alpha \xi) f(\beta,\,\alpha),
\label{DD_Rad}
\ee
where
$\Omega$
is the usual domain in the spectral parameter space
\be
\Omega = \{ |\beta | \le 1\,; \ \ |\alpha | \le 1-|\beta|  \},
\label{Domain_DD}
\ee
and
$f(\beta, \alpha)$
is the double distribution.  Let us perform the change of variables:
$\alpha= \frac{\kappa+\theta}{2}$, $\beta= \frac{\kappa-\theta}{2}$.
This gives:
\be
H(x,\,\xi)= \int_{-1}^1 d \kappa \int_{-1}^1 d \theta  \,
\delta  \! \! \left(
x+ \frac{1-\xi}{2} \, \theta - \frac{1+\xi}{2} \, \kappa
\right) \frac{1}{2} \, F(\kappa,\, \theta)\,,
\label{Spectral_GPD_xi=1}
\ee
where
$ F(\kappa,\, \theta) \equiv f \Big(\frac{\kappa-\theta}{2}, \, \frac{\kappa+\theta}{2} \Big)$.
Instead of the usual factorized Ansatz in
(\ref{DD_Rad})
in the variables
$\{  \beta, \, \alpha \}$,
\be
f(\beta,\alpha)=h(\beta,\alpha) q(\beta),
\ee
let us employ in
(\ref{Spectral_GPD_xi=1})
the following factorized Ansatz in the variables
$\{\kappa, \, \theta \}$:
\be
F(\kappa,\, \theta)= 2 \varphi(\kappa) h(\theta),
\label{Fact_Anz_GPDs}
\ee
with the profile $h(\theta)$ normalized according to
\be
\int_{-1}^1 d \theta h(\theta)=1\,.
\ee
Obviously,
Eq.~(\ref{Spectral_GPD_xi=1}) then
gives
$H(x,\xi=1)=\varphi(x)$.

In order to implement the so-called ``Munich symmetry''
\cite{Mankiewicz:1997uy}
$f(\beta,\alpha)=f(\beta,\,-\alpha)$,
which is the consequence of hermiticity and time-reversal invariance
of non-forward matrix element entering the definition of GPDs, one should require that
\be
h(\theta) \equiv  \varphi(-\theta) \left[ \int_{-1}^1 d \theta \varphi(-\theta) \right]^{-1}.
\ee
Let us emphasize that, in the GPD case, symmetry requirements unambiguously fix the shape of the profile $h$ in the
factorized Ansatz (\ref{Fact_Anz_GPDs}).

Although it is but a toy model, the factorized Ansatz
(\ref{Fact_Anz_GPDs})
may  be applied to the case of quark isovector GPD of a pion
$H_\pi^{u-d}$,
which in the soft-pion limit
\cite{Polyakov:1998ze}
reduces to the pion DA
$\varphi_\pi$:
\be
\lim_{\xi \to 1}H_\pi^{u-d}(x,\xi,t=0)=\varphi_\pi(x)\,.
\ee

\subsection{An alternative form of the spectral representation for $\pi N$ TDAs}
\label{Subsection_Fact_Anz_TDAs}

Let us now apply the trick of previous subsection  for the case of
$\pi N$
TDAs.
According to
\cite{Pire:2010if},
the spectral representation for
$\pi N$
TDAs reads:
\be
&&
H^{(\pi N)}(x_1,\,x_2,\,x_3=2 \xi -x_1-x_2,\,\xi) \nonumber \\ && =
\left[
\prod_{i=1}^3
\int_{\Omega_i} d \beta_i d \alpha_i
\right]
\delta(x_1-\xi-\beta_1-\alpha_1 \xi) \,
\delta(x_2-\xi-\beta_2-\alpha_2 \xi) \,
\nonumber \\ &&
\times
\delta(\beta_1+ \beta_2+ \beta_3)
\delta(\alpha_1+\alpha_2+\alpha_3+1)
 f(\beta_1, \, \beta_2, \, \beta_3, \, \alpha_1, \, \alpha_2, \alpha_3),
\label{Spectral_for_GPDs_x123}
\ee
where
$\Omega_{i}$
denote three copies of the usual domain
(\ref{Domain_DD})
in the spectral parameter space.
The spectral density
$f$
is an arbitrary function of six variables,
which are  subject to two constraints imposed by the
two last delta functions in  eq.~(\ref{Spectral_for_GPDs_x123}).
Therefore,
$f$
is effectively a quadruple distribution. The spectral representation
(\ref{Spectral_for_GPDs_x123})
by the very construction ensures the polynomiality and the support properties
of $\pi N$ TDAs.

Let us employ the particular choice of the
quark-diquark coordinates
$(w_i, \, v_i)$
(\ref{Quark-diquark_coordinates_compact})
and introduce the following combinations of the spectral parameters:
\be
&&
\kappa_i= \alpha_i+ \beta_i \,; \ \ \  \theta_i=\frac{1}{2} \sum_{k,l=1}^3 \varepsilon_{i k l}(\alpha_k+\beta_k); \nonumber \\ &&
\mu_i= \alpha_i- \beta_i\,; \ \ \  \lambda_i= \frac{1}{2} \sum_{k,l=1}^3 \varepsilon_{i k l}(\alpha_k-\beta_k).
\ee
The spectral representation (\ref{Spectral_for_GPDs_x123}) can then be rewritten as:
\be
&&
H(w_i,\,v_i,\,\xi)=
\int_{-1}^1 d \kappa_i \int_{- \frac{1-\kappa_i}{2}}^{ \frac{1-\kappa_i}{2}} d\theta_i
\int_{-1}^1 d \mu_i \int_{- \frac{1-\mu_i}{2}}^{ \frac{1-\mu_i}{2}} d\lambda_i
\, \delta(w_i- \frac{\kappa_i-\mu_i}{2} (1-\xi) - \kappa_i \xi)  \nonumber \\ &&
\times
\delta\left(v_i- \frac{\theta_i-\lambda_i}{2} (1-\xi) - \theta_i \xi \right) \, \frac{1}{4} F_i(\kappa_i, \, \theta_i,\, \mu_i,\, \lambda_i ).
\label{Spectral_for_GPDs_kappa_theta}
\ee
The working formulas for the calculation of $\pi N$ TDAs in the ERBL-like and DGLAP-like domains are summarized in
Appendix~\ref{App_Calc_TDA}.

We suggest   using the following factorized Ansatz for the quadruple distribution
$F_i$
in
(\ref{Spectral_for_GPDs_kappa_theta}):
\be
F_i(\kappa_i, \, \theta_i,\, \mu_i,\, \lambda_i )= 4 V(\kappa_i, \, \theta_i) \, h(\mu_i,\, \lambda_i),
\label{Factorized_ansatz_xi=1}
\ee
with the profile function
$h(\mu_i,\, \lambda_i)$
normalized as
\be
\int_{-1}^1 d \mu_i \int_{- \frac{1-\mu_i}{2}}^{ \frac{1-\mu_i}{2}} d\lambda_i \, h(\mu_i,\, \lambda_i) =1.
\label{Norm_profile}
\ee
Note that the support of the profile function $h$ is that of a baryon DA.

With the help  of the spectral representation
(\ref{Spectral_for_GPDs_kappa_theta}),
one can check that in the limit $\xi \rightarrow 1$
$H_i$  now reduces to
\be
H(w_i,\,v_i,\,\xi=1)= V(w_i, \, v_i).
\ee
We also note that the support properties of
$F_i(\kappa_i, \, \theta_i,\, \mu_i,\, \lambda_i )$ in the
$(\kappa_i, \, \theta_i)$-plane correspond to that of  baryon DAs.
It is thus natural to use
the combination of baryon DAs to which
$\pi N$
TDA reduces in the limit
$\xi \to 1$
due to the soft-pion theorem as input for
$V(w_i, \, v_i)$.

Let us denote the combination of nucleon DAs, to which $\pi N$ TDA $H$ reduces in the
limit $\xi \to 1$, as $V(y_1,y_2,y_3)$.
It is the function of three momentum fractions $y_i$ ($0\le y_i \le 1$)
satisfying the condition $\sum_i y_i=1$.
Then, according to the particular choice of quark-diquark coordinates in (\ref{Spectral_for_GPDs_x123}), one
has to employ in (\ref{Factorized_ansatz_xi=1}):
\be
&&
V(\kappa_1, \theta_1) \equiv  \frac{1}{4} V \left(\frac{\kappa_1+1}{2}, \,  \frac{1-\kappa_1+ 2 \theta_1}{4},\,\frac{1-\kappa_1- 2 \theta_1}{4}  \right);
\nonumber \\ &&
V(\kappa_2, \theta_2) \equiv \frac{1}{4}  V \left(  \frac{1-\kappa_2- 2 \theta_2}{4},\,\frac{\kappa_2+1}{2}, \,\frac{1-\kappa_2+ 2 \theta_2}{4}  \right);
\nonumber \\ &&
V(\kappa_3, \theta_3) \equiv  \frac{1}{4}  V \left( \frac{1-\kappa_3+ 2 \theta_3}{4},\,\frac{1-\kappa_3- 2 \theta_3}{4},\,  \frac{\kappa_3+1}{2} \right).
\ee

The profile function
$h(\mu_i,\, \lambda_i)$
also has the support of a baryon DA:
$-1 \le \mu_i \le 1$;
$-\frac{1-\mu_i}{2} \le \lambda_i \le \frac{1-\mu_i}{2}$.
Contrary to the GPD case, no symmetry constraint from hermiticity and time-reversal invariance
occurs for quadruple distributions. Therefore, we are free to employ an arbitrary shape for the profile function.
For example, we may assume that it is determined by the
asymptotic form of a baryon DA ($120 y_1 y_2 y_3$ with $\sum_i y_i=1$):
\be
h(\mu_i,\, \lambda_i)= 
\frac{15}{16} \, (1+\mu_i) ((1-\mu_i)^2-4 \lambda_i^2)\,.
\label{Profile_h}
\ee
In fact, this is the simplest possible choice for the profile vanishing at the borders of its domain
of definition. The normalization
(\ref{Norm_profile})
is obviously ensured.

\begin{figure}[H]
\begin{center}
 \epsfig{figure=  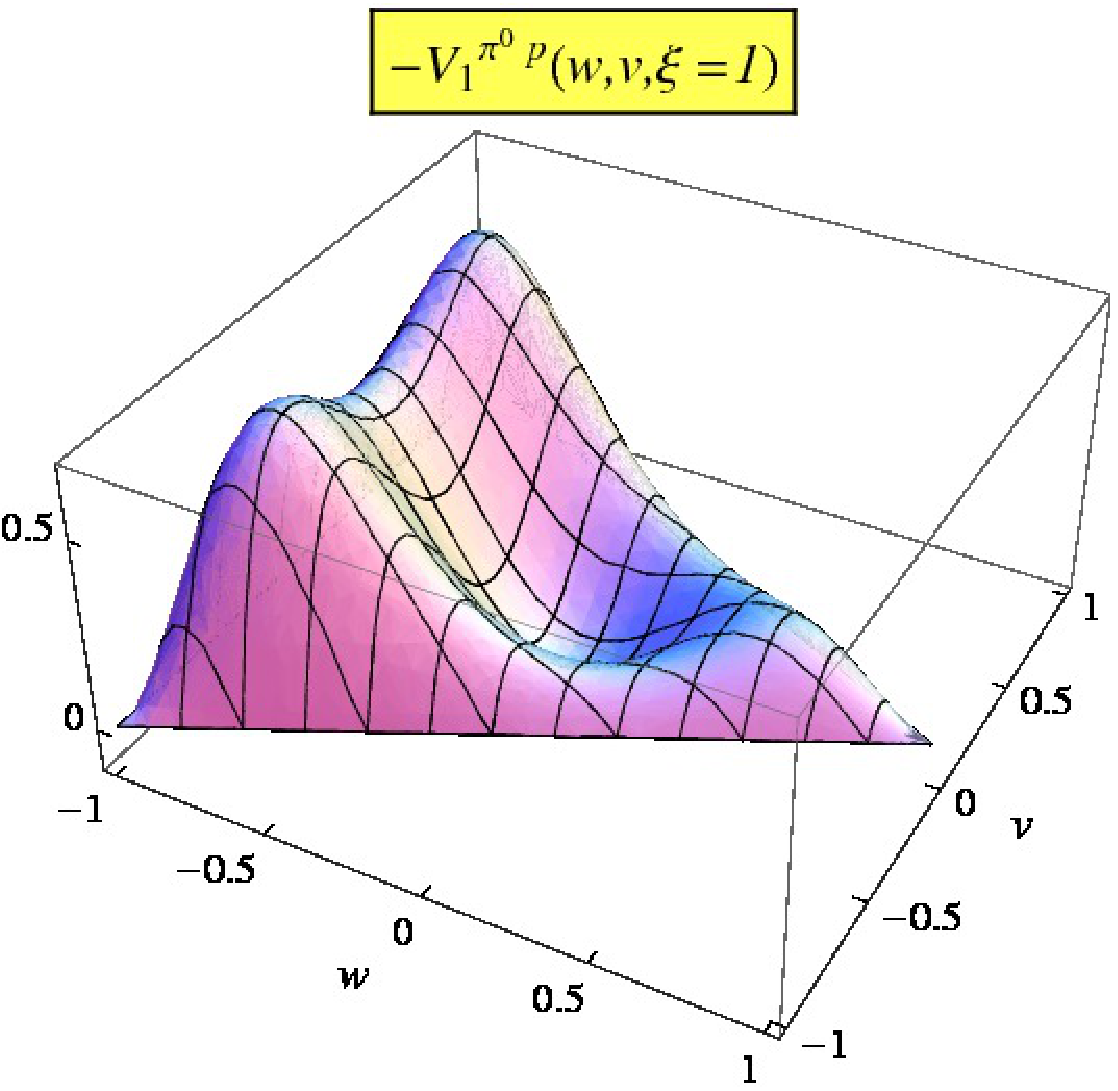 , height=4.5cm}
 \epsfig{figure=  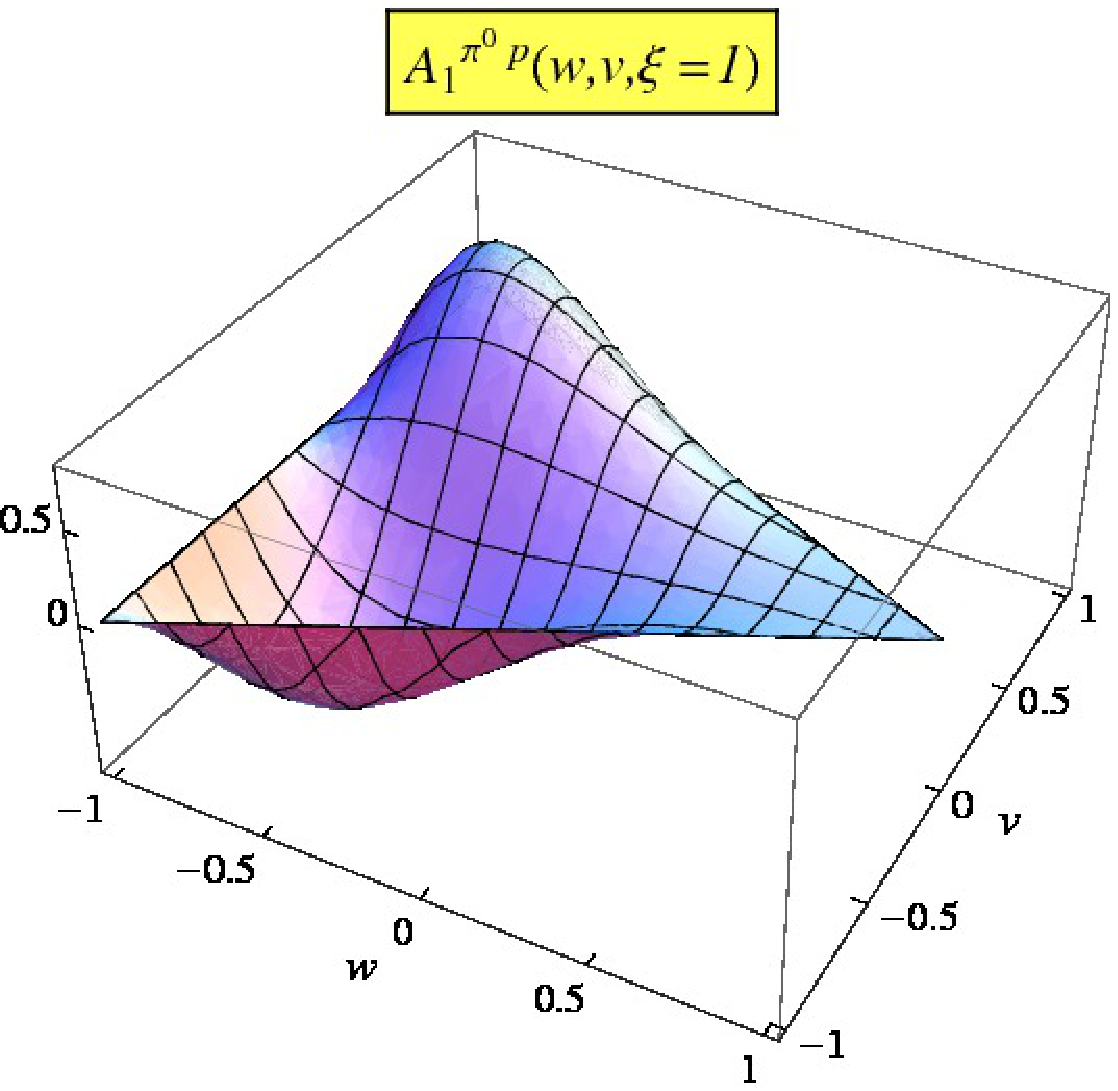 , height=4.5cm}
 \epsfig{figure=  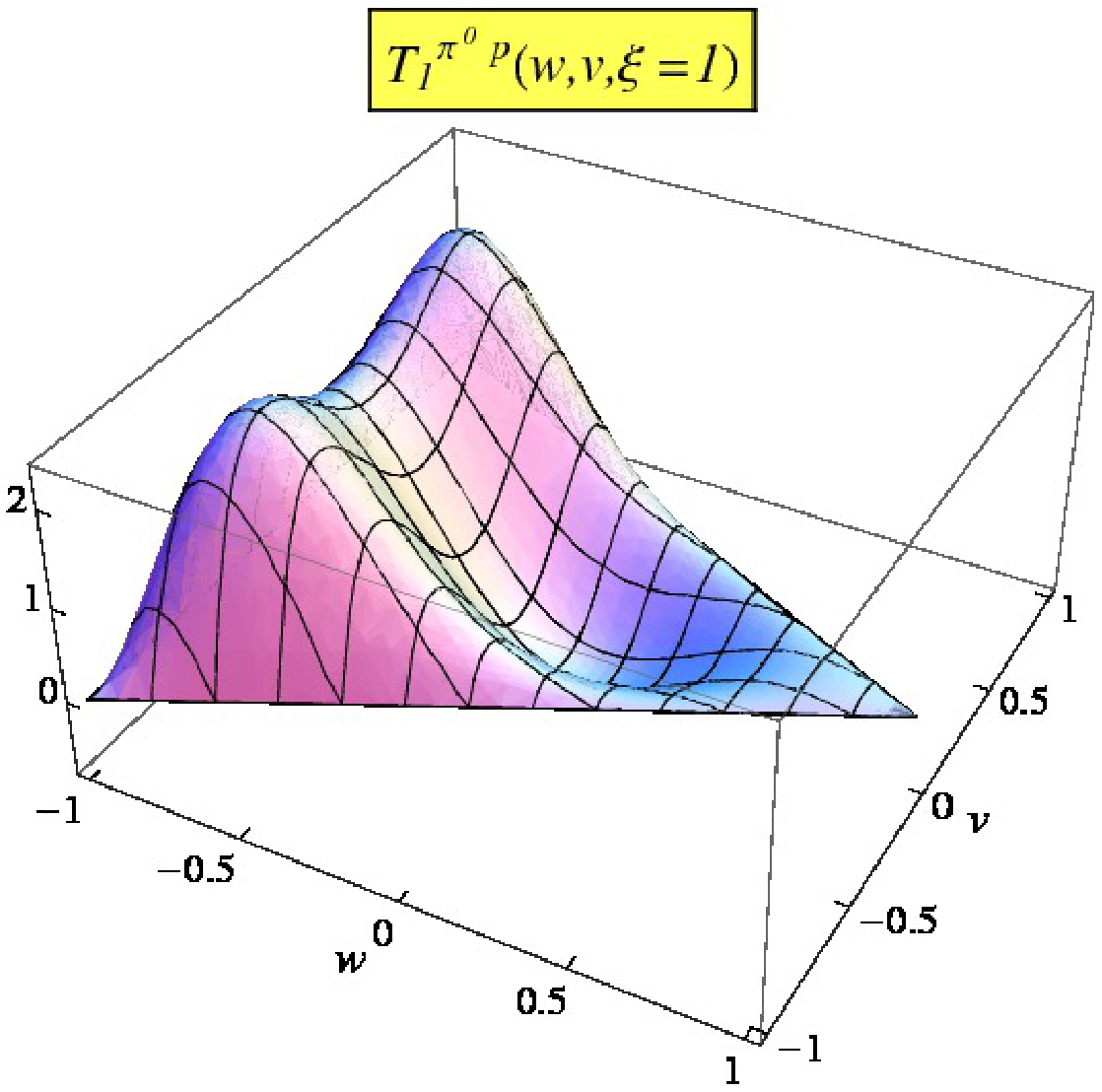 , height=4.5cm}
  \epsfig{figure=  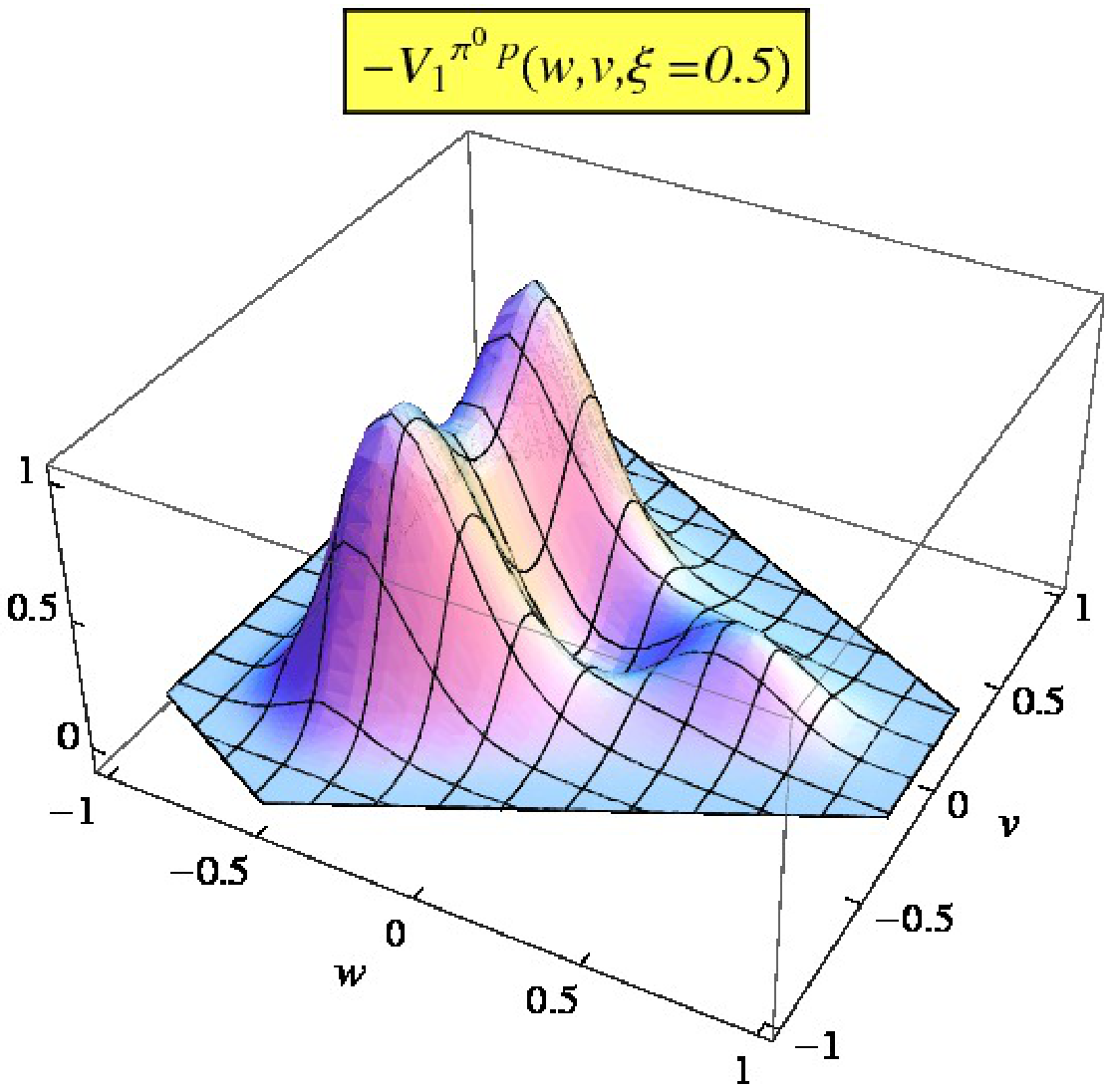 , height=4.5cm}
 \epsfig{figure=  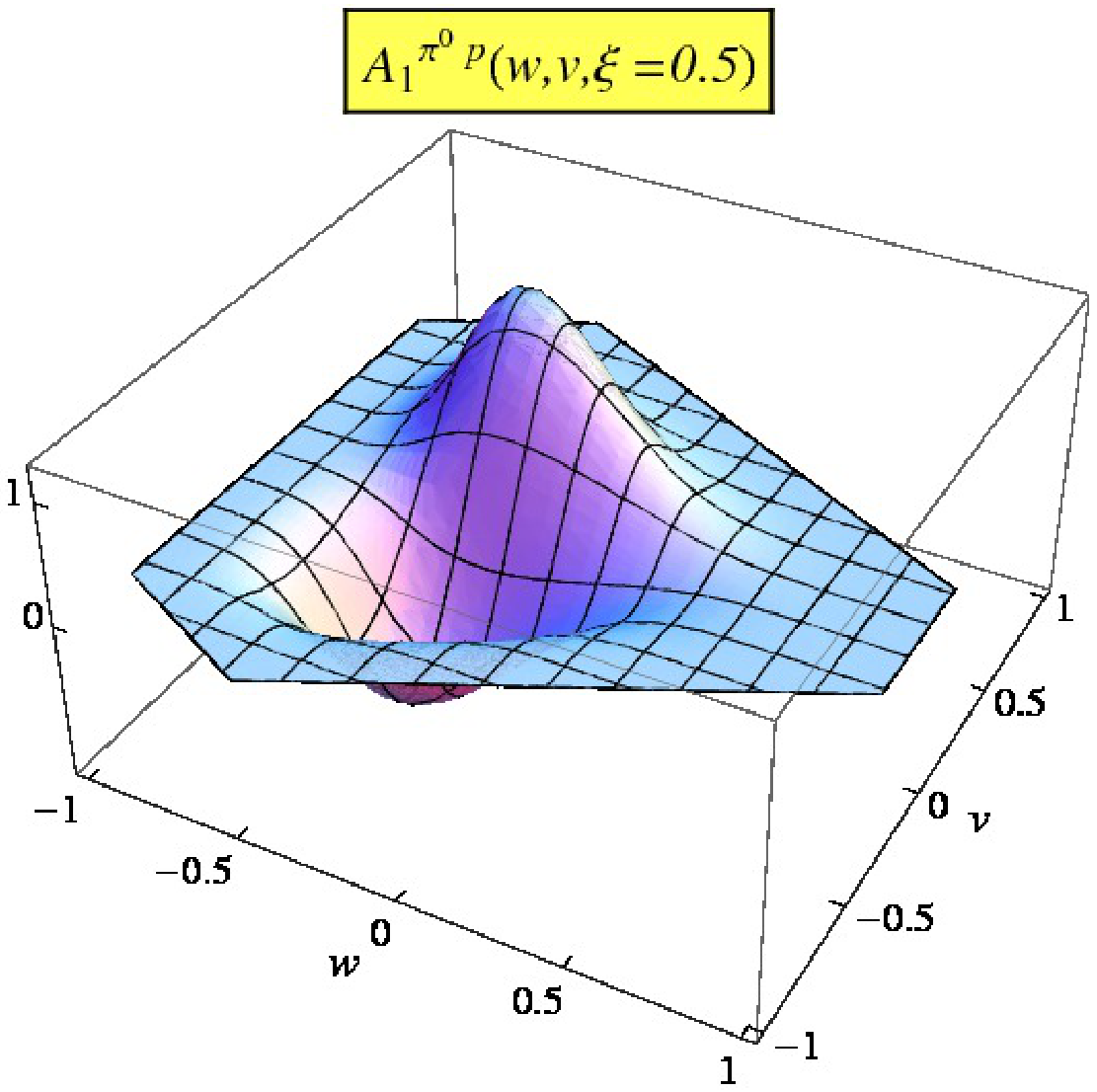 , height=4.5cm}
 \epsfig{figure=  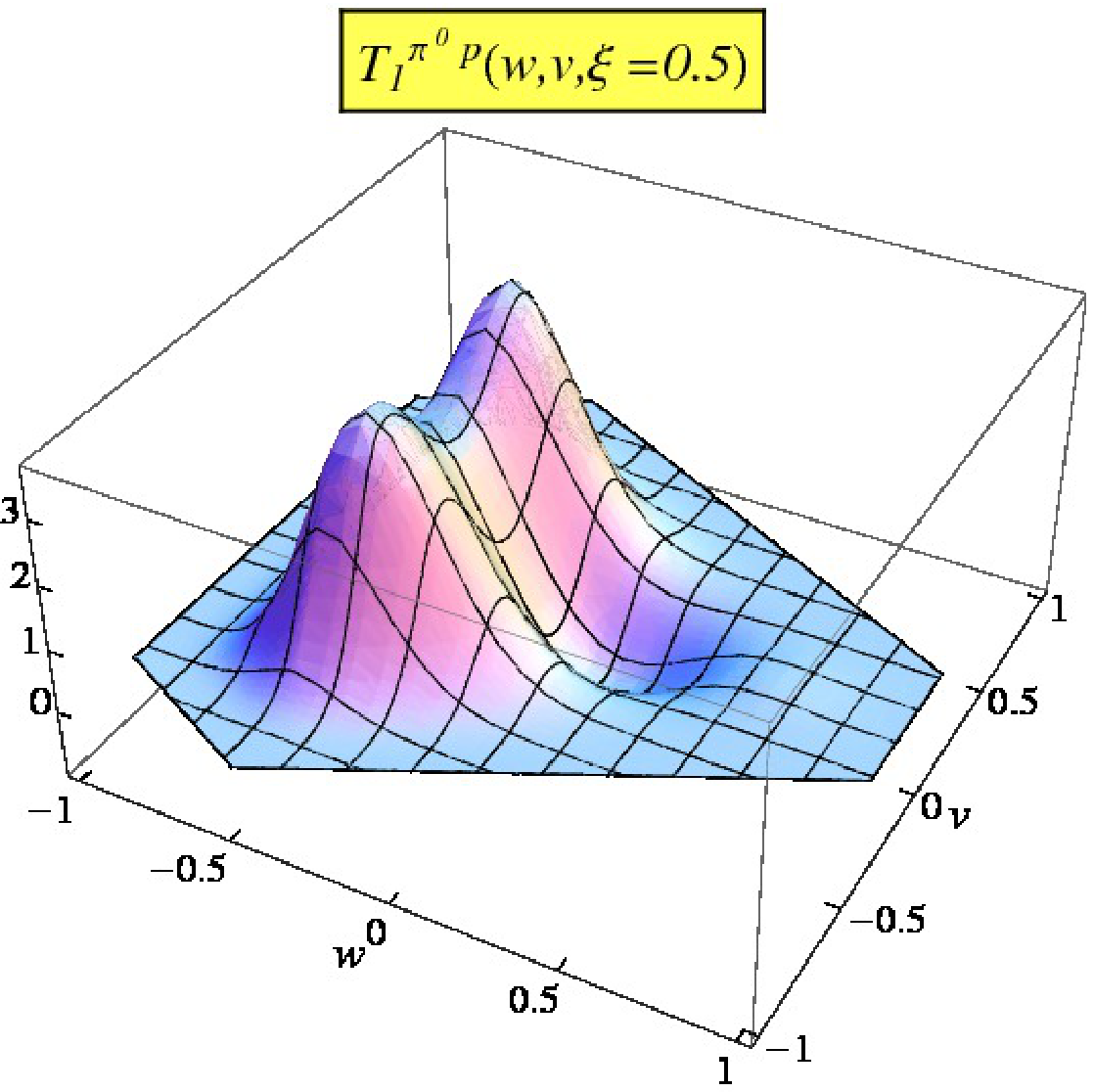 , height=4.5cm}
 \epsfig{figure=  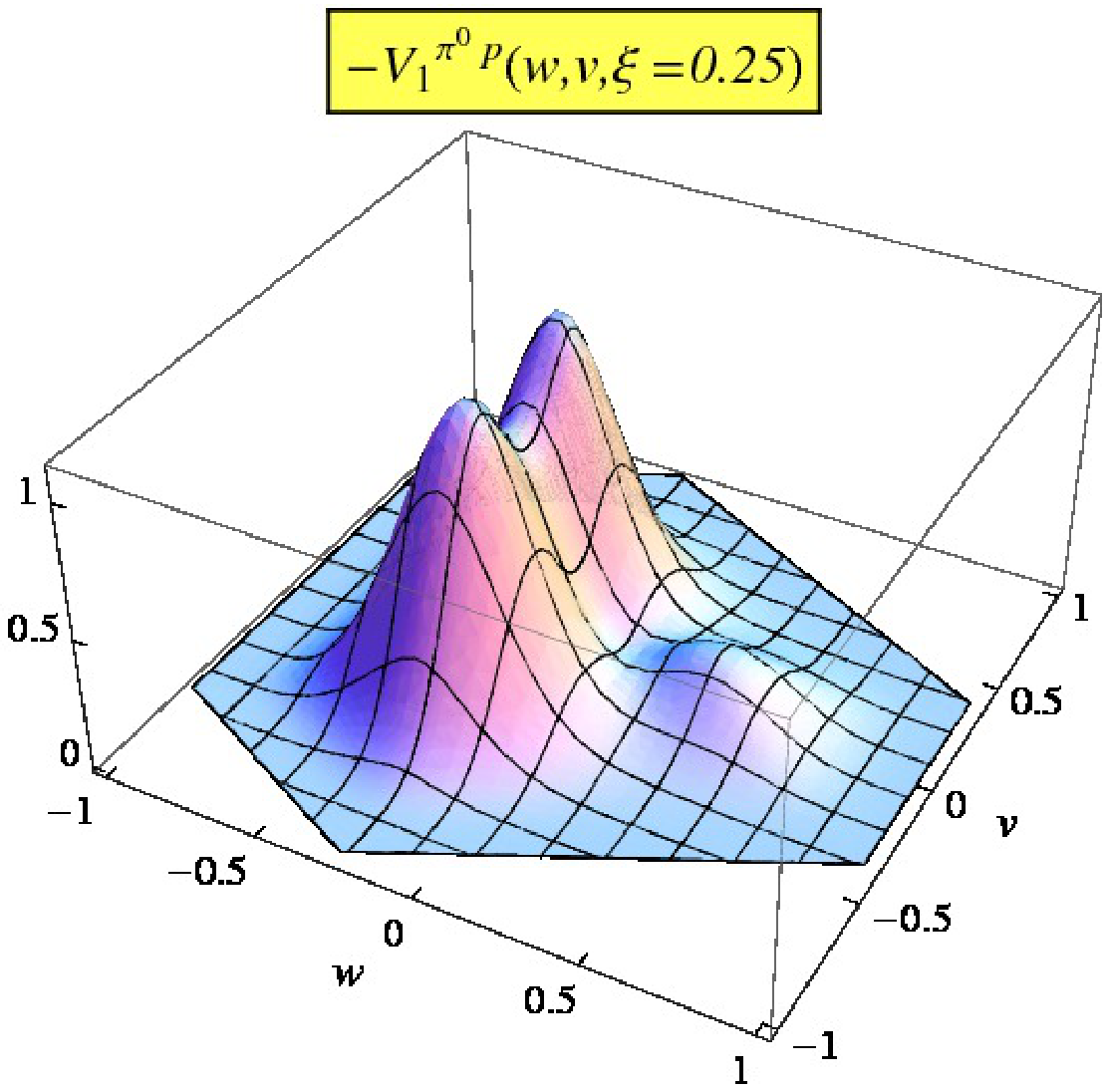 , height=4.5cm}
 \epsfig{figure=  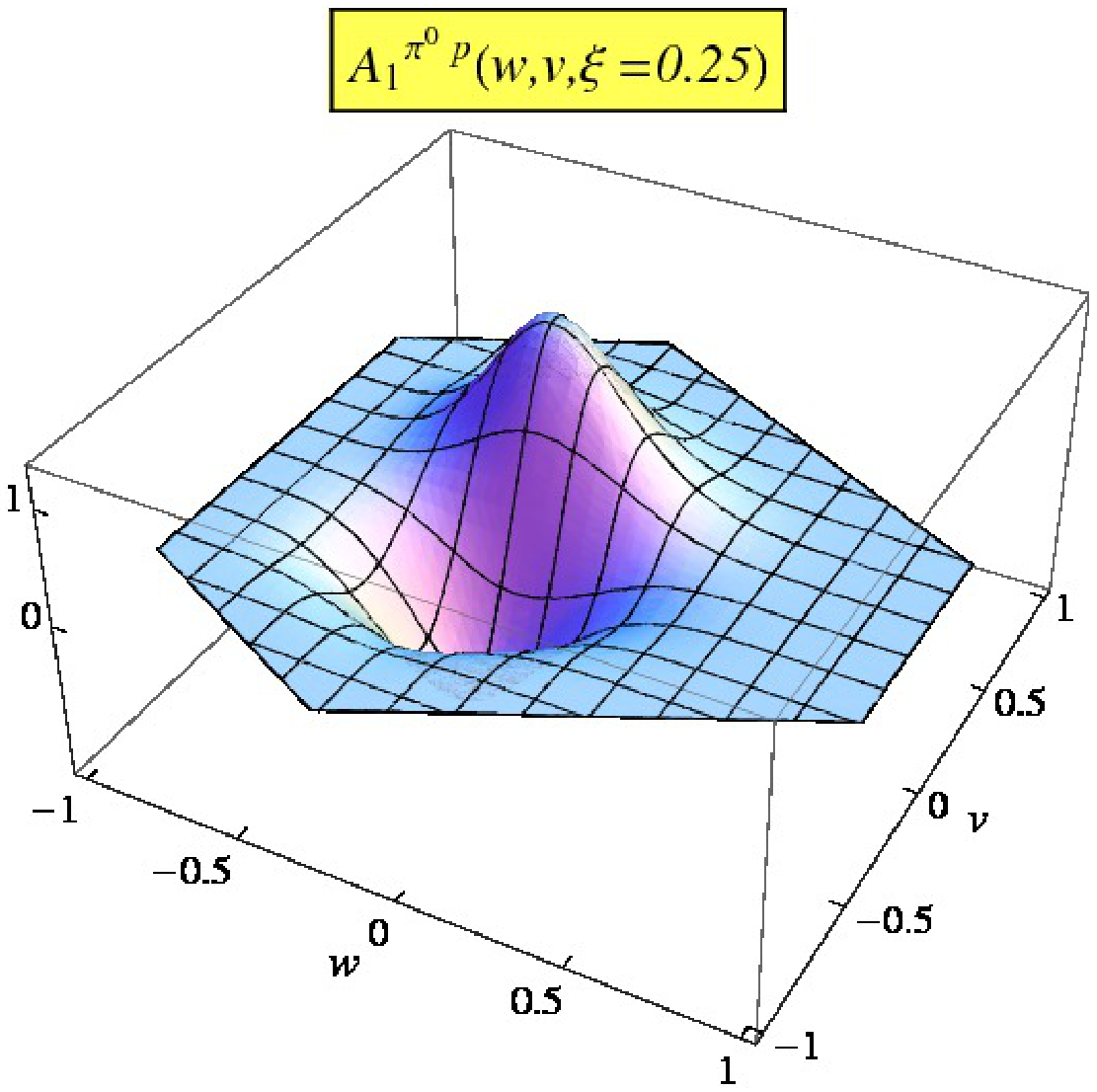 , height=4.5cm}
 \epsfig{figure=  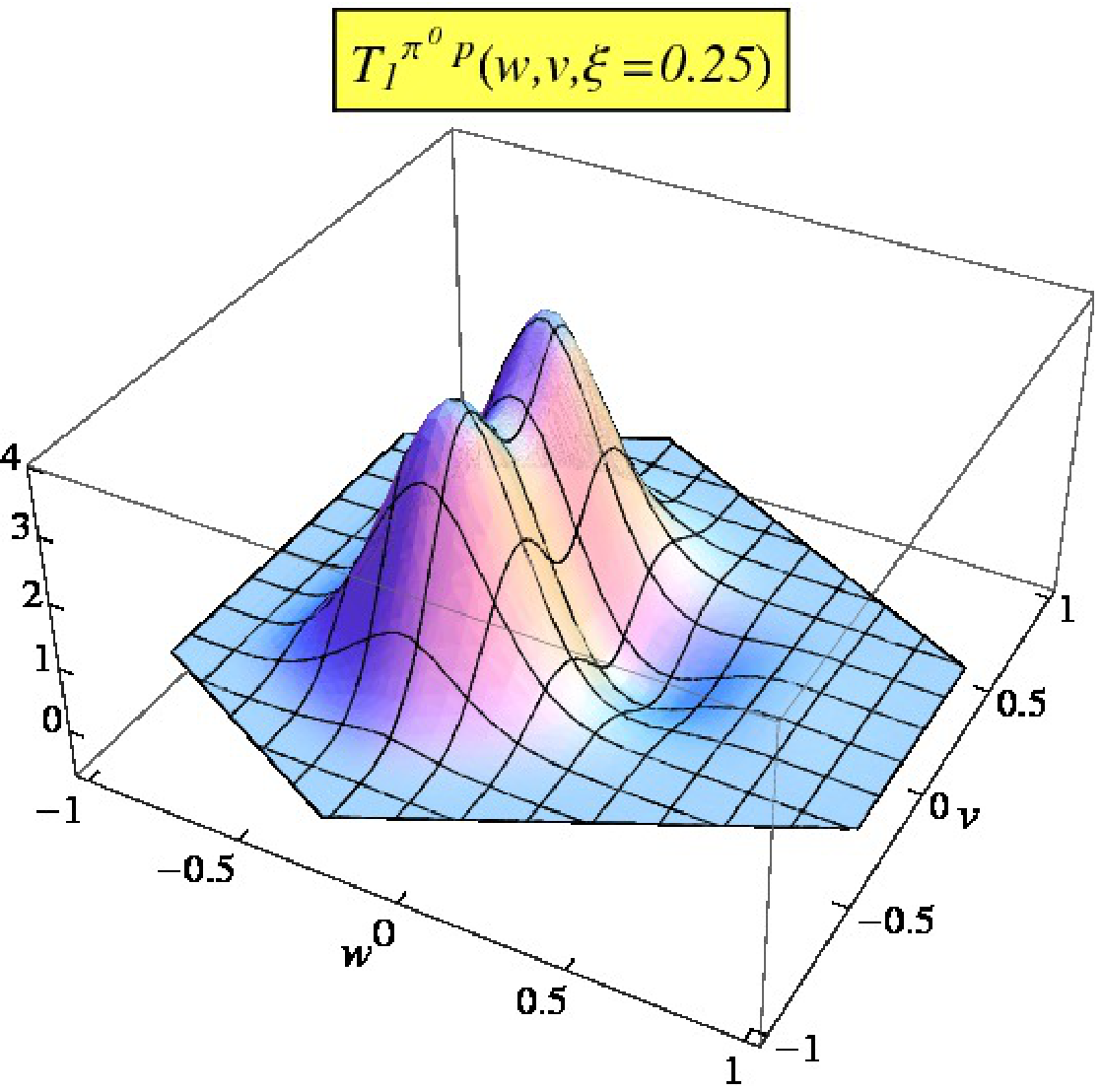 , height=4.5cm}
 \epsfig{figure=  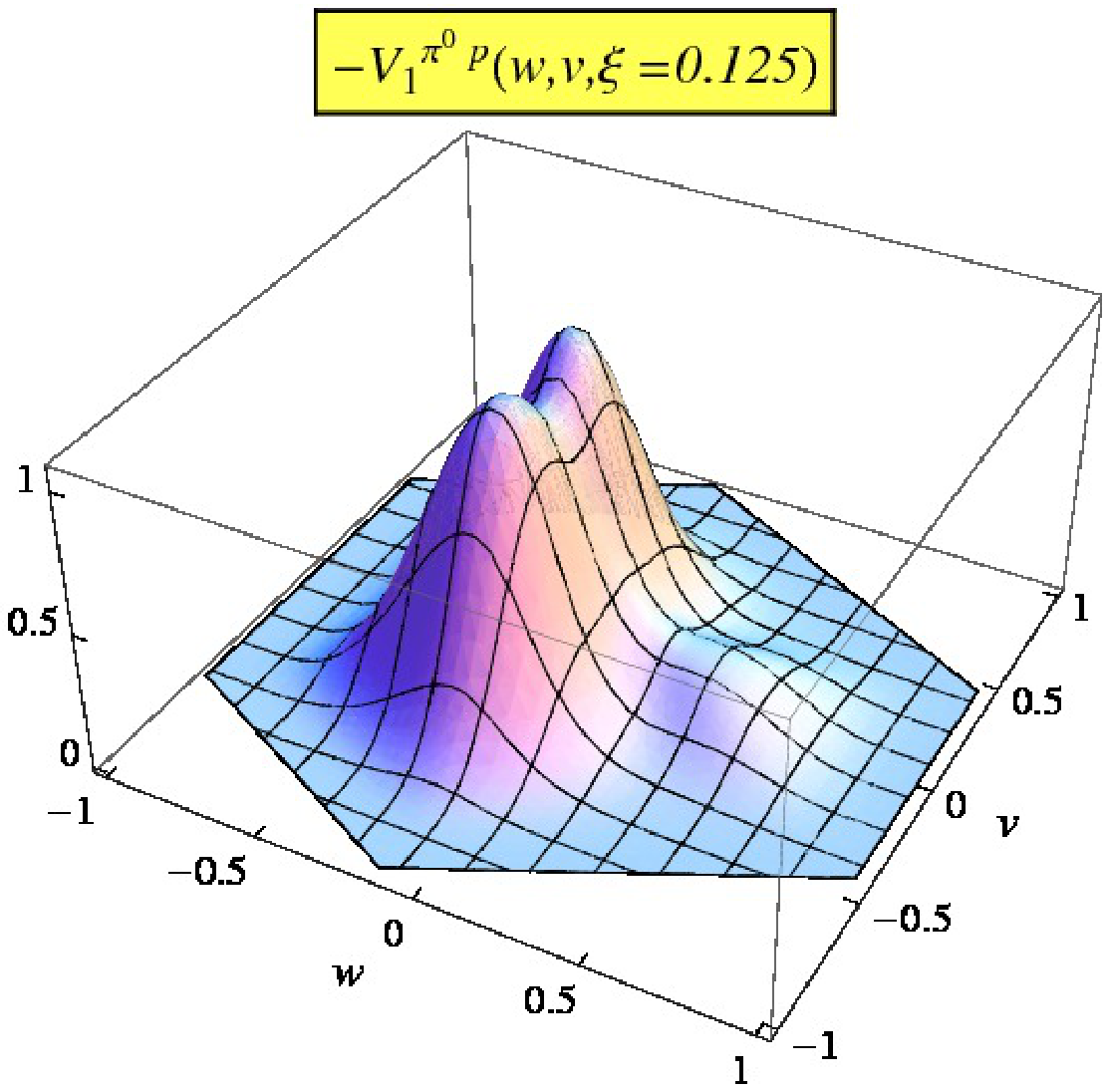 , height=4.5cm}
 \epsfig{figure=  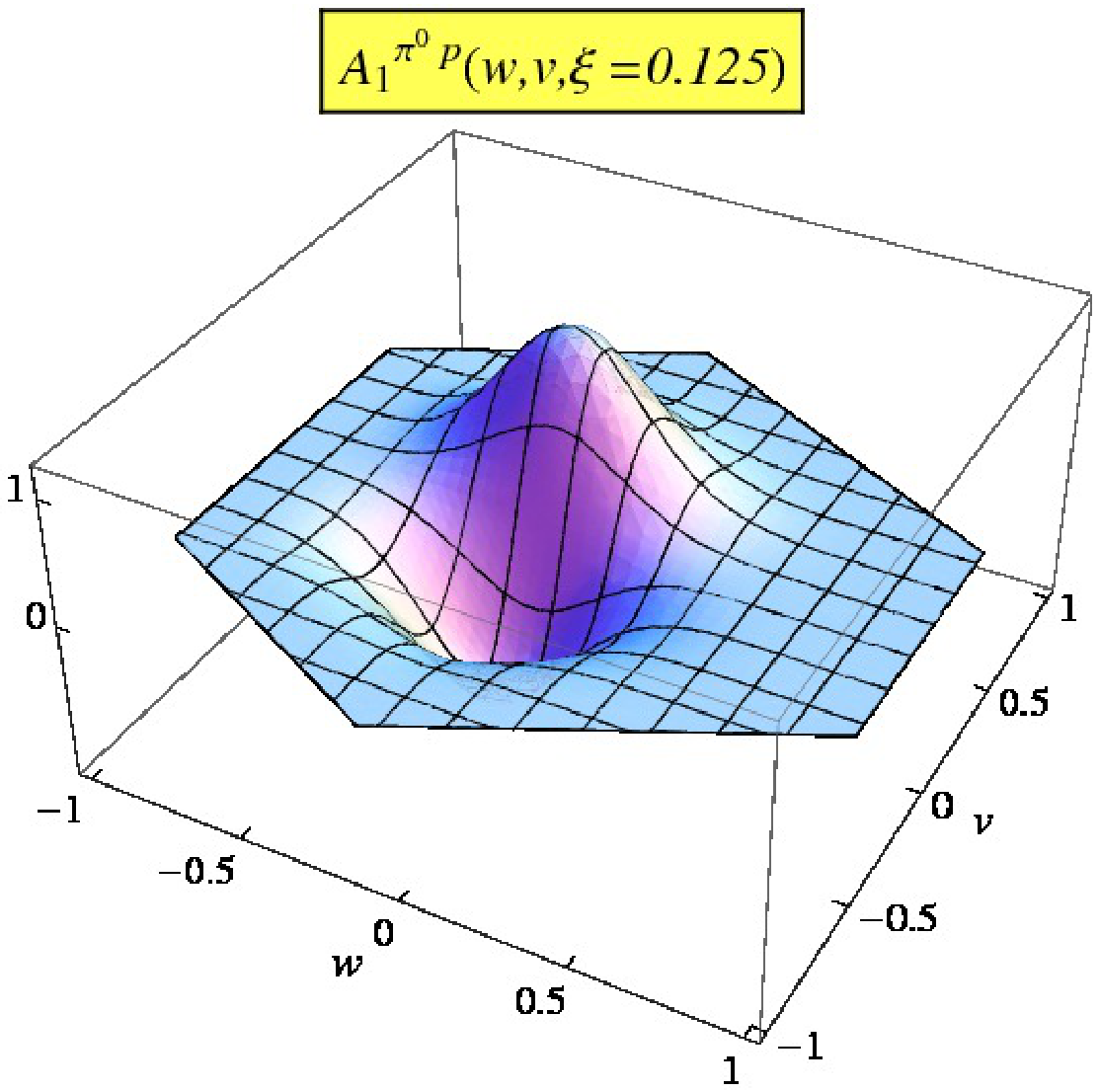 , height=4.5cm}
 \epsfig{figure=  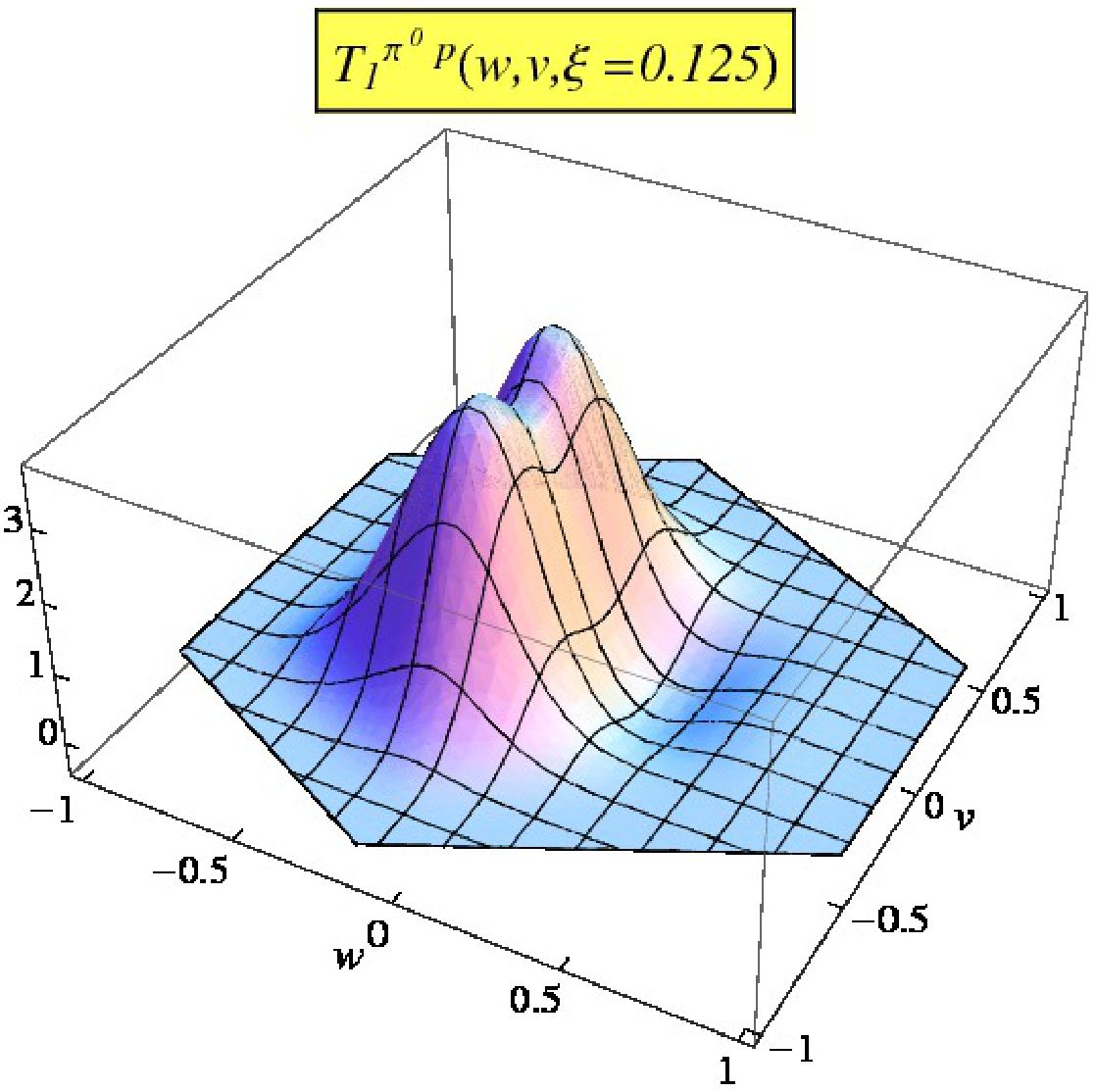 , height=4.5cm}
    \caption{$\pi^0 p$ TDAs $V_1^{ \pi^0 p}$, $A_1^{  \pi^0 p}$ and $T_1^{ \pi^0 p}$ computed in the model based on the factorized Ansatz (\ref{Factorized_ansatz_xi=1}) with the profile function (\ref{Profile_h})
    as   functions of quark-diquark coordinates $w \equiv w_3$, $v \equiv v_3$.
    In the limit $\xi=1$, as required by the soft-pion theorem, TDAs are reduced to the appropriate combinations
    of nucleon DAs $V^p$, $A^p$ and $T^p$ (see eq.~\ref{Soft_pion_th_pi0p}).
    For $\xi<1$, $\pi N$ TDAs are obtained by ``skewing'' $\xi=1$ limit. CZ nucleon DAs \cite{Chernyak:1987nv} are used as numerical input.}
     \label{Fig2}
\end{center}
\end{figure}

On Fig.~\ref{Fig2}, we present the result of the calculation of
$\pi^0 p$ TDAs
from the factorized Ansatz
(\ref{Factorized_ansatz_xi=1}) with the profile function
(\ref{Profile_h})
as a function of quark-diquark coordinates $w \equiv w_3$, $v \equiv v_3$
defined in (\ref{Quark-diquark_coordinates_compact}).
In accordance with the soft-pion theorem, in the
$\xi=1$
limit, our
$\pi^0 p$
TDAs $V^{\pi^0 p}_1$, $A^{\pi^0 p}_1$ and $T^{\pi^0 p}_1$
are reduced to the following combination of nucleon DAs \cite{Pire:2011xv}%
\footnote{Note that  eq.~(11) of \cite{LPS} and eq.~(19) of \cite{Lansberg:2007ec} contain
a sign error for $T^{\pi^0 p}$ as well as erroneous overall factors $2$. This affects the numerical results of these papers.}:
\be
&&
V^{\pi^0 p}_1(x_1,x_2,x_3,\xi=1)= -\frac{1}{2} \times \frac{1}{4} V^{p}\left(\frac{x_1}{2},\frac{x_2}{2},\frac{x_3}{2}\right);
\nonumber \\ &&
A^{\pi^0 p}_1(x_1,x_2,x_3,\xi=1)= -\frac{1}{2} \times  \frac{1}{4} A^{p}\left(\frac{x_1}{2},\frac{x_2}{2},\frac{x_3}{2}\right);
\nonumber \\ &&
T^{\pi^0 p}_1(x_1,x_2,x_3,\xi=1)= \frac{3}{2} \times  \frac{1}{4} T^{p}\left(\frac{x_1}{2},\frac{x_2}{2},\frac{x_3}{2}\right).
\label{Soft_pion_th_pi0p}
\ee
 We employ  the Chernyak-Zhitnitsky (CZ)
nucleon DA \cite{Chernyak:1987nv}
as the numerical input.

Our spectral representation provides a lively $x_i$ and $\xi$ dependence for $\pi N$ TDAs.
However, the suggestion of a reasonable $\Delta^2$ dependence still remains an open question.
The most straightforward solution would be, similarly to early attempts in the GPD case,
to try a sort of factorized form of $\Delta^2$ dependence for quadruple distributions
(\ref{Factorized_ansatz_xi=1}):
\be
F_i(\kappa_i, \, \theta_i,\, \mu_i,\, \lambda_i, \Delta^2 )= 4 V(\kappa_i, \, \theta_i) \, h(\mu_i,\, \lambda_i) \times G(\Delta^2),
\label{Factorized_ansatz_xi=1_Delta2}
\ee
where
$G(\Delta^2)$
is the
$\pi N$
transition form factor of the local three quark operator
$\widehat{O}^{\alpha \beta \gamma}_{\rho \tau \chi}(0,0,0)$
(\ref{oper}).
This leads
to a factorized $\Delta^2$-dependence for $\pi N$ TDAs:
\be
H^{\pi N}(x_i,\xi,\Delta^2)=H^{\pi N}(x_i,\xi) \times G(\Delta^2).
\label{Fact_form_delta2}
\ee
The determination of
$G(\Delta^2)$
goes beyond the scope of the present paper.
Let us however note that such a factorized form of the
$\Delta^2$
dependence is known to be  oversimplified  and was much criticized in the GPD case
(see {\it e.g.} discussion in \cite{arXiv:0805.0152}).

\subsection{$D$-term-like nucleon pole contribution }

\begin{figure}[H]
 \begin{center}
 \epsfig{figure=  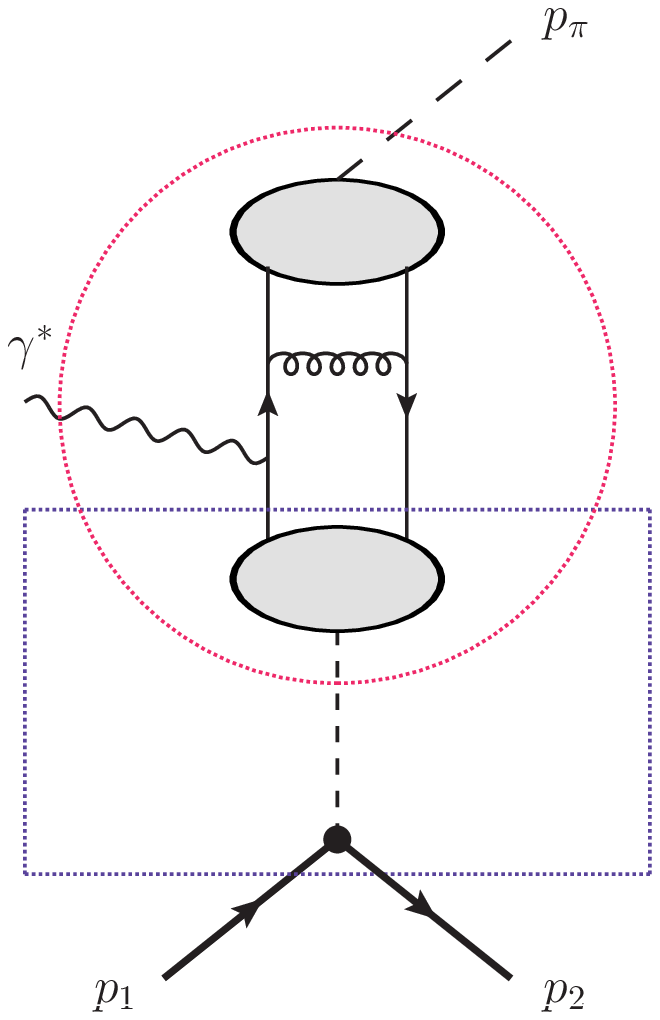 , height=6cm} \ \ \ \ \ \ \ \
  \epsfig{figure= 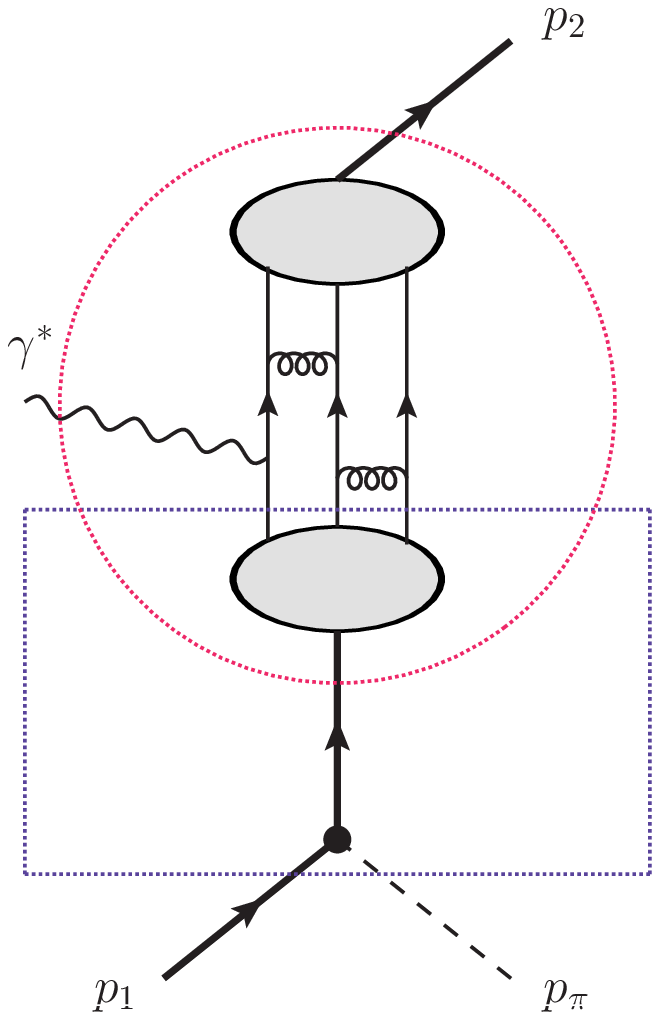 , height=6cm}
     \caption{{\bf Left:} pion pole exchange model for the polarised GPD $\tilde{E}$; lower and upper blobs depict pion DAs;  the dashed circle contains
     a typical LO graph for the pion electromagnetic form factor in perturbative QCD; the rectangle contains the pion pole contribution into GPD.
    {\bf Right:} nucleon pole exchange model for $\pi N$ TDAs;  dashed circle contains a typical LO graph for the nucleon electromagnetic form factor in perturbative QCD; the rectangle contains the nucleon pole contribution into $\pi N$ TDAs.
      }
\label{Fig_Nucleon_pole}
\end{center}
\end{figure}

Similarly to the GPD case,
$\pi N$
TDAs within the spectral representation
(\ref{Spectral_for_GPDs_x123})
do not satisfy the polynomiality condition in its complete form. As it was pointed out in
\cite{Pire:2011xv},
 the spectral representation for
$\pi N$
TDAs
$V_{1,2}^{(\pi N)}$,
$A_{1,2}^{(\pi N)}$,
$T_{1,2}^{(\pi N)}$
has to be complemented by an analogue of the $D$-term.
TDAs $T_{3,4}^{(\pi N)}$ do not require adding the $D$-term.
This $D$-term has a pure ERBL-like support in $x_i$  and hence it contributes only to the
real part of the backward pion electroproduction amplitude
(\ref{helicity ampl}).
In this paper, we employ the simplest possible model for such a $D$-term
which accounts for the contribution of the $u$-channel nucleon exchange
into $\pi N$ TDAs computed in
\cite{Pire:2011xv}.
This model shares many common features with the pion pole model for the polarized nucleon
GPD $\tilde{E}$ suggested in
 Sec.~2.4 of Ref.~\cite{GPV}
(see Fig.~\ref{Fig_Nucleon_pole}).

The explicit expression for the contribution of the $u$-channel nucleon exchange into the isospin-$\frac{1}{2}$ $\pi N$ TDAs
was established in \cite{Pire:2011xv}:
\be
&&
\big\{ V_1, \, A_1 , \, T_1  \big\}^{(\pi N)_{1/2}} (x_1,x_2,x_3,\xi,\Delta^2)\Big|_{N(940)}
\nonumber \\ &&
 =\Theta_{\rm ERBL}(x_1,x_2,x_3) \times  (g_{\pi NN}) \frac{M f_\pi}{\Delta^2-M^2}    \frac{1}{(2 \xi) }
 \big\{ V^p,\,A^p, \,T^p  \big\}\left( \frac{x_1}{2 \xi}, \frac{x_2}{2 \xi}, \frac{x_3}{2 \xi} \right);
  \nonumber \\ &&
\big\{ V_2, \, A_2 , \, T_2  \big\}^{(\pi N)_{1/2}} (x_1,x_2,x_3,\xi,\Delta^2)\Big|_{N(940)}=
\frac{1}{2}
\big\{ V_1, \, A_1 , \, T_1  \big\}^{(\pi N)_{1/2}} (x_1,x_2,x_3,\xi,\Delta^2)\Big|_{N(940)};   \nonumber \\ &&
\big\{  T_3, \, T_4  \big\}^{(\pi N)_{1/2}} (x_1,x_2,x_3,\xi,\Delta^2)\Big|_{N(940)}=0,
\label{Nucleon_exchange_contr_VAT}
\ee
where we employ the notation
\be
\Theta_{\rm ERBL}(x_1,x_2,x_3)  \equiv  \prod_{k=1}^3 \theta(0 \le x_k \le 2 \xi);
\label{theta_ERBL}
\ee
$M$
denotes the nucleon mass;
$f_\pi$
is the pion weak decay constant and
$g_{\pi NN}$
is the pion-nucleon phenomenological coupling
(see {\it e.g. }\cite{EricsonWeise}).

The nucleon pole contribution into $\pi^+ p$ and $\pi^0 p$ expressed through isospin-$1/2$
$\pi N$ TDAs
(\ref{Nucleon_exchange_contr_VAT}) reads:
\be
&&
\{V_{1,2},\,A_{1,2}, T_{1,2}\}^{\pi^+ p}\Big|_{N(940)} \nonumber \\ &&
= -\sqrt{2} \{V_{1,2},\,A_{1,2}, T_{1,2}\}^{\pi^0 p}\Big|_{N(940)} \equiv
-\sqrt{2} \{V_{1,2},\,A_{1,2}, T_{1,2}\}^{(\pi N)_{1/2}}\Big|_{N(940)}; \nonumber \\ &&
\{T_{3,4}\}^{\pi^+ p}\Big|_{N(940)}=\{T_{3,4}\}^{\pi^0 p}\Big|_{N(940)}=0.
\ee

\section{Calculation of $\gamma^* N \to \pi N$ amplitude}
\label{Section_amplitude}

Within the factorized approach,
the leading order (both in $\alpha_s$ and $1/Q$)  amplitude for
hard exclusive
$\gamma^* N \to \pi N$
reaction in the backward region,
$\mathcal{M}^\lambda_{s_1s_2}$,
reads \cite{Lansberg:2007ec}:
\be
&& {\mathcal M}^\lambda_{s_1s_2}  =-i
\frac{(4 \pi \alpha_s)^2 \sqrt{4 \pi \alpha_{em}} f_{N}^2}{ 54 f_{\pi} } \frac{1}{Q^4}
\Big[
{\cal S}^\lambda_{s_1s_2}
{\int \! d^3 x  \int \! d^3y
\Bigg(2\sum_{\alpha=1}^{7} T_{\alpha}+
\sum\limits_{\alpha=8}^{14} T_{\alpha}\Bigg)}
\nonumber
\\ &&
-
{\cal S'}^\lambda_{s_1s_2}
  \int \! d^3 x  \int \! d^3y
\Bigg(2\sum_{\alpha=1}^{7} T'_{\alpha}+
\sum\limits_{\alpha=8}^{14} T'_{\alpha}\Bigg)
\Big],
\label{helicity ampl}
\ee
where
$f_\pi=93$ MeV is the pion weak decay constant and $f_N \sim 5.2 \cdot 10^{-3}$ GeV$^2$ is a constant,
which determines the value of the dimensional nucleon wave function at the origin;
$\alpha_{em}\simeq \frac{1}{137}$ is the electromagnetic fine structure constant; and $\alpha_s$ is the strong coupling constant.
The convolution integrals in $x_i$ and $y_i$  in
(\ref{helicity ampl})
respectively stand over the supports of
$\pi N$
TDAs and nucleon DAs  in $T_\alpha$ and $T'_\alpha$.
The spin structures ${\cal S}$ and ${\cal S}'$ are defined as
\be
{\cal S}^\lambda_{s_1s_2} \equiv \bar U(p_2,s_2)
\hat{\epsilon}(\lambda)
 \gamma^5 U(p_1,s_1)\,; \ \ \
   {\cal S'}^\lambda_{s_1s_2} \equiv \frac{1}{M}\bar U(p_2,s_2)
\hat{ \epsilon }(\lambda) \hat{\Delta}_T
 \gamma^5 U(p_1,s_1),
\ee
where $\epsilon(\lambda)$ denotes the polarization vector of the virtual photon.
We introduce the following notations:
\be
&&
\{\mathcal{I}, \mathcal{I}'\}(\xi,\Delta^2) \equiv { {\int^{1+\xi}_{-1+\xi} }\! \! \! dx_1  {\int^{1+\xi}_{-1+\xi} }\! \! \! dx_2  {\int^{1+\xi}_{-1+\xi} }\! \! \!dx_3 \, \delta(x_1+x_2+x_3-2\xi)
}
\nonumber \\ &&
\times
{{\int^{1}_{0} }\! \! \! dy_1  {\int^{1}_{0} }\! \! \! dy_2  {\int^{1}_{0} }\! \! \!dy_3 \, \delta(y_1+y_2+y_3-1)}
{\Bigg(2\sum_{\alpha=1}^{7} \{ T_{\alpha}, T_{\alpha}'\}+
\sum\limits_{\alpha=8}^{14} \{ T_{\alpha},  T_{\alpha}'\} \Bigg)};  \nonumber \\ &&
\mathcal{C} \equiv
-i
\frac{(4 \pi \alpha_s)^2 \sqrt{4 \pi \alpha_{em}} f_{N}^2}{ 54 f_{\pi} },
\label{Def_IIprime}
\ee
and rewrite (\ref{helicity ampl}) as
\be
{\mathcal M}^\lambda_{s_1s_2}=\mathcal{C} \frac{1}{Q^4} \Big[
{\cal S}^\lambda_{s_1s_2} \mathcal{I}(\xi,\Delta^2)+
{\cal S}'^\lambda_{s_1s_2} \mathcal{I}'(\xi,\Delta^2) \Big].
\label{helicity_ampl_rewr}
\ee

The expressions for the coefficients
$T_\alpha$ and $T'_\alpha$
for the
$\gamma^* p \rightarrow \pi^0 p$ channel
are presented in the Table~I of Ref.~\cite{Lansberg:2007ec}.
The result for
$\gamma^* p \rightarrow \pi^+ n$ channel can be read from the same
Table
with the obvious changes:
\be
&&
\nonumber
Q_u \rightarrow Q_d; \ \ \ Q_d \rightarrow Q_u\,; \\ &&
V^p, \, A^p, \, T^p \, \rightarrow \, V^n, \, A^n, \, T^n \nonumber  \;  \equiv - V^p, \, -A^p, \, -T^p\,; \\ &&
V^{p \pi^0}_{1,2}, \, A^{p \pi^0}_{1,2}, \, T^{p \pi^0}_{1,2,3,4} \,
 \rightarrow
 V^{p \pi^+}_{1,2}, \, A^{p \pi^+}_{1,2}, \, T^{p \pi^+}_{1,2,3,4}\,.
\ee
We note that in  Ref.~\cite{Lansberg:2007ec} a somewhat
inadequate parametrization of $\pi N$ TDAs was employed.
Within this parametrization, $\pi N$ TDAs do not satisfy the polynomiality property in its simple form due to the
appearance of kinematical singularities (see discussion in Ref.~\cite{Pire:2011xv}).
In this paper, we adopt  the parametrization suggested in
\cite{Pire:2011xv}
in which polynomiality is explicit. The relation between the two parameterizations is given by eq. (\ref{Old_to_new}).

As we note, $x_i$ and $y_i$ dependencies in coefficients $T_\alpha$ ($T'_\alpha$) are factorized.
One therefore anticipates that
the
convolution integrals  in eq.~(\ref{helicity ampl})
have the following generic structure%
\footnote{Here
$\alpha=1, ..., 14$
should be understood as a label. No summation over repeating $\alpha$ is implied.}:
\be
&&
 \int d^3x 
 K_\alpha(x_1,x_2,x_3)
\Big[
{\rm combination \ \ of \ \ } \pi N\ \  {\rm TDAs}(x_1,x_2,x_3) \Big]
\nonumber  \\ &&
\times
\int d^3y
R_\alpha(y_1,y_2,y_3)
\Big[
{\rm combination \ \ of \ \ }   N\ \  {\rm DAs}(y_1,y_2,y_3)
\Big].
\label{Convolutions_T_alpha}
\ee
$K_\alpha(x_1,x_2,x_3)$
and
$R_\alpha(y_1,y_2,y_3)$
refer to parts of the singular convolution kernel in
$T_\alpha$ ($T'_\alpha$)
depending on
$x_i$
and
$y_i$
respectively.  These can be read from the Table~I of Ref.~\cite{Lansberg:2007ec}.

The convolution integrals in
$y_i$
in
(\ref{Convolutions_T_alpha})
are similar to those occurring within the perturbative description of the nucleon
electromagnetic form factor. The convolutions with singular kernels
$R_\alpha(y_1,y_2,y_3)$
do not generate any imaginary part since the nucleon DAs have purely ERBL support and vanish at the borders
of their domain of definition. These integrals can be calculated in a straightforward way.

On the contrary, the convolution integrals in
$x_i$
with
$\pi N$
TDAs in
(\ref{Convolutions_T_alpha})
may,
in principle,  generate a  nonzero imaginary part of the amplitude.
$\pi N$ TDAs, indeed,
do not necessarily  vanish on the cross over trajectories
$x_i=0$,
separating ERBL-like and DGLAP-like regimes,
as well as on the lines
$x_i=2 \xi$.

Switching  to quark-diquark coordinates
(\ref{Quark-diquark_coordinates_compact}),
one may show that the following types of convolution kernels
$K_\alpha$
occur in
(\ref{Convolutions_T_alpha}):
\be
&&
K_I^{(\pm, \pm)} (w_i,v_i)= \frac{1}{(w_i\pm \xi \mp i 0)} \frac{1}{(v_i \pm \xi'_i \mp i 0)}, \nonumber
\\ &&
K_{II}^{(-, \pm)}(w_i,v_i)= \frac{1}{(w_i-\xi+ i 0)^2} \frac{1}{(v_i \pm \xi'_i \mp i 0)}. \nonumber
\ee
Throughout the following discussion, we adopt the convention that the first sign in the
$(\pm, \pm)$
index of a quantity
corresponds to the one in the
$w \pm \xi$
denominator while the second sign corresponds to the one in the
$v \pm \xi'$
denominator.

Thus, we have to deal with only two types of integrals:
\be
I_I^{(\pm, \pm)}(\xi)= \int_{-1}^1 dw  \int_{-1+|\xi-\xi'|}^{1-|\xi-\xi'|} dv
\frac{1}{(w\pm \xi \mp i 0)} \frac{1}{(v \pm \xi' \mp i 0)} H(w,v,\xi),
\label{I_I}
\ee
and
\be
I_{II}^{(-, \pm)}(\xi)= \int_{-1}^1 dw  \int_{-1+|\xi-\xi'|}^{1-|\xi-\xi'|} dv
\frac{1}{(w-\xi+ i 0)^2} \frac{1}{(v \pm \xi' \mp i 0)} H(w,v,\xi),
\label{I_II}
\ee
for which we have to develop a method of calculation.
The integration in
(\ref{I_I})
and
(\ref{I_II})
stands over the support  (\ref{The_whole_domain})
of
$\pi N$
TDAs in quark-diquark coordinates.

Using the formulas summarized in Appendix~\ref{App_GenF},
we establish the expression for the real and imaginary parts of
$ I^{(\pm,\pm)}_I(\xi)$:
\be
&&
\re I^{(+,\pm)}_I(\xi)=  \mathcal{P} \!\! \int_{-1}^1 dw \,\frac{1}{(w+\xi)}  \, \mathcal{P} \!\! \int_{-1+|\xi-\xi'|}^{1-|\xi-\xi'|} dv\,
  \frac{1}{(v \pm \xi') }H(w,v,\xi) \pm \pi^2 H(-\xi,\mp \xi,\xi);
\nonumber \\ &&
\re I^{(-,\pm)}_I(\xi)= \mathcal{P} \!\! \int_{-1}^1 dw \, \frac{1}{(w-\xi)}  \mathcal{P} \!\! \int_{-1+|\xi-\xi'|}^{1-|\xi-\xi'|} dv\,
\frac{1}{(v \pm \xi') }H(w,v,\xi) \mp \pi^2 H(\xi,0,\xi);
\nonumber \\
&&
\im  I^{(+,\pm)}_I(\xi)= \mp \pi \mathcal{P} \!\! \int_{-1}^1 dw  \frac{1}{w+\xi } H(w,\mp \xi', \xi)+
\pi \mathcal{P} \!\! \int_{-1 }^{1 }dv \frac{1}{v \pm \xi} H(-\xi,v,\xi);
\nonumber \\ &&
\im  I^{(-,\pm)}_I(\xi)= \mp \pi \mathcal{P} \!\! \int_{-1}^1 dw  \frac{1}{w-\xi} H(w,\mp \xi', \xi)-
\pi \mathcal{P}  \!\! \int_{-1+\xi }^{1 -\xi}dv \frac{1}{v  } H(\xi,v,\xi).
\ee

Let us now consider the second type of integrals by rewriting it as:
\be
I_{II}^{(-, \pm)}(\xi)
=\int_{-1}^1 dw \frac{1}{(w-\xi+ i 0)^2}
\Bigg\{ \mp i \pi H(w,\mp \xi',\xi)+
J^{(\pm)}(w,\xi)\Bigg\},
\label{I2_start}
\ee
where
we introduced the notation:
\be
&&
J^{(\pm)}(w,\xi)=\mathcal{P} \!\! \int_{-1+|\xi-\xi'|}^{1-|\xi-\xi'|} dv
\frac{1}{v \pm \xi'} H(w,v,\xi)\,.
\label{Def_J}
\ee

Let us emphasize that in
(\ref{I2_start})
we are dealing with convolution of the  product of two  generalized functions
with the test function $H(w\,,v,\,\xi)$.
In order to assign meaning to this ill defined expression as it is done in eq.~(\ref{I2_start}),
$H(w,\mp \xi',\xi)$
and
$J^{(\pm)}(w,\xi)$
and their first derivatives in
$w$
should be  continuous
in the vicinity of
$w=\xi$.
One can check that these assumptions are justified by the use of the
spectral representation
(\ref{Spectral_for_GPDs_x123}) with continuous input quadruple distributions
vanishing at the borders of their domain of definition.

We obtain the following  contributions  to the real and imaginary parts of the amplitude from (\ref{I2_start}):
\be
&&
\re I_{II}^{(-, \pm)}(\xi) \nonumber \\ && = \pm \pi^2  \left( \frac{d H(w,\,\mp \xi',\xi)}{d w} \right)_{w=\xi}-2 J^{(\pm)}(\xi,\,\xi)+
\mathcal{P} \!\! \int_{-1}^1 dw  \frac{1}{(w-\xi)} \frac{ \left( J^{(\pm)}(w,\xi)-J^{(\pm)}(\xi,\xi) \right)}{(w-\xi)};
\nonumber \\ &&
\nonumber \\ &&
\im I_{II}^{(-, \pm)}(\xi) \nonumber \\ &&= \pm 2 \pi H(\xi,0,\xi) \mp \pi \mathcal{P} \!\! \int_{-1}^1 dw \frac{1}{(w-\xi)}
\frac{(H(w,\,\mp \xi',\xi)- H(\xi,0,\xi))}{(w-\xi)} %
- \pi \left(
\frac{d J^{(\pm)}(w,\xi)}{dw}
\right)_{w=\xi}.
\nonumber \\ &&
\label{ReIm_I2}
\ee

The formulas for the calculation of the real and the imaginary parts of
$I_{I}^{(\pm, \pm)}(\xi)$
and
$I_{II}^{(-, \pm)}(\xi)$
in  the model based on the factorized Ansatz for quadruple distributions
(\ref{Factorized_ansatz_xi=1})
with input from the
soft-pion theorem at
$\xi=1$
and with the profile function
$h$
given by
(\ref{Profile_h})
are summarized in  Appendix~\ref{App_Calc_Ampl}.

We are going now to present the results of calculation of $\mathcal{I}(\xi)$
in our composite model for $\pi N$ TDAs of Sec.~\ref{Section_Spectral_Rep}.
As it was already pointed out, the coefficients
$T_\alpha$, $T'_\alpha$ (\ref{Def_IIprime}),
which can be read from the Table~I of Ref.~\cite{Lansberg:2007ec}, are defined
with respect to the alternative parametrization of $\pi N$ TDAs.
The relation between that parametrization and the one employed
in the present paper is summarized in
Appendix~\ref{App_A}.
Within the parametrization of Ref.~\cite{Lansberg:2007ec},
$\mathcal{I}(\xi)$ receives contributions only from
$\pi N$ TDAs
$\{ V_1,A_1,T_1,T_4 \}_{\text{\cite{Lansberg:2007ec}}}$
while
$\mathcal{I}'(\xi)$
receives contributions only from
$\pi N$ TDAs
$\{ V_2,A_2,T_2,T_3 \}_{\text{\cite{Lansberg:2007ec}}}$.

One may establish
the following relations for the spectral part of  $\pi N$ TDA model  based on
the factorized Ansatz (\ref{Factorized_ansatz_xi=1}) with input from the soft-pion theorem:
\be
\left.{ \{V_{1},A_{1},T_{1}\}^{\pi N}}(x_1,x_2,x_3,\xi) \right|^{\text{\cite{Lansberg:2007ec}}}_{\rm spectral \; part}
=
\left.{ \{V_{1},A_{1},T_{1}\}^{\pi N}}(x_1,x_2,x_3,\xi) \right|^{\text{\cite{Pire:2011xv}  \&  \text{this paper}}}_{\rm spectral \; part}
\ee
and
\be
\left.{ \{V_{2},A_{2},T_{2},T_{3},T_4\}^{\pi N}}(x_1,x_2,x_3,\xi)  \right|^{\text{\cite{Lansberg:2007ec}}}_{\rm spectral \; part}
=
0\,.
\ee

Now we consider the nucleon pole part
(\ref{Nucleon_exchange_contr_VAT})
of  the two component model for $\pi N$
TDAs
\be
&&
\{V_{1},A_{1},T_{1}\}^{\pi N}
(x_1,x_2,x_3,\xi)
\big|_{N (940)}^{\text{\cite{Lansberg:2007ec}}}=
\frac{1-\xi}{1+\xi}
\{V_{1},A_{1},T_{1}\}^{\pi N}
(x_1,x_2,x_3,\xi)
\big|_{N (940)}^{\text{\cite{Pire:2011xv}  \&  \text{this paper}}};
\nonumber \\ &&
\{V_{2},A_{2} \}^{\pi N}
(x_1,x_2,x_3,\xi)
\big|_{N (940)}^{\text{\cite{Lansberg:2007ec}}}=
\{V_{1},A_{1},T_{1}\}^{\pi N}
(x_1,x_2,x_3,\xi)
\big|_{N (940)}^{\text{\cite{Pire:2011xv}  \&  \text{this paper}}};
\nonumber \\ &&
(T_{2}+T_{3} )^{\pi N}
(x_1,x_2,x_3,\xi)
\big|_{N (940)}^{\text{\cite{Lansberg:2007ec}}}=
2 T_{1}^{\pi N}
(x_1,x_2,x_3,\xi)
\big|_{N (940)}^{\text{\cite{Pire:2011xv}  \&  \text{this paper}}}.
\ee
Consequently,  in our model, the following relation holds for the nucleon pole contribution into
$\mathcal{I}$ and $\mathcal{I}'$:
\be
{\rm Re} \, \mathcal{I} (\xi ,\Delta^2) \big|_{\rm N (940)}=\frac{1-\xi}{1+\xi} \, {\rm Re} \, \mathcal{I}' (\xi ,\Delta^2)\big|_{\rm N (940)}.
\ee

\begin{figure}[H]
 \begin{center}
 \epsfig{figure=  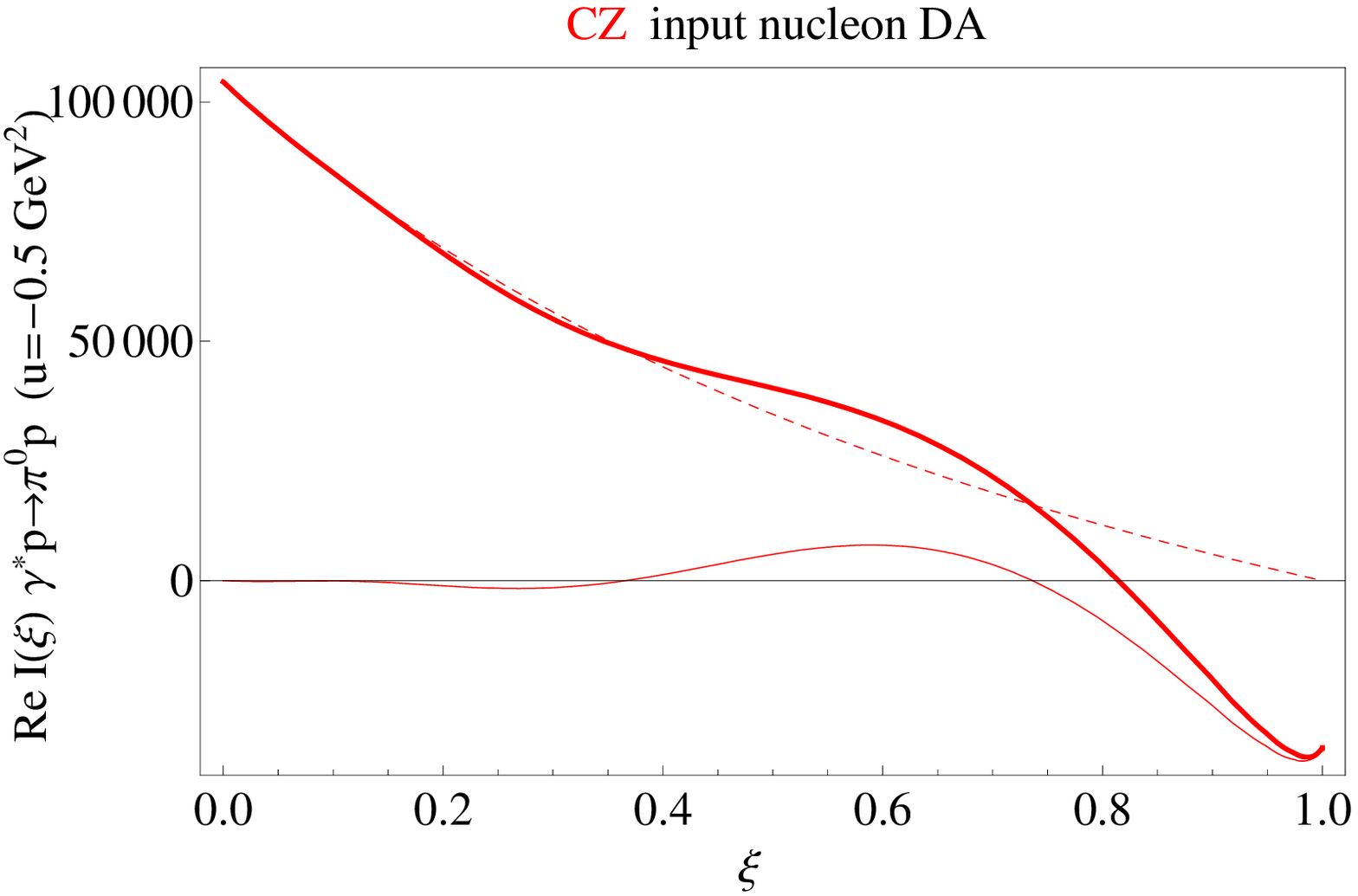 , height=5cm}
  \epsfig{figure= 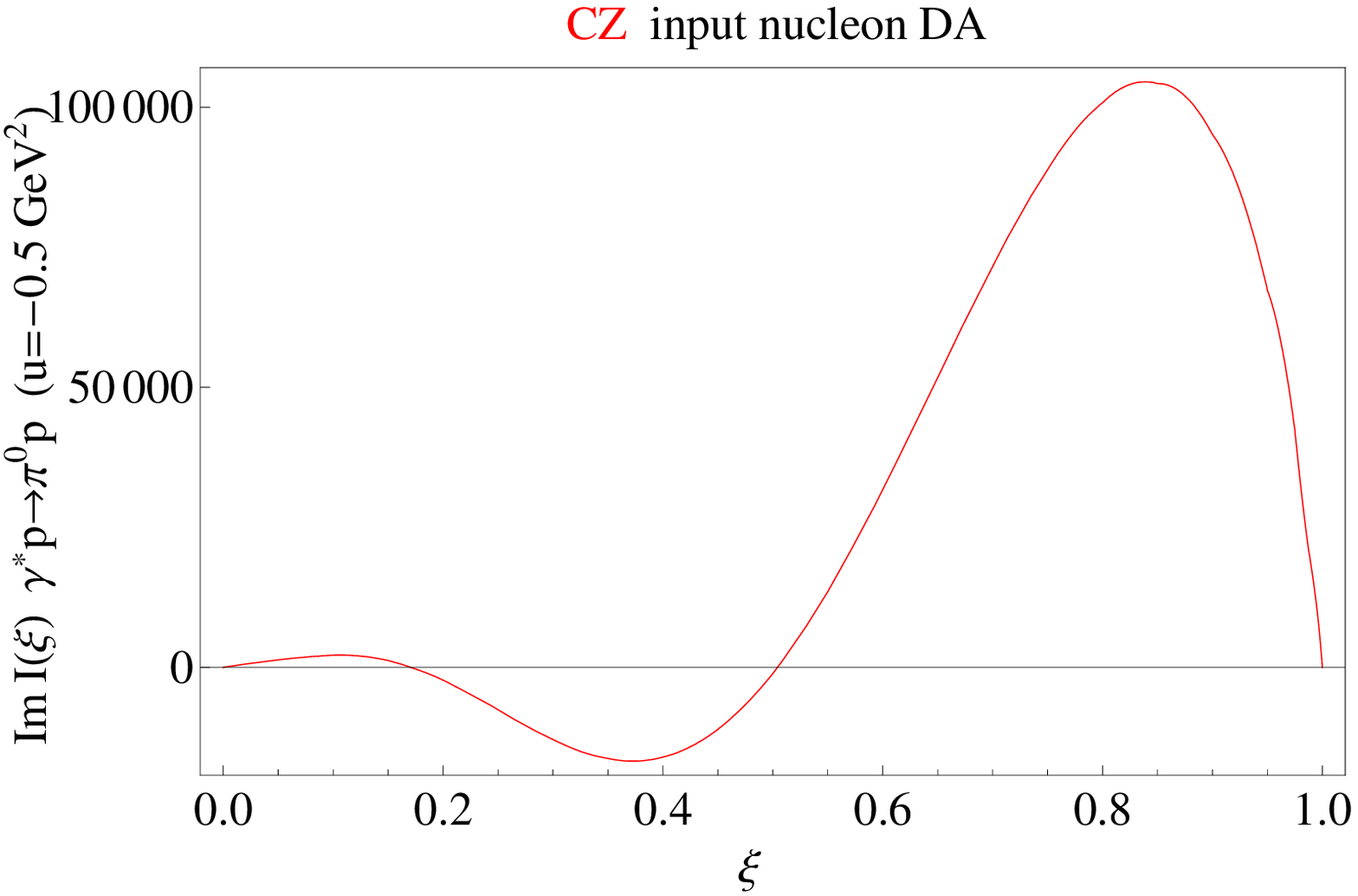 , height=5cm}
  \epsfig{figure=  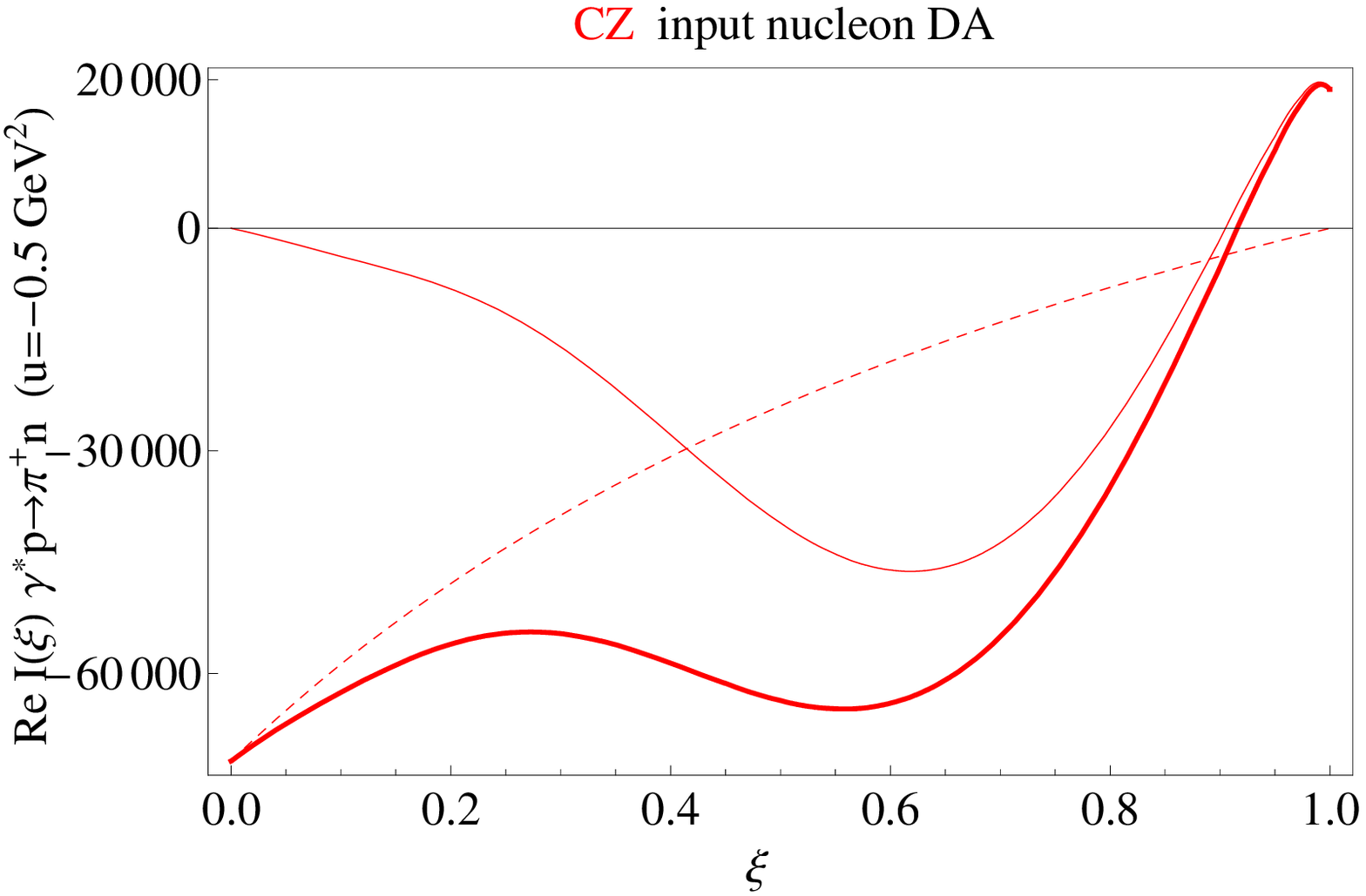 , height=5cm}
  \epsfig{figure= 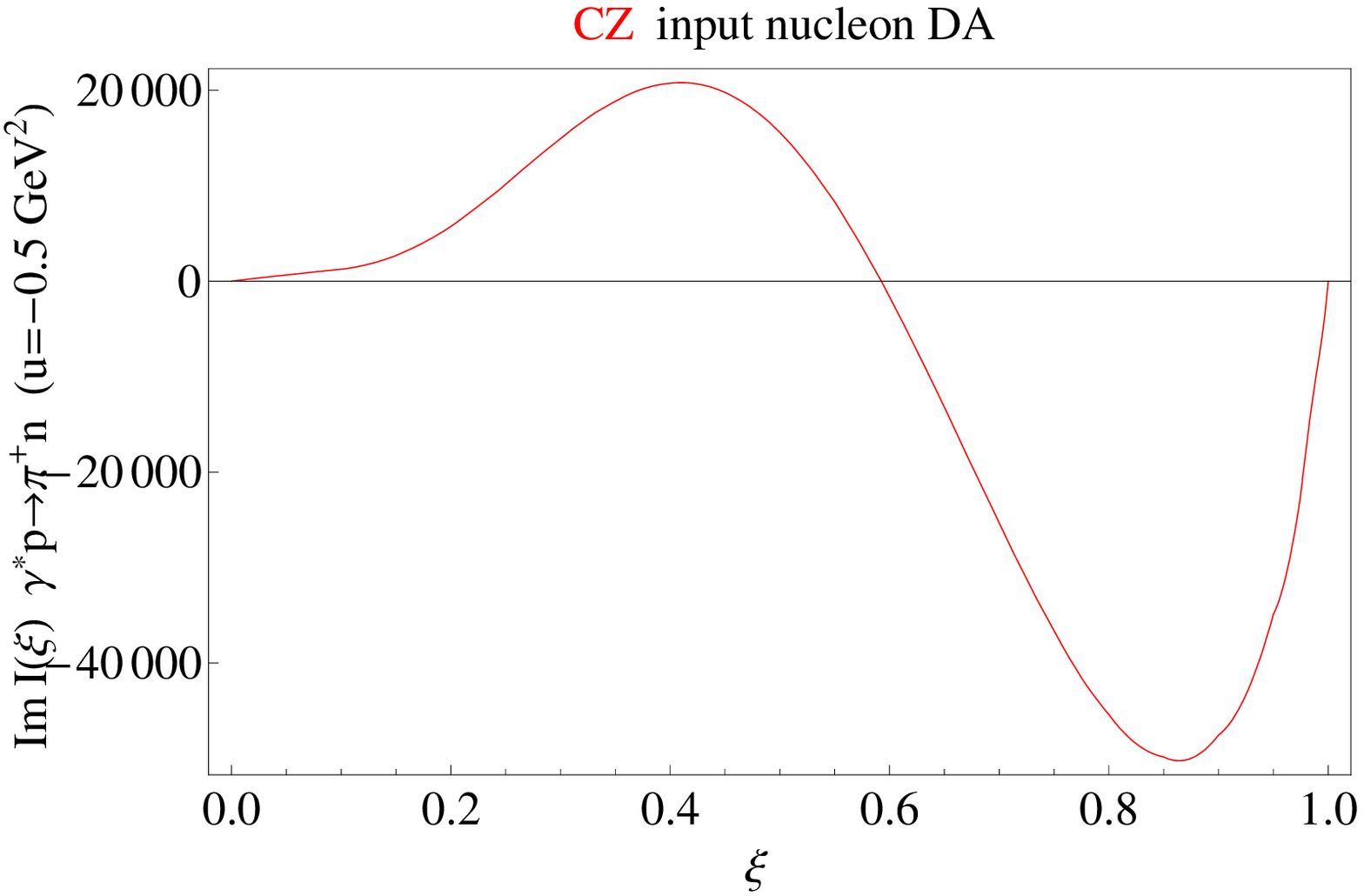 , height=5cm}
     \caption{ Real and imaginary parts of $\mathcal{I}(\xi)$
     for $\gamma^* p \to \pi^0 p$ and $\gamma^* p \to \pi^+ n$ backward production
     as functions of $\xi$ computed in our two component model for $\pi N$ TDAs. Dashed lines:
     nucleon pole contribution into $\re \mathcal{I}(\xi)$; thin solid line: spectral representation
     with input from the soft-pion theorem; solid line: sum of two contributions (for the real part).
      }
\label{Fig_ReIm_I}
\end{center}
\end{figure}

On Fig.~\ref{Fig_ReIm_I} we present the results in our model for the real and imaginary parts of $\mathcal{I}(\xi)$
for backward production of $\pi^0$ (two upper panels) and $\pi^+$  (two lower panels) off proton showing separately
the spectral part, the pole part and their sum. The CZ phenomenological solution
\cite{Chernyak_Nucleon_wave}
for the nucleon DA is used as the numerical input
for our model.
For small $\xi$ ($\xi  \lesssim 0.3 \div 0.5$), the real part is dominated by the contribution of the nucleon pole. The contribution
of the spectral part to the real part becomes relatively more important for larger $\xi$.
The nucleon pole contribution is purely real. The apparence of a significant imaginary part stemming
from the spectral component of the model is a distinctive feature of our approach. This is crucial for the non-vanishing
of the  transverse target single spin asymmetry for backward pion electroproduction discussed in
Sec.~\ref{Sec_CS_AS}.

\section{Unpolarized cross section and  transverse target single spin asymmetry for backward  pion production}
\label{Sec_CS_AS}

Let us first specify our conventions for the backward pion electroproduction cross section.
In the one photon exchange approximation, the
unpolarized cross section of hard leptoproduction of a pion off a nucleon (\ref{reaction})
can be decomposed as follows
\cite{Kroll:1995pv}:
\be
&&
\frac{d^4 \sigma}{ds dQ^2 d \varphi dt}= \frac{\alpha_{\rm em} (s-M^2)}{4 (2 \pi)^2 ( {k}_0^L)^2 M^2 Q^2 (1-\varepsilon)}
\nonumber \\ &&
\times
\Big(
{
\frac{d \sigma_T}{dt}
+ \varepsilon \frac{d \sigma_L}{dt}+
\varepsilon \cos 2\varphi \frac{d \sigma_{TT}}{dt}+
\sqrt{2 \varepsilon (1+\varepsilon)} \cos \varphi \frac{d \sigma_{LT}}{dt} }\
\Big),
\label{Def_CS_Kroll}
\ee
where
$\varphi$
is the angle between the leptonic and hadronic planes;
$s=(p_1+q)^2 \equiv W^2$
and
$t=(p_2-p_1)^2$
are the Mandelstam variables;
${k}_0^L$
is the initial state electron energy in the laboratory (LAB) frame  (beam energy).
$\varepsilon$
is the polarization parameter of the virtual photon:
\be
\varepsilon=\Big[1+2\frac{\big({k}_0^L-{k'}_0^L \big)^2+Q^2}{Q^2} \tan^2 \frac{\theta^L_e }{2}\Big]^{-1},
\ee
where
${k'}_0^L $
is the energy of the final state electron in the LAB frame and
$\theta^L_e $
is the electron scattering angle in the LAB frame.

\begin{figure}[H]
 \begin{center}
\epsfig{figure= 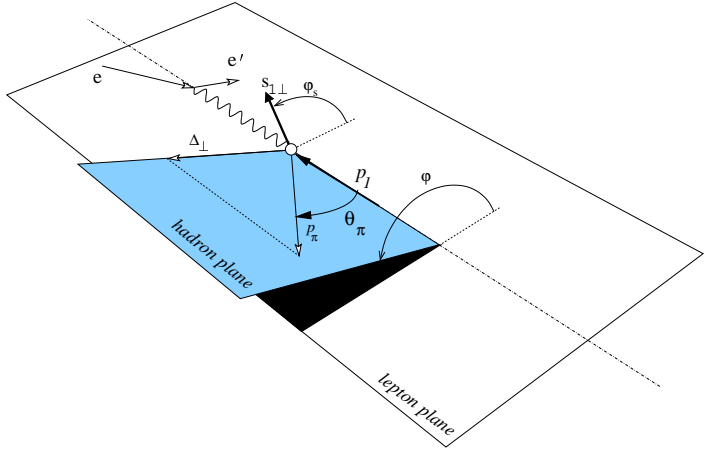, height=6cm}
      \caption{ Kinematics of electroproduction of a pion off a nucleon in the CMS frame of
      $\gamma^*$ nucleon.
}
\label{Fig_Kin}
\end{center}
\end{figure}

Within the suggested factorization mechanism for backward pion leptoproduction, only the
transverse cross section
$\frac{d \sigma_T}{dt}$
receives a contribution at the leading twist level.
Using the explicit expression relating
scattering amplitudes of leptoproduction to those for
virtual photoproduction
(eq.~(2.12) of Ref.~\cite{Kroll:1995pv}),
we express  $\frac{d^2 \sigma_T}{d \Omega_\pi}$
in the center of mass (CMS)
system of the pion and final nucleon
through
$\gamma^* N \rightarrow N \pi$
helicity amplitudes
${\cal M}^\lambda_{s_1s_2}$
defined in
(\ref{helicity ampl}):
\be
&&
\frac{d^5 \sigma}{d E' d \Omega_{e'} d \Omega_\pi}
\nonumber \\ &&
= \Gamma  \times \frac{\Lambda(s,m^2,M^2)}{128 \pi^2 s (s-M^2)}
\sum_{s_1, \, s_2}
\Big\{
\frac{1}{2}
\big(
|{\cal M}^1_{s_1s_2}|^2+
|{\cal M}^{-1}_{s_1s_2}|^2
\big)
+...
\Big\}
= \Gamma \times \Big(  \frac{d^2 \sigma_T}{d \Omega_\pi} +...\Big).
\nonumber \\ &&
\label{CS_working}
\ee
Here,
$\Omega_{e'}$
is the differential solid angle for the scattered electron in the LAB frame;
$\Omega_\pi$
is  the differential solid angle of the produced pion in
$N' \pi$
CMS frame
(see Fig.~\ref{Fig_Kin}
for the definition of angular variables);
by dots, we denote the subleading twist terms supressed by powers of $1/Q$;
$\Lambda$
is the usual Mandelstam function
\be
\Lambda(x,y,z)= \sqrt{x^2+y^2+z^2-2xy-2xz-2yz}.
\label{Def_lambda}
\ee
$\Gamma$
is the virtual photon flux factor in Hand's convention
\cite{Hand_9388}
given by
\be
\Gamma = \frac{\alpha_{\rm em}}{2 \pi^2} \frac{{k'}_{0 }^L}{k_{0 }^L} \frac{s-M^2}{2 M Q^2} \frac{1}{1-\varepsilon}.
\ee

Our present goal is to establish the expression for the LO transverse
cross section through the helicity amplitudes defined in (\ref{helicity_ampl_rewr}).
We rewrite our formula (\ref{helicity_ampl_rewr}) as
\be
\mathcal{M}_{\lambda}^{s_1 s_2}=\mathcal{C} \frac{1}{Q^4} \bar{U}(p_2,s_2) \,  \Gamma_H \, U(p_1,s_1),
\label{Def_hel}
\ee
where
\be
\Gamma_H= \hat{\epsilon}(\lambda) \gamma_5 \mathcal{I}-\hat{\epsilon}(\lambda) \frac{\hat{\Delta}_T}{M} \gamma_5 \mathcal{I}'.
\ee
Let us now square the amplitude and sum over the transverse polarizations of the virtual photon and over the spin of
outgoing nucleon:
\be
&&
|\mathcal{M}_{T}^{s_1}|^2=|\mathcal{C}|^2 \frac{1}{Q^8}
\sum_{\lambda_T}
{\rm Tr}
\Big\{
(\hat{p}_2+M) \Gamma_H \frac{1+ \gamma_5 \hat{s}_1}{2} (\hat{p}_1+M) \gamma_0 \Gamma_H^\dag \gamma_0
\Big\}.
\label{Trace_to_compute}
\ee
Let us first consider the  trace
\be
&&
\sum_{\lambda_T} {\rm Tr}
\Big\{
(\hat{p}_2+M) \Gamma_H  (\hat{p}_1+M) \gamma_0 \Gamma_H^\dag \gamma_0
\Big\}=
\\ &&
-\sum_{\lambda_T}  \epsilon^\nu(\lambda)  \epsilon^{\mu *}(\lambda)
 {\rm Tr}
\Big\{
(\hat{p}_2+M) \big(\gamma^\nu \gamma_5 \, \mathcal{I} - \gamma^\nu \frac{\hat{\Delta}_T}{M} \gamma_5 \mathcal{I}' \big)
(\hat{p}_1+M)
\big(\gamma_5 \gamma^\mu  \, \mathcal{I}^* -  \gamma_5 \frac{\hat{\Delta}_T}{M} \gamma^\nu  (\mathcal{I}')^* \big)
\Big\}\,.
\nonumber
\ee
We employ the relation
\be
\sum_{\lambda_T}  \epsilon^\nu(\lambda) \epsilon^{\mu *}(\lambda)=-g^{\mu \nu}+\frac{1}{(p \cdot n)}(p^\mu n^\nu+p^\nu n^\mu)
\ee
to sum over the transverse polarizations of the virtual photon.
We use the backward kinematics for the reaction (\ref{reaction})
summarized in
\cite{Lansberg:2007ec}:
\be
&&
p_1 \cdot n= \frac{1+\xi}{2}; \ \ \ p_1 \cdot p= \frac{M^2}{2(1+\xi)};
\nonumber \\ &&
p_2 \cdot n=O(1/Q^2); \ \ \ p_2 \cdot p=\frac{Q^2}{4 \xi}+O(Q^0).
\ee
Then for the part which is independent of the nucleon spin we get:
\be
&&
\sum_{\lambda_T} {\rm Tr}
\Big\{
(\hat{p}_2+M) \Gamma_H  (\hat{p}_1+M) \gamma_0 \Gamma_H^\dag \gamma_0
\Big\}\nonumber \\ &&
=\frac{2 Q^2 (1+\xi)}{\xi} |\mathcal{I}|^2
-\frac{2 Q^2 (1+\xi)}{\xi} \frac{\Delta_T^2}{M^2} |\mathcal{I}'|^2 +O \left (   Q^0    \right) \,.
\ee

Now we turn to the nucleon spin dependent part of the trace
(\ref{Trace_to_compute}).
\be
&&
\sum_{\lambda_T} {\rm Tr}
\Big\{
(\hat{p}_2+M) \Gamma_H \gamma_5 \hat{s}_1  (\hat{p}_1+M) \gamma_0 \Gamma_H^\dag \gamma_0
\Big\} \nonumber \\ &&
=\sum_{\lambda_T}  \epsilon^\nu(\lambda) \epsilon^{\mu *}(\lambda)
\Big\{
(\hat{p}_2+M) \gamma^\nu   \hat{s}_1  (\hat{p}_1+M) \gamma_5 \frac{\hat{\Delta}_T}{M} \gamma^\mu
\Big\} \mathcal{I} (\mathcal{I}')^*
\nonumber \\ &&
+
\sum_{\lambda_T} \epsilon^\nu(\lambda) \epsilon^{\mu *}(\lambda)
\Big\{
(\hat{p}_2+M) \gamma^\nu \frac{\hat{\Delta}_T}{M}   \hat{s}_1  (\hat{p}_1+M) \gamma_5  \gamma^\mu
\Big\} \mathcal{I}' (\mathcal{I})^*
\nonumber \\ &&
=4 \frac{Q^2(1+\xi)}{M \xi} \varepsilon(n, p, s_1, \Delta_T)
 \big( -i \mathcal{I} (\mathcal{I}')^*+i \mathcal{I}' (\mathcal{I})^* \big)
\nonumber \\ &&
=-4 \frac{Q^2(1+\xi)}{ \xi} \frac{|\Delta_T|}{M}|\vec{s}_1| \sin(\varphi -\varphi_s) \rm Im(\mathcal{I}'(\mathcal{I})^*).
\label{Trace_2}
\ee
In the last line of (\ref{Trace_2}), we consider  $s_1$ as being purely transverse
and choose the reference frame so that $s_1$ has only an $x$-component.
Then
\be
\varepsilon(n, p, s_1, \Delta_T)= \frac{1}{2} |\Delta_T| |\vec{s}_1| \sin(\varphi -\varphi_s),
\ee
where
$\varphi$
is the angle between the leptonic and hadronic planes and
$\varphi_s$
is the angle between the leptonic plane the transverse
target spin (see Fig.~\ref{Fig_Kin}).
We employ the conventions in which
$  \varepsilon^{0123}=  1$ with $\gamma_5=- \frac{i}{4!} \varepsilon^{\mu \nu \rho \sigma} \gamma_\mu \gamma_\nu \gamma_\rho \gamma_\sigma $ .
Finally, we conclude that
\be
|\mathcal{M}_{T}^{s_1}|^2=|\mathcal{C}|^2 \frac{1}{Q^6} \frac{  (1+\xi)}{\xi}
\left(
 |\mathcal{I}|^2
- \frac{\Delta_T^2}{M^2} |\mathcal{I}'|^2-2 \frac{|\Delta_T|}{M}|\vec{s}_1|  \, {\rm Im} (\mathcal{I}'(\mathcal{I})^*) \sin \tilde{\varphi}
\right)+ O(1/Q^8),
\ee
where
$\tilde{\varphi} \equiv \varphi-\varphi_s$.

Hence, we establish the following formula for the LO unpolarized cross section
(\ref{CS_working})
through the coefficients
$\mathcal{I}$, $\mathcal{I}'$,
introduced in
(\ref{helicity_ampl_rewr}):
\be
\frac{d^2 \sigma_T}{d \Omega_\pi}= |\mathcal{C}|^2 \frac{1}{Q^6}
\frac{\Lambda(s,m^2,M^2)}{128 \pi^2 s (s-M^2)} \frac{1+\xi}{\xi}
\big(
|\mathcal{I}|^2
-  \frac{\Delta_T^2}{M^2} |\mathcal{I}'|^2
\big).
\label{Work_fla_CS}
\ee
Within our kinematics
\be
\Delta_T^2= \frac{(1 - \xi) \left (\Delta^2 - 2 \xi \left( \frac{M^2}{1 + \xi} - \frac{m^2} {1-\xi} \right) \right)}{1+\xi}\,,
\ee
where $m$ is the pion mass.

\begin{figure}[H]
 \begin{center}
 \epsfig{figure=  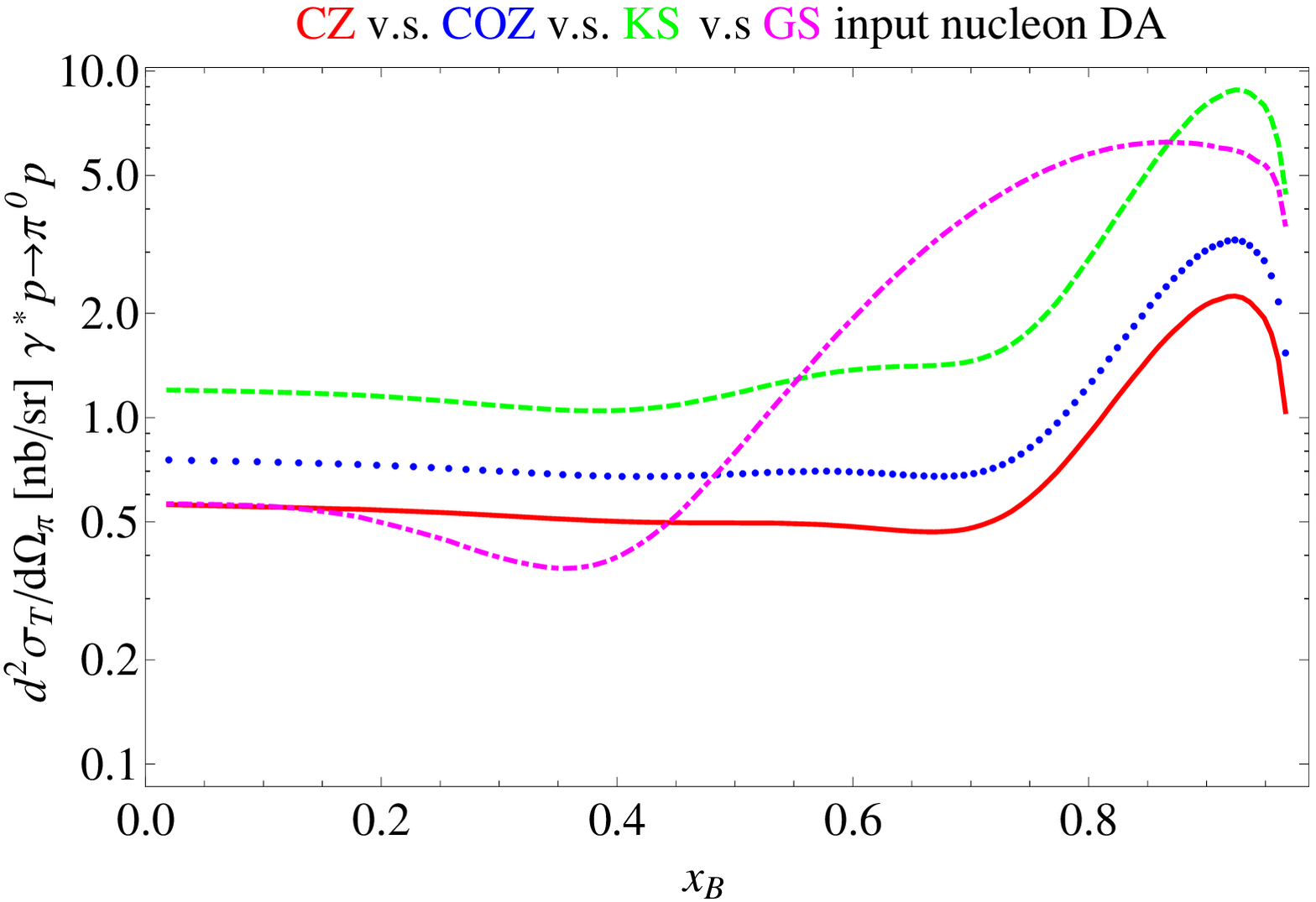 , height=7cm} \ \ \
 \epsfig{figure=  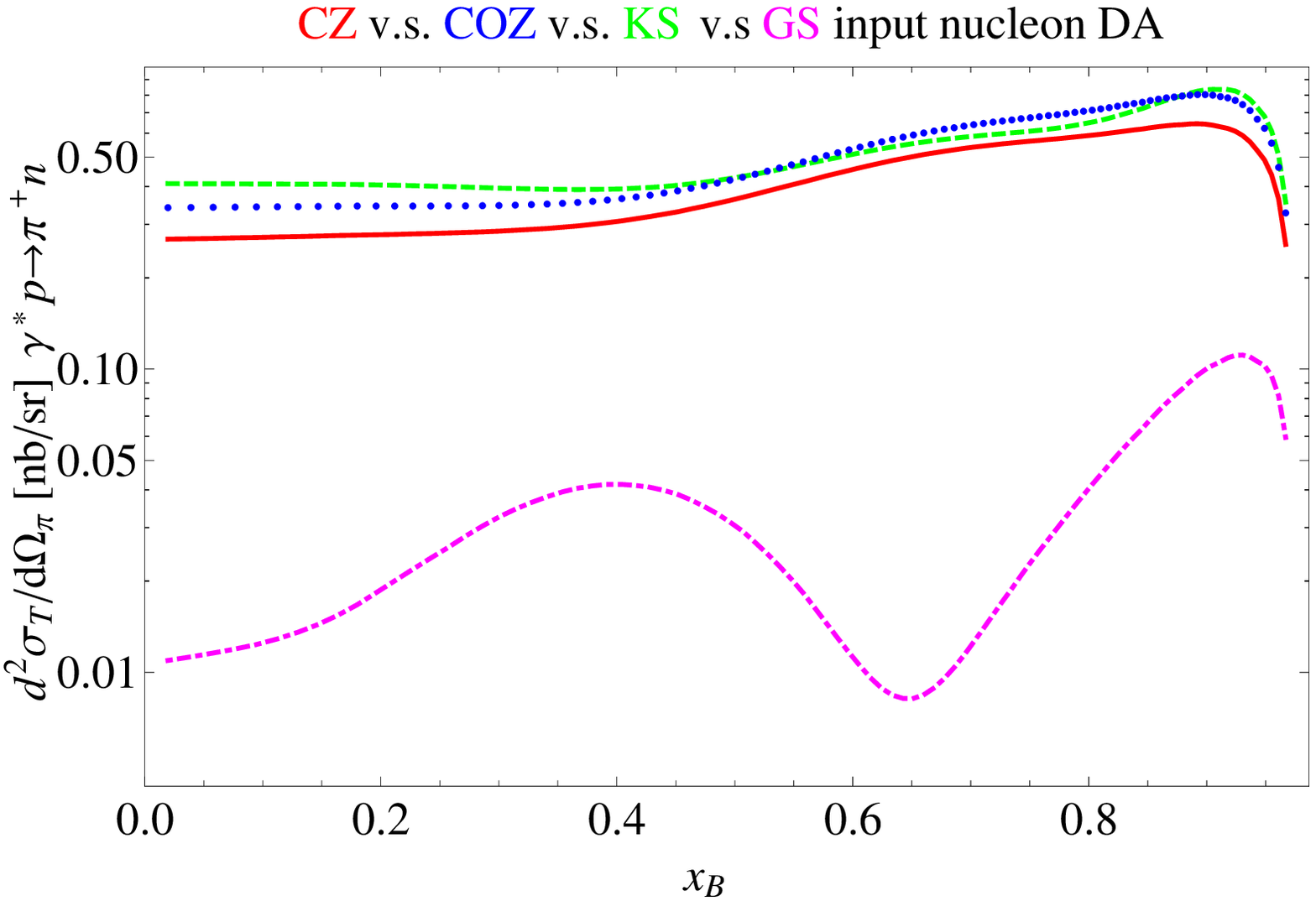 , height=7cm}
     \caption{
  Unpolarized cross section $\frac{d^2 \sigma_T}{d \Omega_\pi}$ (in ${\rm nb}/{\rm sr}$)
  for backward $\gamma^*p \to p \pi^0$  {\bf (upper panel)} and for backward $\gamma^*p \to n \pi^+$ {\bf (lower panel)}
  as the function of $x_{\rm B}$
  computed in the two component
  model for $\pi N$ TDAs for $Q^2=10 \, {\rm GeV}^2$, $u=-0.5 \, {\rm GeV}^2$
  as a function of $x_{\rm B}$.
  CZ \cite{Chernyak_Nucleon_wave} (solid lines), COZ \cite{Chernyak:1987nv} (dotted lines), KS \cite{King:1986wi}  (dashed lines)
and GS \cite{Gari:1986ue} (dash-dotted lines) nucleon DAs
   were used as inputs for our model.
}
\label{Fig_CS}
\end{center}
\end{figure}

On Fig.~\ref{Fig_CS}, we present our estimates for the unpolarized cross section
$\frac{d^2 \sigma_T}{d \Omega_\pi}$
(\ref{Work_fla_CS})
of backward production of $\pi^0$ and $\pi^+$ off protons for $Q^2=10 {\rm GeV}^2$
and $u=-0.5 {\rm GeV}^2$
in ${\rm nb}/{\rm sr}$ as function of $x_{B}$.

We use the two component model for $\pi N$ TDAs presented in Secs.~\ref{Section_Spectral_Rep} and \ref{Section_amplitude}.
In order to quantify the sensitivity of our model prediction on the input nucleon DAs,
we show the cross section for the case of four phenomenological solutions fitting the
nucleon electromagnetic form factor:
CZ \cite{Chernyak_Nucleon_wave} (solid lines),
Chernyak-Ogloblin-Zhitnitsky (COZ) \cite{Chernyak:1987nv} (dotted lines),
King and Sachrajda (KS) \cite{King:1986wi}  (dashed lines)
and Gari and Stefanis (GS) \cite{Gari:1986ue} (dash-dotted lines).
The magnitude of these cross sections is large enough for a detailed investigation to be carried
at high luminosity experiments such as J-lab@12GeV and EIC.
We recall that the scaling law for the cross section (\ref{Work_fla_CS}) is $1/Q^8$.

\begin{figure}[H]
 \begin{center}
 \epsfig{figure=  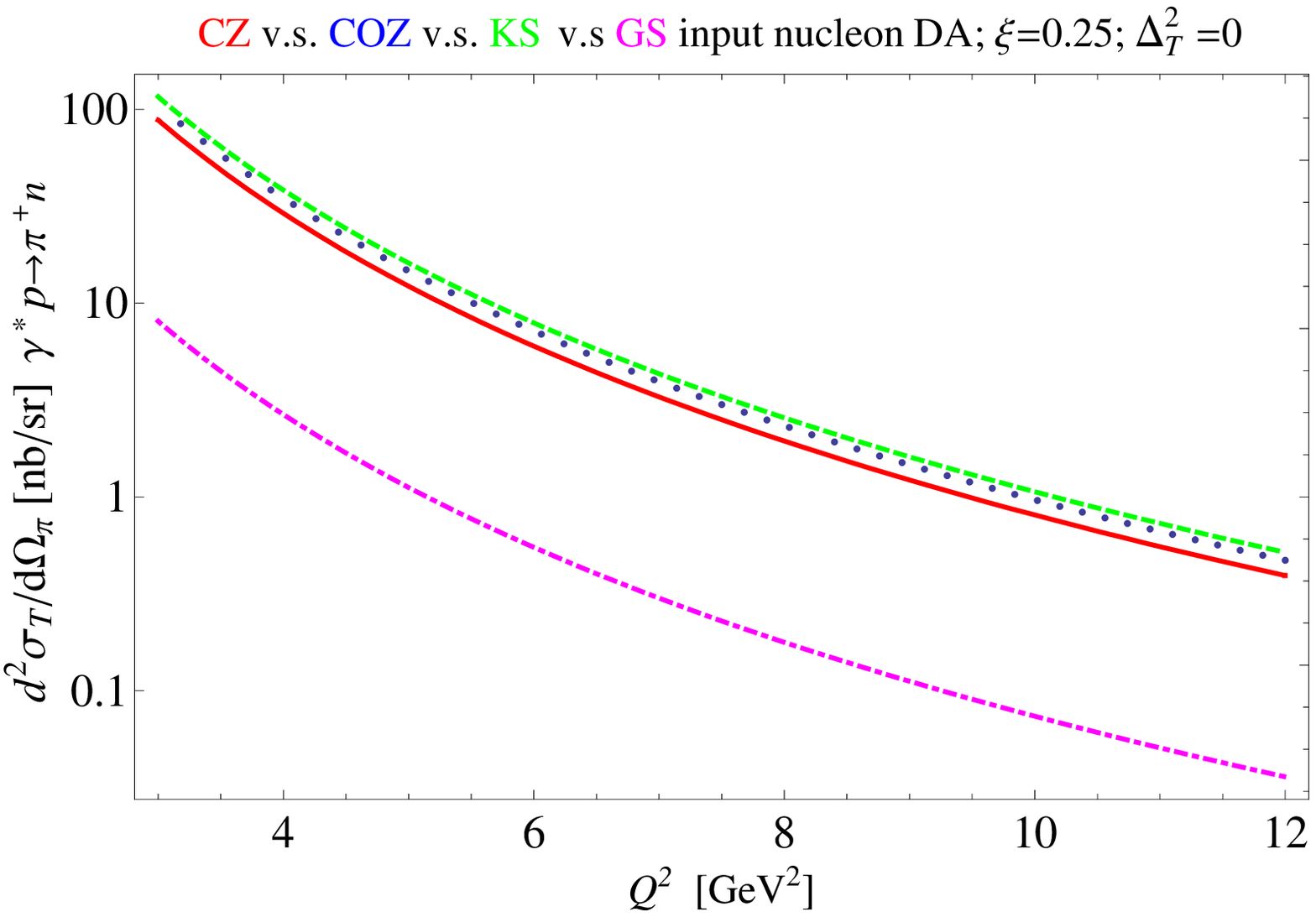 , height=7cm} \ \ \
 \epsfig{figure=  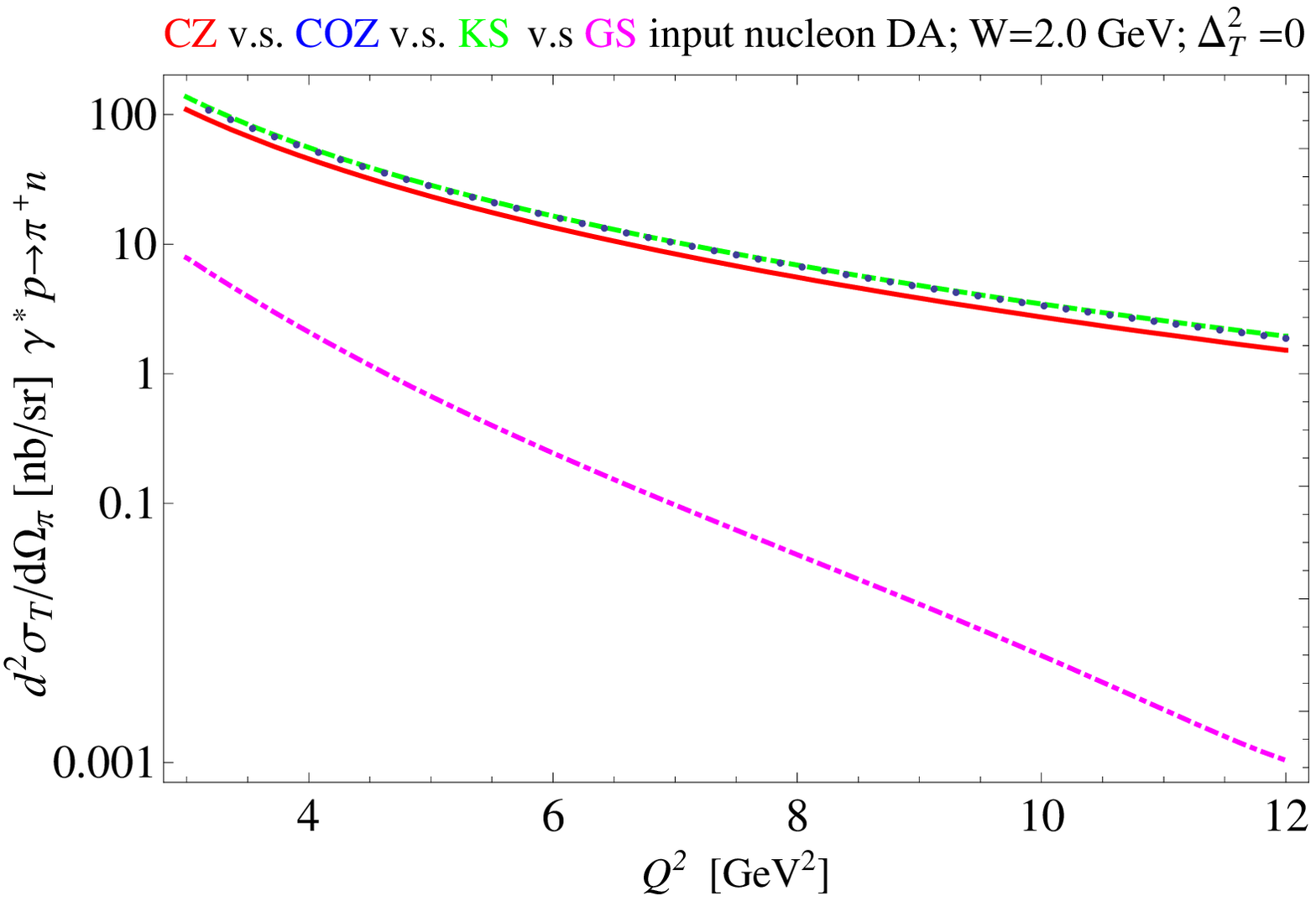 , height=7cm}
     \caption{
  {\bf Upper panel:} unpolarized cross section $\frac{d^2 \sigma_T}{d \Omega_\pi}$ (in ${\rm nb}/{\rm sr}$)
  for backward $\gamma^*p \to n \pi^+$ for fixed $\xi=0.25$ and $\Delta_T^2=0$ as a function of $Q^2$
  in the two component model for $\pi N$ TDAs.
  {\bf Lower panel:} unpolarized cross section $\frac{d^2 \sigma_T}{d \Omega_\pi}$ (in ${\rm nb}/{\rm sr}$)
  for backward $\gamma^*p \to n \pi^+$ for fixed $W=2.0$~GeV and $\Delta_T^2=0$ as a function of $Q^2$
  in the two component model for $\pi N$ TDAs.
   CZ \cite{Chernyak_Nucleon_wave} (solid lines), COZ \cite{Chernyak:1987nv} (dotted lines), KS \cite{King:1986wi}  (dashed lines)
and GS \cite{Gari:1986ue} (dash-dotted lines) nucleon DAs
  are used as inputs for the model.
}
\label{Fig_CS_WXI}
\end{center}
\end{figure}

On the upper panel of Fig.~\ref{Fig_CS_WXI}, we show the
$Q^2$
dependence of the unpolarized differential cross section
of
$\gamma^*p \to n \pi^+$
for fixed
$\xi=0.25$
which is characteristic for the J-lab kinematics and for
$\Delta_T^2=0$.
The  plot exhibits the expected  universal
$1/Q^8$
scaling behavior; the shape of
$\pi N$ TDAs, indeed, affects only the overall normalization.
On the lower panel of Fig.~\ref{Fig_CS_WXI}, similarly to
\cite{arXiv:0709.1946},
we show instead the same cross section as the function of
$Q^2$
for fixed
$W=2.0$~GeV and $\Delta_T^2=0$.
Let us emphasize that the scaling behavior in the latter case
is shadowed due to the fact that, for fixed
$W$, the running of
$Q^2$
also imposes variation of the scaling variable
$\xi$.

Asymmetries, being ratios of the cross sections,
are less sensitive to perturbative corrections.
Therefore, they are usually considered to be  more reliable to test the factorized description of
hard reactions.
For the backward pion electroproduction,
an evident candidate is the  transverse target single spin asymmetry (STSA)
\cite{Lansberg:2010mf}
defined as:
\be
&&
\mathcal{A}= \frac{1}{|\vec{s}_1|}
\left(
\int_0^\pi d \tilde{\varphi} |\mathcal{M}_{T}^{s_1}|^2 - \int_\pi^{2\pi} d \tilde{\varphi} |\mathcal{M}_{T}^{s_1}|^2
\right) \left(
\int_0^{2\pi} d \tilde{\varphi} |\mathcal{M}_{T}^{s_1}|^2
\right)^{-1}
 \nonumber \\ &&
 = -\frac{4}{\pi} \frac{\frac{|\Delta_T|}{M}  \, {\rm Im} (\mathcal{I}'(\mathcal{I})^*)}{|\mathcal{I}|^2
- \frac{\Delta_T^2}{M^2} |\mathcal{I}'|^2}.
\label{Def_asymmetry}
\ee
As argued in Sec.~\ref{Section_amplitude},
within the two component model for $\pi N$ TDAs,
the non-vanishing of the numerator in the last equality of
(\ref{Def_asymmetry})
is achieved due to the  interference of  the spectral part contribution into
$\im \mathcal{I}(\xi)$
and of the nucleon pole part contribution into
$\re \mathcal{I}'(\xi)$.

\begin{figure}[H]
 \begin{center}
 \epsfig{figure=  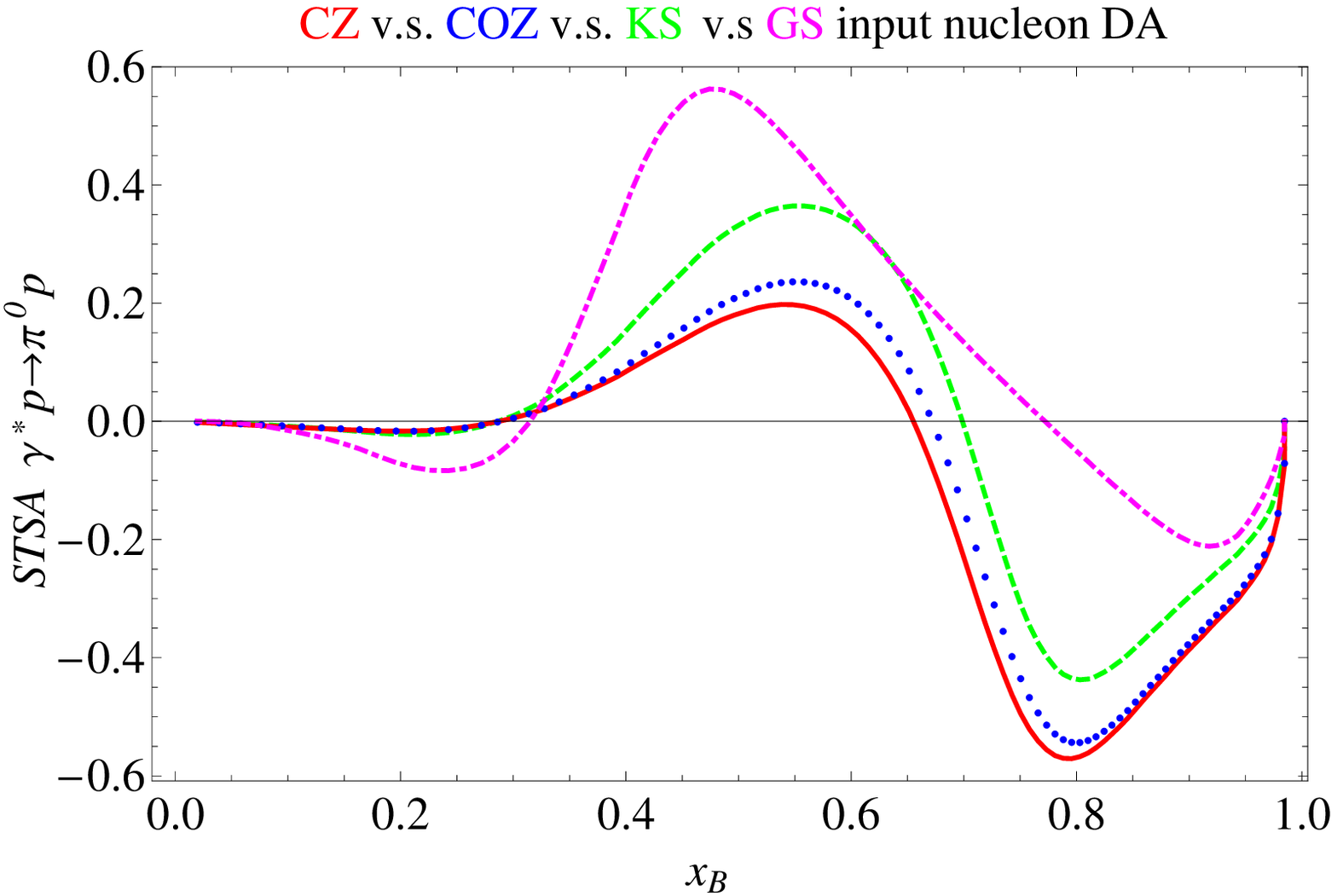 , height=7cm} \ \ \
 \epsfig{figure=  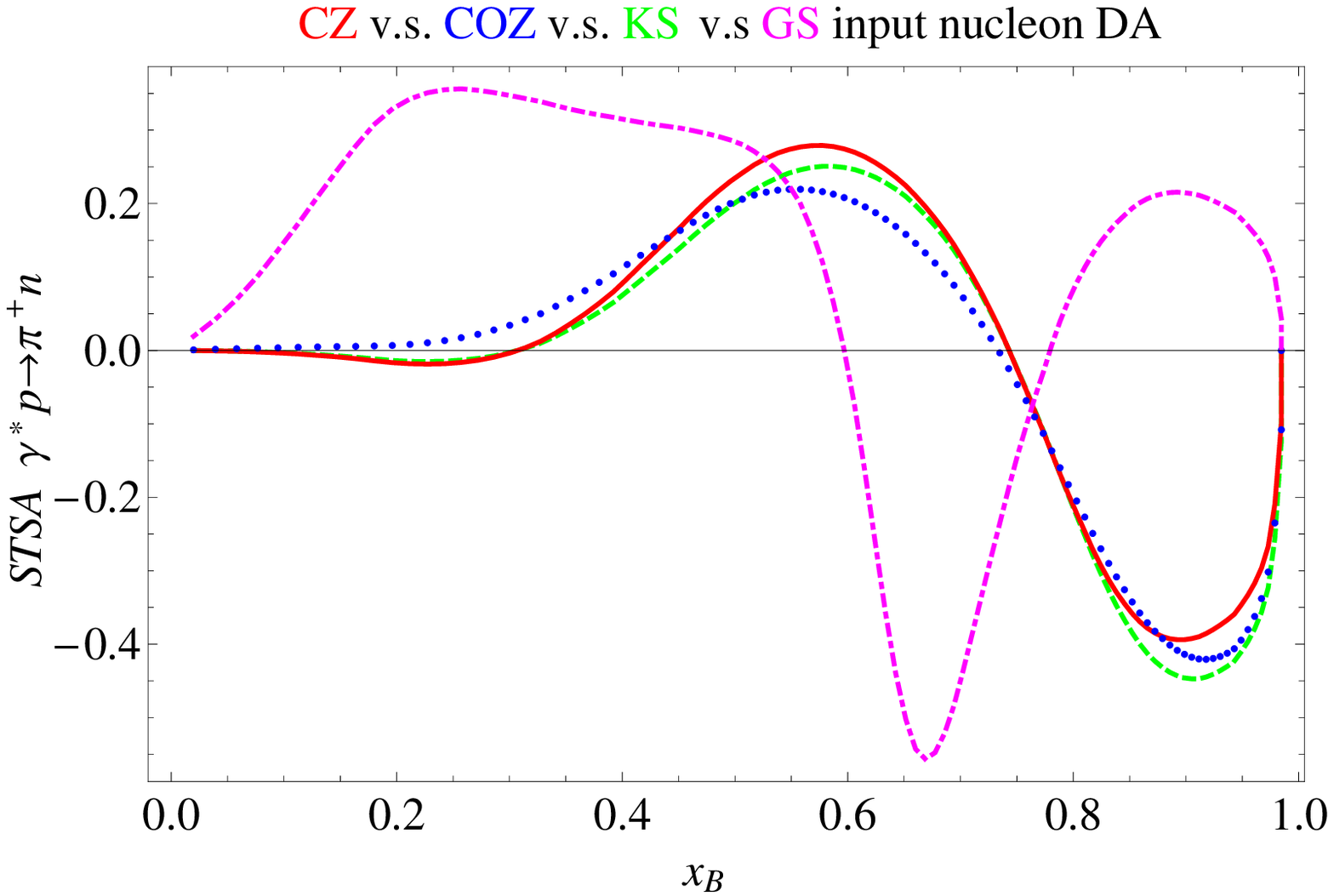 , height=7cm}
     \caption{
  Transverse target single spin asymmetry (\ref{Def_asymmetry})
  for backward $\gamma^*p^{\uparrow} \to p \pi^0$  {\bf (upper panel)} and for backward $\gamma^*p^{\uparrow} \to n \pi^+$ {\bf (lower panel)}
  as a function of $x_{ B}$
  computed with the two component 
  model for $\pi N$ TDAs for $Q^2=10 \, {\rm GeV}^2$, $u=-0.5 \, {\rm GeV}^2$
  as a function of $x_{ B}$. We show the results of our model with different input nucleon DAs: CZ (solid line), COZ (dotted line), KS (dashed line)  and GS (dot dashed line) used as input.
}
\label{Fig_AS}
\end{center}
\end{figure}

On Fig.~\ref{Fig_AS}, we show the result of our calculation of the
STSA for backward $\pi^0$ and $\pi^+$ electroproduction off protons for $Q^2=10 \, {\rm GeV}^2$ and $u=-0.5 \, {\rm GeV}^2$.
CZ \cite{Chernyak_Nucleon_wave}, COZ \cite{Chernyak:1987nv} , KS \cite{King:1986wi} and GS \cite{Gari:1986ue} nucleon DAs are used as phenomenological input  for our model.
We conclude that STSA turns out to be sizable in the valence region. Its measurement should therefore be
considered as a crucial test of the applicability of our collinear factorized scheme for
backward pion electroproduction.

\section{Conclusions}
\label{Section_conclusions}

For the first time, we have managed to build a consistent model
of $\pi N$ TDAs in their whole  domain of definition. It satisfies
general constraints imposed by the underlying QCD such as isospin symmetry, the Lorentz
invariance manifested through the polynomiality  property of the Mellin moments of
$\pi N$ TDAs in the light-cone momentum fractions, as well as the chiral properties.
We used this model in the estimates of the unpolarized
cross section and the transverse target single  spin asymmetry
for backward $\pi^+$ and $\pi^0$ electroproduction off protons.
Our results make us hope for bright experimental prospects for measuring baryon to meson TDAs with high luminosity lepton
beams such as COMPASS, J-lab@ 12 GeV and EIC
\cite{Boer:2011fh}.
Experimental data from J-lab@ 6 GeV on backward
$\pi^+$, $\pi^0$, $\eta$
and
$\omega$
meson production are currently being analyzed
\cite{J-lab}.
We eagerly wait for the first evidences of the factorized picture  for backward electroproduction
reactions as suggested in our approach.

\section*{Acknowledgements}

We are   thankful to
Aurore Courtoy,
Michel Guidal,
Valery Kubarovsky,
C\`{e}dric Lorc\`{e},
Kijun Park,
Barbara Pasquini,
Paul Stoler, Mark Strikman,
Samuel~Wallon and Christian Weiss
for many discussions and helpful comments.

This work is supported in part by the Polish NCN grant DEC-2011/01/B/ST2/03915
and by the French-Polish Collaboration Agreement Polonium.

\setcounter{section}{0}
\setcounter{equation}{0}
\renewcommand{\thesection}{\Alph{section}}
\renewcommand{\theequation}{\thesection\arabic{equation}}

\section{Parametrization of leading twist $\pi N$ TDAs}
\label{App_A}

The  parametrization of the leading twist-$3$ $\pi N$ TDAs of given flavor contents
suggested in \cite{Pire:2011xv} which we employ in this paper reads:
\be
&&
4 (P \cdot n)^3 \int   \left[ \prod_{j=1}^3 \frac{d \lambda_j}{2 \pi}   \right]
e^{i \sum_{k=1}^3 x_k \lambda_k (P \cdot n)} \langle \pi (p_\pi) | \widehat{O}_{\rho \, \tau \, \chi}( \lambda_1 n, \,\lambda_2 n, \, \lambda_3 n )| N (p_1) \rangle \nonumber \\ && =
\delta(x_1+x_2+x_3-2 \xi)
 \times i \frac{f_N}{f_\pi M}     \big[ V_1^{ \pi N}(x_1,x_2,x_3,\xi,\Delta^2)  (\hat{P}C)_{\rho \, \tau } (\hat{P} U)_\chi
 \nonumber \\ &&
 +
   A_1^{\pi N} (x_1,x_2,x_3,\xi,\Delta^2) (\hat{P} \gamma^5 C)_{\rho \, \tau } (\gamma^5 \hat{P} U)_\chi 
+
  T_1^{\pi N} (x_1,x_2,x_3,\xi,\Delta^2) (\sigma_{P \mu} C)_{\rho \, \tau } (\gamma^\mu \hat{P} U)_\chi
\nonumber \\ &&
+    V_2^{\pi N} (x_1,x_2,x_3,\xi,\Delta^2) (\hat{P}C)_{\rho \, \tau }  (\hat{\Delta}  U )_\chi
+
  A_2^{\pi N} (x_1,x_2,x_3,\xi,\Delta^2)(\hat{P} \gamma^5 C)_{\rho \, \tau } (\gamma^5 \hat{\Delta}  U )_\chi \nonumber \\ &&
+
 T_2^{\pi N} (x_1,x_2,x_3,\xi,\Delta^2) (\sigma_{P \mu} C)_{\rho \, \tau } (\gamma^\mu \hat{\Delta}  U )_\chi
+
\frac{1}{M}  T_3^{\pi N} (x_1,x_2,x_3,\xi,\Delta^2) (\sigma_{P \Delta } C)_{\rho \, \tau } (\hat{P} U)_\chi
 \nonumber \\ && +
\frac{1}{M}  T_4^{\pi N} (x_1,x_2,x_3,\xi,\Delta^2) (\sigma_{P \Delta } C)_{\rho \, \tau } (\hat{\Delta}   U )_\chi \big],
\label{Decomposition_piN_TDAs_new}
\ee
where
$f_\pi $   
is the pion weak decay constant and 
$f_N $ 
is a constant, which determines the value of the dimensional nucleon wave function 
at the origin; $U$ is the usual Dirac spinor and $C$ is the charge conjugation matrix.
We employ Dirac's ``hat'' notation:
$\hat{a} \equiv \gamma_{\mu} a^{\mu}$
and  adopt the conventions:
$\sigma^{\mu \nu}= \frac{1}{2} [\gamma^\mu, \, \gamma^\nu]$;
$\sigma^{v \nu} \equiv v_\mu \sigma^{\mu \nu}$,
where 
$v_\mu$ 
is an arbitrary 
$4$-vector.

The relation of the parametrization 
(\ref{Decomposition_piN_TDAs_new}) 
for 
$\pi N$ 
TDAs to that of
\cite{Lansberg:2007ec}
is given by:
\be
&&
\left.{ \{V_{1},A_{1},T_{1}\}^{\pi N}} \right|_{\text{\cite{Lansberg:2007ec}}}=
 \left(
\frac{1}{1+\xi} \{V_{1},A_{1},T_{1}\}^{\pi N}  -
\frac{2 \xi}{1+\xi} \{V_{2},A_{2},T_{2}\}^{\pi N}
\right)
\Big|_{\text{\cite{Pire:2011xv}  \&  \text{this paper}}} \,;  \nonumber \\  &&
\left.{\{V_{2},A_{2}\}^{\pi N}} \right|_{\text{\cite{Lansberg:2007ec}}}=
\left. \big(  {\{V_{2},A_{2}\}^{\pi N}}  +
\frac{1}{2}   {\{V_{1},A_{1}\}^{\pi N}} \big) \right|_{\text{\cite{Pire:2011xv}  \&  \text{this paper}}}\,;
 \nonumber \\  &&
\left.{T_{3}^{\pi N}} \right|_{\text{\cite{Lansberg:2007ec}}}=  
 \left.  {T_{2}^{\pi N}} \right|_{\text{\cite{Pire:2011xv}  \&  \text{this paper}}}+
\frac{1}{2}  \left.  {T_{1}^{\pi N}} \right|_{\text{\cite{Pire:2011xv}  \&  \text{this paper}}}
\nonumber \\ &&
\left.{T_{2}^{\pi N}} \right|_{\text{\cite{Lansberg:2007ec}}}=   \left.
\left(
\frac{1}{2} T_1^{\pi N}+T_2^{\pi N}+T_3^{\pi N}-2\xi T_4^{\pi N}
\right)
\right|_{\text{\cite{Pire:2011xv}  \&  \text{this paper}}}\,;
\nonumber \\ &&
\left.{T_{4}^{\pi N}} \right|_{\text{\cite{Lansberg:2007ec}}}=   \left. \left(  \frac{1+\xi}{2}  T_{3}^{\pi N}    +
 (1+\xi)  {T_{4}^{\pi N}} \right) \right|_{\text{\cite{Pire:2011xv}  \&  \text{this paper}}}\,.
 \label{Old_to_new}
\ee

\section{An alternative form of spectral representation for GPDs and baryon to meson TDAs}
\subsection{GPD in the ERBL and DGLAP regions}

From (\ref{Spectral_GPD_xi=1}), one can derive the following expressions for GPD in the DGLAP and the ERBL regions:
\begin{enumerate}
\item for $-1 \le x \le -\xi$ (DGLAP~$1$ region):
\be
H(x,\xi)= \frac{1}{1-\xi} \int_{-1}^{\frac{1-\xi+2x}{1+\xi}} d \kappa F \Big(\kappa, \frac{\kappa(1+\xi)-2x}{1-\xi} \Big);
\ee
\item For $-\xi \le x \le \xi$ (ERBL region):
\be
H(x,\xi)= \frac{1}{1-\xi} \int_{\frac{-1+\xi+2x}{1+\xi}}^{\frac{1-\xi+2x}{1+\xi}} d \kappa F \Big(\kappa, \frac{\kappa(1+\xi)-2x}{1-\xi} \Big);
\ee
\item For $\xi \le x \le 1$ (DGLAP~$2$ region):
\be
H(x,\xi)= \frac{1}{1-\xi} \int_{\frac{-1+\xi+2x}{1+\xi}}^{1} d \kappa F \Big(\kappa, \frac{\kappa(1+\xi)-2x}{1-\xi} \Big).
\ee
\end{enumerate}

\subsection{Set of working formulas for $\pi N$ TDAs in the ERBL-like and DGLAP-like regions}
\label{App_Calc_TDA}
In order to be able to compute $\pi N$ TDAs from the spectral
representation
(\ref{Spectral_for_GPDs_kappa_theta}),
we   perform  integrals over 
$\mu$ 
and 
$\lambda$  
with the help of two $\delta$-functions.
We omit the index 
$i$ 
referring to the choice of quark-diquark coordinates in the formulas 
of this Appendix.

The resulting domain of integration in 
$(\kappa, \, \theta)$ 
is defined by the inequalities:
\be
&&
-1 \le \kappa \le 1\,; \ \ \ \ \ \ -\frac{1-\kappa}{2} \le \theta \le \frac{1-\kappa}{2}\,;  \nonumber \\  &&
\frac{-1+\xi+2 w}{1+\xi} \, \le \, \kappa \le \frac{1-\xi+2 w}{1+\xi}\,; \nonumber \\  &&
\frac{\kappa}{2} - \frac{1}{1+\xi}(w-2v+ \frac{1-\xi}{2}) \, \le \, \theta  \, \le \,
-\frac{\kappa}{2}+ \frac{1}{1+\xi}(w+2v+ \frac{1-\xi}{2}).
\ee

Below, we summarize the explicit expressions for $\pi N$ TDAs from the spectral representation
(\ref{Spectral_for_GPDs_kappa_theta})
in the ERBL-like and DGLAP-like regions. Let us introduce the following notation for
the integrand:
\be
F(....) \equiv F \left(\kappa, \, \theta,\,\frac{\kappa(1+\xi)-2w}{1-\xi}, \, \frac{\theta(1+\xi)-2v}{1-\xi} \right).
\ee
\begin{enumerate}
\item For $w \in [-1;\, -\xi]$ and $v \in [\xi'; \, 1-\xi'+\xi]$ (DGLAP-like type I domain):
\be
H(w,v,\xi) 
 = \frac{1}{(1-\xi)^2}
\int_{-1}^{\frac{1-2v+w}{1+\xi}} d \kappa
\int_{\frac{\kappa}{2} - \frac{1}{1+\xi}(w-2v+ \frac{1-\xi}{2})}^{\frac{1-\kappa}{2}} d \theta
F(....)
\label{TDA_skew1_DGLAP-like_type_I_1}
\ee

\item For $w \in [-1;\, -\xi]$ and $v \in [-\xi'; \,  \xi' ]$ (DGLAP-like type II domain):
\be
H(w,v,\xi)
 =
\frac{1}{(1-\xi)^2}
\int_{-1}^{ \frac{1-\xi+2 w}{1+\xi} } d \kappa
\int_{\frac{\kappa}{2} - \frac{1}{1+\xi}(w-2v+ \frac{1-\xi}{2})}^{-\frac{\kappa}{2}+ \frac{1}{1+\xi}(w+2v+ \frac{1-\xi}{2})} d \theta
F(....)
\label{TDA_skew1_DGLAP-like_type_II_1}
\ee

\item For $w \in [-1;\, -\xi]$ and $v \in [-1+\xi'-\xi; \,  -\xi' ]$ (DGLAP-like type I domain):
\be
H(w,v,\xi) 
=
\frac{1}{(1-\xi)^2}
\int_{-1}^{ \frac{1+2v+ w}{1+\xi} } d \kappa
\int_{-\frac{1-\kappa}{2} }^{-\frac{\kappa}{2}+ \frac{1}{1+\xi}(w+2v+ \frac{1-\xi}{2})} d \theta
F(....). 
\label{TDA_skew1_DGLAP-like_type_I_2}
\ee

\item For $w \in [-\xi;\, \xi]$ and $v \in [\xi';\,1-\xi+\xi']$ (DGLAP-like type II domain):
\be
H(w,v,\xi) 
 =
\frac{1}{(1-\xi)^2}
\int_{ \frac{-1+\xi+2w}{1+\xi}}^{ \frac{1-2v+ w}{1+\xi} } d \kappa
\int_{\frac{\kappa}{2} - \frac{1}{1+\xi}(w-2v+ \frac{1-\xi}{2})}^{\frac{1-\kappa}{2} } d \theta
F (....).  
\label{TDA_skew1_DGLAP-like_type_II_2}
\ee

\item For $w \in [-\xi;\, \xi]$ and $v \in [-\xi';\,  +\xi']$ (ERBL-like   domain):
\be
H(w,v,\xi) \nonumber 
=
\frac{1}{(1-\xi)^2}
\int_{ \frac{-1+\xi+2w}{1+\xi}}^{ \frac{1-\xi+ 2w}{1+\xi} } d \kappa
\int_{\frac{\kappa}{2} - \frac{1}{1+\xi}(w-2v+ \frac{1-\xi}{2})}^{-\frac{\kappa}{2}+ \frac{1}{1+\xi}(w+2v+ \frac{1-\xi}{2})} d \theta
F (....)
. 
\label{TDA_skew1_ERBL-like}
\ee

\item For $w \in [-\xi;\, \xi]$ and $v \in [-1+\xi-\xi';\, -\xi']$ (DGLAP-like type II domain):
\be
H(w,v,\xi) 
=
\frac{1}{(1-\xi)^2}
\int_{ \frac{-1+\xi+2w}{1+\xi}}^{ \frac{1+2v+ w}{1+\xi} } d \kappa
\int_{ - \frac{1-\kappa}{2}}^{-\frac{\kappa}{2}+ \frac{1}{1+\xi}(w+2v+ \frac{1-\xi}{2})} d \theta
F(....)
.
\label{TDA_skew1_DGLAP-like_type_II_3}
\ee

\item For $w \in [\xi;\, 1]$ and $v \in [ -\xi';\,  1-\xi+\xi']$, the result coincides with (\ref{TDA_skew1_DGLAP-like_type_II_2})
as it certainly should be, since this is the part of the same DGLAP-like type II domain.
\item For $w \in [\xi;\, 1]$ and $v \in [ \xi';\, -\xi']$ (DGLAP-like type II domain):
\be
H(w,v,\xi)= 
\frac{1}{(1-\xi)^2}
\int_{ \frac{-1+\xi+2w}{1+\xi}}^{ 1 } d \kappa
\int_{ - \frac{1-\kappa}{2}}^{\frac{1-\kappa}{2}} d \theta
F (....)
.  
\label{TDA_skew1_DGLAP-like_type_I_3}
\ee

\item For $w \in [\xi;\, 1]$ and $v \in [ -1+\xi-\xi';\, \xi']$, the result coincides with (\ref{TDA_skew1_DGLAP-like_type_II_3}),
since this is the part of the same DGLAP-like type II domain.

\end{enumerate}

\setcounter{equation}{0}
\section{On the relevant generalized functions}
\label{App_GenF}

Sohotsky's formula (see {\it e.g.} Chapter~II of
\cite{Vladimirov})
reads:
\be
\frac{1}{x \pm i 0}= \mp i \pi \delta (x)+ \mathcal{P} \frac{1}{x}\,,
\label{Soh_formula}
\ee
where
$\mathcal{P} $
stands for the Cauchy principal value prescription.
The generalized function
$\mathcal{P}\frac{1}{x^2}$
is then defined as
$
\frac{d}{dx} \mathcal{P}\frac{1}{x}=-\mathcal{P}\frac{1}{x^2}
$.
For an arbitrary test function $\varphi(x)$,
\be
\left(\mathcal{P} \frac{1}{x^2}, \, \varphi(x) \right)=
\mathcal{P} \! \! \int dx \frac{\varphi(x)-\varphi(0)}{x^2}\,.
\label{Px2_def}
\ee
Employing
(\ref{Soh_formula})
and
(\ref{Px2_def})
one can establish the familiar relation:
\be
\frac{d}{dx} \frac{1}{x \pm i 0}=
\mp i \pi \delta'(x)- \mathcal{P} \frac{1}{x^2}\,.
\label{Vlad_fla}
\ee

The formula
(\ref{Vlad_fla})
concerns the conventional generalized functions dealing with the class of test functions
defined on $(-\infty; \infty)$ and sufficiently fast decreasing at
the infinity. In our case, we have to consider a different class of
generalized functions dealing with the test functions
defined on the interval
$[A; B]$ ($A<0$, $B>0$
so that the singularity point
$x=0$
belongs to the interval).
Let
$\varphi(x)$
be a test function defined on the interval
$[A; B]$.
Then
\be
&&
\int_{A}^B  \left( \frac{d}{dx} \left( \mathcal{P} \frac{1}{x} \right), \varphi(x) \right)
\equiv
\left(  \frac{d}{dx} \left( \mathcal{P} \frac{1}{x } \right), \, \varphi(x) \right)_{[A,\,B]}
\nonumber \\ &&
=
\left.  \frac{1}{x} \varphi(x) \right|_{x= A}^{x=B}-
\lim_{\epsilon \rightarrow 0}
\left(\int_{ A}^{-\epsilon}+ \int_{\epsilon}^{B} \right)
dx \frac{\varphi'(x)}{x}.
\label{Conv_test_F}
\ee
Let us  consider the  integral term at the r.h.s of eq.~(\ref{Conv_test_F}):
\be
&&
\lim_{\epsilon \rightarrow 0}
\left(\int_{A}^{-\epsilon}+ \int_{\epsilon}^{B} \right)
dx 
\frac{\varphi'(x)}{x}
=
\lim_{\epsilon \rightarrow 0}
\left(\int_{A}^{-\epsilon}+ \int_{\epsilon}^{B} \right)
dx 
\frac{\frac{d}{dx} (\varphi(x)-\varphi(0))}{x}
\nonumber \\ &&
=\left.  \frac{1}{x}  (\varphi(x)-\varphi(0)) \right|_{x=A}^{x=B}+
\mathcal{P} \int_{A}^B dx \frac{1}{x} \frac{\varphi(x)-\varphi(0)}{x}\,.
\ee
Thus we conclude that
\be
&&
\left( \mathcal{P} \frac{1}{x^2}, \, \varphi(x) \right)_{[A,\,B]} \equiv
- \left(  \frac{d}{dx} \left( \mathcal{P} \frac{1}{x } \right), \, \varphi(x) \right)_{[A,\,B]}
\nonumber \\ &&
=
-\left. \varphi(0) \frac{1}{x} \right|_{x=A}^{x=B}
+
\mathcal{P}\int_{A}^B dx \frac{1}{x} \frac{\varphi(x)-\varphi(0)}{x}\,. \nonumber \\ &&
\ee
So finally we establish the following relation:
\be
&&
\left(\frac{1}{(x \pm i 0)^2},\, \varphi(x) \right)_{[A,B]}=
\pm i \pi \left(\delta'(x),\, \varphi(x) \right)_{[A,B]}+\left( \mathcal{P} \frac{1}{x^2}, \, \varphi(x) \right)_{[A,\,B]}
\nonumber \\ &&
= \mp i \pi \varphi'(0)+ \varphi(0) \frac{(B-A)}{AB}+
\mathcal{P}\int_{A}^B dx \frac{1}{x} \frac{\varphi(x)-\varphi(0)}{x}\,,
\ee
that represents the version of
(\ref{Vlad_fla})
adopted for the use on the finite interval.

\setcounter{equation}{0}
\section{Calculation of the convolution integrals}
\label{App_Calc_Ampl}
In the calculation of
$\re I^{(\pm,\pm)}_I(\xi)$
and
$\re I^{(-,\pm)}_{II}(\xi)$,
we encountered the following double principal value integrals
\be
&&
\mathcal{P} \!\! \int_{-1}^1 dw \, \frac{1}{(w \pm \xi)}  \mathcal{P} \!\! \int_{-1+|\xi-\xi'|}^{1-|\xi-\xi'|} dv\,
\frac{1}{(v \pm \xi') }H(w,v,\xi);
\label{Master_integral_PV1} \\ &&
\mathcal{P} \!\! \int_{-1}^1 dw \, \frac{1}{(w - \xi)^2}  \mathcal{P} \!\! \int_{-1+|\xi-\xi'|}^{1-|\xi-\xi'|} dv\,
\frac{1}{(v \pm \xi') }H(w,v,\xi).
\label{Master_integral_PV2}
\ee

We propose here a strategy of computation of these integrals once $H(w,v,\xi)$
is parameterized with the help of the spectral representation
(\ref{Spectral_for_GPDs_kappa_theta})
with the use of the factorized Ansatz
(\ref{Factorized_ansatz_xi=1}).
The procedure generalizes the  way of proceeding with the principal value
integrals when computing the real part of the elementary DVCS amplitude with GPDs parameterized
through Radyushkin's factorized Ansatz.
The following steps are to be performed:
\begin{enumerate}
\item By interchanging the order of integration in
(\ref{Master_integral_PV1}) and (\ref{Master_integral_PV2}),
$w$ and $v$ integrals may be computed using the two delta functions. One is left
with four integrations over the spectral parameters.
\item After suitable change of variables, two principal value integrations
can be performed analytically.
\item The double integration over the remaining two spectral parameters
is performed numerically. The corresponding integrands   possess
only logarithmic singularities which are perfectly integrable. In this way, we managed to
reduce the problem of performing highly singular principal value integrals
(\ref{Master_integral_PV1}),
(\ref{Master_integral_PV2})
to much less singular integration. This allows us to construct a stable and reliable
numerical procedure.
\end{enumerate}

We present below the results for the principal value integral
(\ref{Master_integral_PV1})
for the case of the factorized Ansatz (\ref{Factorized_ansatz_xi=1})
with the profile
$h(\mu,\, \lambda)$, given by eq.~(\ref{Profile_h}).
\be
&&
\mathcal{P} \!\! \int_{-1}^1 dw \, \frac{1}{(w \pm \xi)}  \mathcal{P} \!\! \int_{-1+|\xi-\xi'|}^{1-|\xi-\xi'|} dv\,
\frac{1}{(v \pm \xi') }H(w,v,\xi)  \nonumber  \\ &&
=
\frac{  60 \eta^{(\pm,\pm)}}{(1-\xi)^5} \int_{-1}^1 d \kappa \int_{- \frac{1-\kappa}{2}}^{\frac{1-\kappa}{2}}  d \theta \, \,
  Z_1 \!\! \left( a^{(\pm,\pm)}(\kappa,\,\theta,\,\xi), \,b^{(\pm,\pm)}(\kappa,\,\theta,\,\xi),\, c^{(\pm)}(\kappa,\,\xi)\right)
  V(\kappa,\,\theta).
  \nonumber  \\ &&
  \label{PV_Int_1_computed}
\ee
Here $\eta$ is the sign factor:
\be
\eta^{(\pm,+)}=1\,; \ \ \ \ \eta^{(\pm,-)}=-1\,.
\label{Def_eta}
\ee
The coefficient functions
$a^{(\pm,\pm)}$, $b^{(\pm,\pm)}$
are defined as follows:
\be
&&
a^{(\pm,+)}(\kappa,\,\theta,\, \xi) 
=\frac{1}{2} \left(\frac{1-\kappa}{2} + \theta \right) (1+\xi)\,;
\nonumber \\ &&
a^{(\pm,-)}(\kappa,\,\theta,\, \xi) 
= \frac{1}{2} \left(\frac{1-\kappa}{2} - \theta \right) (1+\xi)\,;
\label{Def_a}
\ee
\be
&&
b^{(-,+)}(\kappa, \, \theta, \, \xi) 
= -\frac{1}{2} \left(\frac{1-\kappa}{2} - \theta \right) (1+\xi)\,;
\nonumber \\ &&
b^{(-,-)}(\kappa, \, \theta, \, \xi) 
=-\frac{1}{2} \left(\frac{1-\kappa}{2} + \theta \right) (1+\xi)\,;
\nonumber \\ &&
b^{(+,+)}(\kappa, \, \theta, \, \xi) 
= 2 \xi -\frac{1}{2} \left(\frac{1-\kappa}{2} - \theta \right) (1+\xi)\,;
\nonumber \\ &&
b^{(+,-)}(\kappa, \, \theta, \, \xi) 
=2 \xi -\frac{1}{2} \left(\frac{1-\kappa}{2} + \theta \right) (1+\xi)\,.
\label{Def_b}
\ee
The first (second) sign in the indices of
$\eta^{(\pm,\pm)}$,
$a^{(\pm,\pm)}$,
$b^{(\pm,\pm)}$
corresponds to that in the
$w \pm \xi$
($v \pm \xi'$)
denominator in
(\ref{Master_integral_PV1})
respectively.
The coefficient functions
$c^{(\pm)}$
are defined as
\be
c^{(\pm)}(\kappa,\,\xi) 
= \frac{1}{2} \Big( \kappa(1+\xi)+(1-\xi) \pm 2 \xi \Big)\,.
\label{Def_c}
\ee
The sign in the index corresponds to the one in the
$w \pm \xi$
denominator of
(\ref{Master_integral_PV1}).

The explicit expression for
$Z_1(a,b,c)$
in
(\ref{PV_Int_1_computed})
reads:
\be
&&
Z_1(a,b,c)\nonumber  =
-\frac{a^3}{6}+b a^2+\frac{3 c a^2}{2}-\frac{b^2 a}{2}+\frac{3 c^2 a}{2}-\frac{b^3}{3}-\frac{c^3}{6}+b c^2-\frac{b^2 c}{2}
\nonumber \\ && +
\left(\frac{a b^2}{2}-\frac{a c^2}{2}\right) \log \left(\frac{(b-c)^2}{a^2}\right)-\left(\frac{a^2 c}{2}-\frac{b^2 c}{2}\right) \log
   \left(\frac{(a-b)^2}{c^2}\right)
\nonumber \\ &&
+
abc \left(
\frac{1}{2} \log \left(\frac{a^2}{b^2}\right) \log \left(\frac{(a-b)^2}{c^2}\right)
\right.
\nonumber \\ &&
\left.
-\log \left(1-\frac{a}{b}\right) \log \left(\frac{a^2}{b^2}\right)-2 {\rm Li}_2\left(\frac{a}{b}\right)+
\log \left(\frac{(b-c)^2}{b^2}\right) \log \left(\frac{c}{b}\right)+2  {\rm Li}_2\left(1-\frac{c}{b}\right)
\right)\,,  \nonumber \\ &&
\label{Int_dilog_master}
\ee
where $ {\rm Li}_2(z)$ is the usual dilogarithm function
\be
{\rm Li}_2(z)= - \int_0^z dz \frac{\log(1-z)}{z}\,.
\ee
One may check, that for real
$a$, $b$
and
$c$,
no imaginary part appears in
(\ref{Int_dilog_master})
as it  should be, since it is the result of integration of a real function over a real interval.
The imaginary part occurring from dilogarithms for
$\frac{a}{b}>1$
and
$\frac{c}{b}<0$
is exactly canceled by the imaginary parts stemming from logarithms in the last line of
(\ref{Int_dilog_master})
due to the well known property of dilogarithm:
\be
\im ( {\rm Li}_2(z+i 0) )= -\pi \log z \ \ \ {\rm for} \ \ \ z>1\,.
\ee

We now turn to the second principal value integral
(\ref{Master_integral_PV2}).
For the factorized Ansatz
(\ref{Factorized_ansatz_xi=1}),
with the profile
$h(\mu,\, \lambda)$,
given by
eq.~(\ref{Profile_h})
the result reads:
\be
&&
\mathcal{P} \!\! \int_{-1}^1 dw \, \frac{1}{(w - \xi)^2}  \mathcal{P} \!\! \int_{-1+|\xi-\xi'|}^{1-|\xi-\xi'|} dv\,
\frac{1}{(v \pm \xi') }H(w,v,\xi)
\nonumber \\ &&
=
\frac{  60 \eta^{(-,\pm)}}{(1-\xi)^5} \int_{\kappa_0}^1 d \kappa \int_{- \frac{1-\kappa}{2}}^{\frac{1-\kappa}{2}}  d \theta \, \,
  Z_2 \!\! \left( a^{(-,\pm)}(\kappa,\,\theta,\,\xi), \,b^{(-,\pm)}(\kappa,\,\theta,\,\xi),\, c^{(-)}(\kappa,\,\xi)\right)
  V(\kappa,\,\theta)
\nonumber \\ &&
+ \frac{  60 \eta^{(-,\pm)}}{(1-\xi)^5} \int^{\kappa_0}_{-1} d \kappa \int_{- \frac{1-\kappa}{2}}^{\frac{1-\kappa}{2}}  d \theta \, \,
  \tilde{Z}_2 \!\! \left( a^{(-,\pm)}(\kappa,\,\theta,\,\xi), \,b^{(-,\pm)}(\kappa,\,\theta,\,\xi),\, c^{(-)}(\kappa,\,\xi)\right)
  V(\kappa,\,\theta),
  \nonumber \\ &&
\ee
where
$\eta^{(-,\pm)}$,
$a^{(-,\pm)}$, $b^{(-,\pm)}$, $c^{(-)}$
are given by
(\ref{Def_eta}) -- (\ref{Def_c})
and
$\kappa_0$
is defined by the equation
\be
c^{(-)} (\kappa_0,\xi)=0: \ \ \
\kappa_0= \frac{-1+3 \xi}{1+\xi}.
\ee
The explicit expressions for
$Z_2(a,\,b,\,c)$
and
$\tilde{Z}_2(a,\,b,\,c)$
read
\be
&&
Z_2(a,\,b,\,c) \nonumber \\ &&
=-\frac{1}{2} (a-b+c) (5 a+3 b+c)  +
(a-b) c \log \left((a-b)^2\right)-\frac{a c \left(a \log \left(a^2\right)-b \log \left(b^2\right)\right)}{a-b}+\nonumber \\ && \frac{1}{2} \left(a^2-b^2\right) \log
   \left(\frac{(a-b)^2}{c^2}\right)+ a b \log \left(\frac{b^2}{c^2}\right)+b c \log \left(c^2\right)+a (b-c) \log \left(\frac{a^2 c^2}{(b-c)^4}\right)
\nonumber \\ &&
-\frac{1}{2} a (b+c) \log \left(\frac{a^2}{b^2}\right) \log \left(\frac{(b-a)^2}{c^2}\right)
\nonumber \\ &&
+ a (b+c) \left(\log \left(1-\frac{a}{b}\right) \log \left(\frac{a^2}{b^2}\right)-\log \left(\frac{c}{b}\right) \log
   \left(\left(1-\frac{c}{b}\right)^2\right)+2  {\rm Li}_2\left(\frac{a}{b}\right)-2  {\rm Li}_2\left(1-\frac{c}{b}\right)\right);
   \nonumber \\ &&
\tilde{Z}_2(a,\,b,\,c)
  =  Z_2(a,\,b,\,c) +  \left( \frac{1}{a-b}+\frac{1}{c}\right) c \left(b^2-a^2+a b \log \left(\frac{a^2}{b^2}\right) \right).
\ee

To complete the calculation of the
real and imaginary parts of $I_{II}^{(-,\pm)}$
(\ref{ReIm_I2}),
we also present the explicit expression  for
$\left( \frac{d H(w,\, \mp\xi',\xi)}{d w} \right)_{w= \xi}$
and
$
\left(
\frac{d J^{(\pm)}(w,\xi)}{dw}
\right)_{w=\xi}$
(see eq.~(\ref{Def_J})).

Using the formulas summarized in the Appendix~\ref{App_Calc_TDA}
one may check that
\be
&&
\left( \frac{d H(w,\, \mp\xi',\xi)}{d w} \right)_{w= \xi}
\nonumber \\ &&
=
\frac{1}{(1-\xi )^3} \int _{\kappa_0}^1 d\kappa  \, \int _{-\frac{1-\kappa}{2}}^{\frac{1-\kappa }{2}}
d\theta  \,
4 
V(\kappa ,\theta )
\left\{ \mp {h^{(0,1)}}\left(\frac{\kappa  (1+\xi ) -2 \xi}{1-\xi
   },\frac{\theta(1+ \xi)}{1-\xi }\right)
   \right.
   \nonumber \\ &&
   \left.
   -2  {h^{(1,0)}}\left(\frac{\kappa  (1+\xi ) -2 \xi}{1-\xi
   },\frac{\theta(1+ \xi)}{1-\xi }\right) \right\} , 
   \label{partial_der_H_w}
\ee
where
$h^{(1,0)}(\mu,\,\lambda) \equiv \frac{\partial}{\partial \mu} h(\mu,\,\lambda)$;
$h^{(0,1)}(\mu,\,\lambda) \equiv \frac{\partial}{\partial \lambda} h(\mu,\,\lambda)$
and we use the fact that $h$
vanishes at the border of its domain of definition: $h(-1, \lambda)=0$.

Finally,
\be
&&
\left(
\frac{d J^{(\pm)}(w,\xi)}{dw}
\right)_{w=\xi}
\nonumber \\ &&
=
\frac{60 \eta^{(-,\pm)}}{(1-\xi)^5}
\int_{\kappa_0}^1 d \kappa \int_{-\frac{1-\kappa}{2}}^{\frac{1-\kappa}{2}} d \theta \, V(\kappa, \theta) Z_3(
a^{(-,\pm)}(\kappa,\theta,\xi),
b^{(-,\pm)}(\kappa,\theta,\xi),
c^{(-)}(\kappa,\xi)), \nonumber \\ &&
\ee
where
\be
Z_3(a,b,c)= (a-b) (a+b+2 c)-a (b+c) \log \left(\frac{a^2}{b^2}\right)\,.
\ee


\begin{thebibliography}{99}
\bibitem{Collins:1996fb}
  J.~C.~Collins, L.~Frankfurt and M.~Strikman,
  Phys.\ Rev.\  D {\bf 56}, 2982 (1997)
  [arXiv:hep-ph/9611433].

\bibitem{Radyushkin:1997ki}
  A.~V.~Radyushkin,
  Phys.\ Rev.\  D {\bf 56}, 5524 (1997)
  [arXiv:hep-ph/9704207].

\bibitem{Frankfurt:1999fp}
  L.~L.~Frankfurt, P.~V.~Pobylitsa, M.~V.~Polyakov and M.~Strikman,
  Phys.\ Rev.\  D {\bf 60}, 014010 (1999)
  [arXiv:hep-ph/9901429].

\bibitem{Frankfurt:2002kz}
  L.~Frankfurt, M.~V.~Polyakov, M.~Strikman, D.~Zhalov and M.~Zhalov,
  arXiv:hep-ph/0211263.

\bibitem{Radyushkin:1977gp}
  A.~V.~Radyushkin,
  arXiv:hep-ph/0410276.

\bibitem{Efremov:1978rn}
  A.~V.~Efremov and A.~V.~Radyushkin,
  Theor.\ Math.\ Phys.\  {\bf 42}, 97 (1980)
  [Teor.\ Mat.\ Fiz.\  {\bf 42}, 147 (1980)].

\bibitem{Lepage:1980}
  G.~P.~Lepage and S.~J.~Brodsky, Phys. Rev. D {\bf 22}, 2157 (1980).

\bibitem{Chernyak:1983ej}
  V.~L.~Chernyak and A.~R.~Zhitnitsky,
  Phys.\ Rept.\  {\bf 112}, 173 (1984).

\bibitem{Chernyak_Nucleon_wave}
V.L. Chernyak and I.R. Zhitnitsky, Nucl. Phys. {\bf B 246}, 52 (1984).


\bibitem{Lansberg:2007ec}
  J.~P.~Lansberg, B.~Pire and L.~Szymanowski,
  Phys.\ Rev.\  D {\bf 75}, 074004 (2007)
  [Erratum-ibid.\  D {\bf 77}, 019902 (2008)]
  [arXiv:hep-ph/0701125].


\bibitem{Pire:2010if}
   B.~Pire, K.~Semenov-Tian-Shansky and L.~Szymanowski,
  Phys.\ Rev.\  D {\bf 82}, 094030 (2010)
  [arXiv:1008.0721 [hep-ph]].

\bibitem{Pire:2011xv}
B.~Pire, K.~Semenov-Tian-Shansky, L.~Szymanowski,
  Phys.\ Rev.\ D {\bf 84}, 074014 (2011)
  [arXiv:1106.1851 [hep-ph]].



\bibitem{Pire:2005mt}
  B.~Pire and L.~Szymanowski,
  PoS {\bf HEP2005}, 103 (2006)
  [arXiv:hep-ph/0509368].

\bibitem{Pire:2005ax}
  B.~Pire and L.~Szymanowski,
  Phys.\ Lett.\  B {\bf 622}, 83 (2005)
  [arXiv:hep-ph/0504255].

\bibitem{LPS}
  J.~P.~Lansberg, B.~Pire and L.~Szymanowski,
  Phys.\ Rev.\  D {\bf 76}, 111502 (2007)
  [arXiv:0710.1267 [hep-ph]].






\bibitem{Pasquini:2009ki}
  B.~Pasquini, M.~Pincetti and S.~Boffi,
  Phys.\ Rev.\  D {\bf 80}, 014017 (2009)
  [arXiv:0905.4018 [hep-ph]].

\bibitem{Burkardt:2000za}
  M.~Burkardt,
  Phys.\ Rev.\  D {\bf 62}, 071503 (2000)
  [arXiv:hep-ph/0005108].

\bibitem{Burkardt:2002ks}
  M.~Burkardt,
  Phys.\ Rev.\  D {\bf 66}, 114005 (2002)
  [arXiv:hep-ph/0209179].


\bibitem{Diehl:2002he}
  M.~Diehl,
  Eur.\ Phys.\ J.\  C {\bf 25}, 223 (2002)
  [arXiv:hep-ph/0205208].


\bibitem{Ralston:2001xs}
  J.~P.~Ralston and B.~Pire,
  Phys.\ Rev.\  D {\bf 66}, 111501 (2002)
  [arXiv:hep-ph/0110075].

\bibitem{Strikman:2009bd}
  M.~Strikman and C.~Weiss,
  Phys.\ Rev.\  D {\bf 80}, 114029 (2009)
  [arXiv:0906.3267 [hep-ph]].



\bibitem{Stefanis:1999wy}
  N.~G.~Stefanis,
  Eur.\ Phys.\ J.\ direct C {\bf 7}, 1 (1999)
  [arXiv:hep-ph/9911375].

\bibitem{Braun:1999te}
  V.~M.~Braun, S.~E.~Derkachov, G.~P.~Korchemsky and A.~N.~Manashov,
  Nucl.\ Phys.\  B {\bf 553}, 355 (1999)
  [arXiv:hep-ph/9902375].

\bibitem{DualVSRad}
 M.~V.~Polyakov and K.~M.~Semenov-Tian-Shansky,
  Eur.\ Phys.\ J.\  A {\bf 40}, 181 (2009)
  [arXiv:0811.2901 [hep-ph]].
\bibitem{arXiv:1001.2711}
  K.~M.~Semenov-Tian-Shansky,
  Eur.\ Phys.\ J.\ A\ {\bf 45}, 217  (2010)
  [arXiv:1001.2711 [hep-ph]].


\bibitem{hep-ph/0509204}
  D.~Mueller and A.~Schafer,
  Nucl.\ Phys.\ B\ {\bf 739}, 1  (2006)
  [hep-ph/0509204].

\bibitem{Polyakov:1999gs}
  M.~V.~Polyakov and C.~Weiss,
  Phys.\ Rev.\  D {\bf 60}, 114017 (1999)
  [arXiv:hep-ph/9902451].

\bibitem{RDDA1}
  A.~V.~Radyushkin,
  Phys.\ Rev.\  D {\bf 56}, 5524 (1997)
  [arXiv:hep-ph/9704207].
%
\bibitem{RDDA2}
  A.~V.~Radyushkin,
  Phys.\ Lett.\  B {\bf 449}, 81 (1999)
  [arXiv:hep-ph/9810466].
%
\bibitem{RDDA3}
A.~V.~Radyushkin,
Phys.\ Rev.\  D {\bf 59}, 014030 (1999)
[arXiv:hep-ph/9805342].
\bibitem{RDDA4}
  I.~V.~Musatov and A.~V.~Radyushkin,
  Phys.\ Rev.\  D {\bf 61}, 074027 (2000)
  [arXiv:hep-ph/9905376].



\bibitem{Polyakov:1998ze}
  M.~V.~Polyakov,
  Nucl.\ Phys.\  B {\bf 555}, 231 (1999)
  [arXiv:hep-ph/9809483].


\bibitem{Pobylitsa:2001cz}
  P.~V.~Pobylitsa, M.~V.~Polyakov and M.~Strikman,
  Phys.\ Rev.\ Lett.\  {\bf 87}, 022001 (2001)
  [arXiv:hep-ph/0101279].

\bibitem{BLP1}
  V.~M.~Braun, D.~Y.~Ivanov, A.~Lenz and A.~Peters,
  Phys.\ Rev.\  D {\bf 75}, 014021 (2007)
  [arXiv:hep-ph/0611386].
\bibitem{BLP2}
V.~M.~Braun, D.~Y.~Ivanov and A.~Peters,
  Phys.\ Rev.\  D {\bf 77}, 034016 (2008).




\bibitem{Kivel:2002ia}
  N.~Kivel and M.~V.~Polyakov,
  arXiv:hep-ph/0203264.

\bibitem{Mankiewicz:1997uy}
  L.~Mankiewicz, G.~Piller, T.~Weigl,
  Eur.\ Phys.\ J.\  {\bf C5}, 119-128 (1998).
  [hep-ph/9711227].


\bibitem{arXiv:0805.0152}
  K.~Kumericki, D.~Mueller and K.~Passek-Kumericki,
  Eur.\ Phys.\ J.\ C\ {\bf 58}, 193  (2008)
  [arXiv:0805.0152 [hep-ph]].


\bibitem{GPV} K.Goeke, M.V.Polyakov and M.Vanderhaeghen, Progr. Part. Nucl.
Phys. Vol.47, No 2,  401-515 (2001) [arXiv:hep-ph/0106012].




\bibitem{Vladimirov}
V.~S.~Vladimirov,  {\em Equations of mathematical physics} , MIR, Moscow  (1984)  (Translated from Russian).





\bibitem{Chernyak:1987nv}
  V.~L.~Chernyak, A.~A.~Ogloblin and I.~R.~Zhitnitsky,
  Z.\ Phys.\  C {\bf 42}, 583 (1989)
  [Yad.\ Fiz.\  {\bf 48}, 1398 (1988)]
  [Sov.\ J.\ Nucl.\ Phys.\  {\bf 48}, 889 (1988)].


\bibitem{King:1986wi}
  I.~D.~King and C.~T.~Sachrajda,
  Nucl.\ Phys.\  B {\bf 279}, 785 (1987).

\bibitem{Gari:1986ue}
  M.~Gari and N.~G.~Stefanis,
  Phys.\ Lett.\  B {\bf 175}, 462 (1986).


\bibitem{EricsonWeise}
T.~Ericson, W.~Weise, {\it Pions and Nuclei}
(Clarendon Press, Oxford, 1988).


\bibitem{Kroll:1995pv}
  P.~Kroll, M.~Schurmann, P.~A.~M.~Guichon,
  Nucl.\ Phys.\  {\bf A598}, 435-461 (1996).
  [hep-ph/9507298].

\bibitem{Hand_9388}
  L.~N.~Hand,
  Phys.\ Rev.\ {\bf 129}, 1834  (1963).


\bibitem{Lansberg:2010mf}
  J.~P.~Lansberg, B.~Pire, L.~Szymanowski,
  J.\ Phys.\ Conf.\ Ser.\  {\bf 295}, 012090 (2011).
  [arXiv:1011.6635 [hep-ph]].











\bibitem{Boer:2011fh}
  D.~Boer {\it et al.},
polarization,
  arXiv:1108.1713 [nucl-th].





\bibitem{arXiv:0709.1946}
  K.~Park {\it et al.} [CLAS Collaboration],
  Phys.\ Rev.\ C\ {\bf 77}, 015208  (2008)
  [arXiv:0709.1946 [nucl-ex]].


\bibitem{J-lab}
K.~Park,  {\it private communication}; V.~Kubarovsky, P.~Stoler,  {\it private communication}.







\end{thebibliography}
\end{document}